\documentclass[fleqn,usenatbib]{mnras}

\usepackage{amsmath}	
\usepackage{amssymb} 
\usepackage{txfonts}
\usepackage{mathptmx}
\usepackage{mathrsfs}
\usepackage[T1]{fontenc}
\usepackage{ae,aecompl}

\usepackage{xcolor}
\usepackage{graphicx}	

\title[Photometric Reverberation Mapping]{Modeling photometric reverberation mapping data for the next generation of big data surveys. Quasar accretion disks sizes with the LSST}

\author[F. Pozo Nu\~nez et al.]{
F. Pozo Nu\~nez$^{1}$\thanks{E-mail: francisco.pozonunez@h-its.org}\href{https://orcid.org/0000-0002-6716-4179}{\includegraphics[scale=0.9]{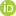}},
C. Bruckmann$^{2}$,
S. Desamutara$^{1}$,
B. Czerny$^{3}$\href{https://orcid.org/0000-0001-5848-4333}{\includegraphics[scale=0.9]{orcidicon.png}},
S. Panda$^{4}$\thanks{CNPq fellow}\href{https://orcid.org/0000-0002-5854-7426}{\includegraphics[scale=0.9]{orcidicon.png}},
\newauthor ~A.P. Lobban$^{5}$, G. Pietrzy\'{n}ski$^{6}$, and K. L. Polsterer$^{1}$\\
$^{1}$Astroinformatics, Heidelberg Institute for Theoretical Studies, Schloss-Wolfsbrunnenweg 35, 69118 Heidelberg, Germany\\
$^{2}$I. Physikalisches Institut, Universit\"at zu K\"oln, Z\"ulpicher Str. 77, D-50937 K\"oln, Germany\\
$^{3}$Center for Theoretical Physics, Polish Academy of Sciences, Al. Lotnik\'ow 32/46, 02-668 Warsaw, Poland\\
$^{4}$Laborat\'{o}rio Nacional de Astrof\'{i}sica, MCTIC, Rua dos Estados Unidos, 154, Bairro das Na\c c\~oes, Itajub\'a, MG 37501-591, Brazil\\
$^{5}$European Space Agency (ESA), European Space Astronomy Centre (ESAC), E-28691 Villanueva de al Ca\~nada, Madrid, Spain\\
$^{6}$Centrum Astronomiczne im. Mikolaja Kopernika, Polish Academy of Sciences, Bartycka 18, 00-716 Warsaw, Poland
\\
}

\date{Accepted 2023 January 23. Received 2022 December 17; in original form 2022 October 14}

\pubyear{2023}

\begin{document}
\label{firstpage}
\pagerange{\pageref{firstpage}--\pageref{lastpage}}
\maketitle


\begin{abstract}

Photometric reverberation mapping can detect the radial extent of the accretion disc (AD) in Active Galactic Nuclei by measuring the time delays between light curves observed in different continuum bands. Quantifying the constraints on the efficiency and accuracy of the delay measurements is important for recovering the AD size-luminosity relation, and potentially using quasars as standard candles.  We have explored the possibility of determining the AD size of quasars using next-generation Big Data surveys. We focus on the Legacy Survey of Space and Time (LSST) at the Vera C. Rubin Observatory, which will observe several thousand quasars with the Deep Drilling Fields and up to 10 million quasars for the main survey in six broadband filter during its 10-year operational lifetime. We have developed extensive simulations that take into account the characteristics of the LSST survey and the intrinsic properties of the quasars. The simulations are used to characterise the light curves from which AD sizes are determined using various algorithms. We find that the time delays can be recovered with an accuracy of 5 and 15\% for light curves with a time sampling of 2 and 5 days, respectively. The results depend strongly on the redshift of the source and the relative contribution of the emission lines to the bandpasses. Assuming an optically thick and geometrically thin AD, the recovered time-delay spectrum is consistent with black hole masses derived with 30\% uncertainty.

\end{abstract}




\section{Introduction}

Quasars are known to be powered by the accretion of matter onto supermassive black holes (SMBHs) with variable luminosity on different time scales and wavelengths. 
Because their inner regions are extremely compact (ranging from a few to several hundred light-days), they cannot be resolved with conventional imaging techniques, so indirect methods are needed to study their physics.

Photometric reverberation mapping (PRM; \citealt{1973ApL....13..165C}; \citealt{2011A&A...535A..73H}; \citealt{2012A&A...545A..84P}; \citealt{2012ApJ...747...62C}) of active galactic nuclei (AGN\footnote{In the course of this work, we will use the terms AGN and quasar generically for all galaxies hosting a supermassive accreting black hole.}) uses a combination of narrow- and broad-band observational data to estimate the time delay $\tau$ between the triggering continuum variations of the accretion disc (AD) and the response of the broad line region (BLR) emission lines (e.g., H$\beta$, H$\alpha$). 
The time delay leads to an estimate of the effective BLR radius, $R_{\rm BLR} = \tau\ c$ ($c$ is the speed of light). 
Since the BLR gas clouds are in virialised motion around the black hole (\citealt{1988ApJ...325..114G}; \citealt{1989ApJ...345..637K}; \citealt{1991ApJ...371..541K}; 
\citealt{1999ApJ...521L..95P}; \citealt{2001ApJ...551...72K}; \citealt{2002ApJ...572..746O}; \citealt{2003A&A...407..461K}), the BLR radius together with the velocity dispersion $\sigma_{v}$ of the emitting gas allows the estimation of the black hole masses $M_{\rm BH}$ by the virial product $M_{\rm BH} = f\ R_{\rm BLR}\ \sigma_{v}^{2}/G$, where $G$ is the gravitational constant and the factor $f$ depends on the kinematics and geometric distribution of the BLR clouds (e.g. \citealt{1977SvAL....3....1D};  \citealt{1988ApJ...325..114G}; \citealt{2004ApJ...615..645O}; \citealt{2021arXiv210310961W}).

PRM was able to estimate the BLR size and black hole masses in several local Seyfert-1 galaxies (\citealt{2012A&A...545A..84P,2013A&A...552A...1P,2015A&A...576A..73P}; \citealt{2012ApJ...756...73E}; \citealt{2015A&A...581A..93R,2018A&A...620A.137R}; \citealt{2019ApJ...884..103K}) with accuracy comparable to spectroscopic reverberation mapping (SRM) of the BLR (\citealt{1982ApJ...255..419B}; \citealt{1986ApJ...305..175G}; \citealt{1993PASP..105..247P}; \citealt{2012ApJ...755...60G}). 
The feasibility of PRM has also been tested recently in high-redshift quasars (\citealt{2020MNRAS.492.3940R}). 
Starting from feature-rich and well-sampled light curves, PRM allows us to infer the basic geometry of the BLR, i.e. whether it is spherical or disc-shaped, by analysing the RM transfer functions (\citealt{1991ApJ...379..586W}; \citealt{2003SPIE.4854..262H}), thus constraining the geometric factor $f$ needed to convert the time delay and velocity width into the black hole mass (e.g. \citealt{2011ApJ...730..139P,2012ApJ...754...49P}; \citealt{2014A&A...568A..36P}).

One of the most important discoveries from the RM campaigns is the correlation between the H$\beta$ BLR size with the optical luminosity ($R_{\rm{BLR}} \propto L_{\rm{AGN}}^{\alpha}$). 
This correlation was first assumed and used by \cite{1977SvAL....3....1D} who adopted a value of $\alpha = 0.33$. Assuming that the continuum shape and density stay the same, photoionization models predict $\alpha = 0.5$ (\citealt{1979RvMP...51..715D}; \citealt{1990agn..conf...57N}). From RM \cite{1991ApJ...370L..61K} found empirically that $\alpha$ was consistent with 0.5. The study by \cite{2000ApJ...533..631K} from a sample of 17 low redshift quasars ($z < 0.3$) suggested $\alpha = 0.7$. More recent, larger reverberation mapped samples give $\alpha$ close to 0.5 (e.g., \citealt{2009ApJ...705..199B}; \citealt{2013ApJ...767..149B}) for H$\beta$ and the optical flux.

Similarly relevant to the BLR-luminosity relation is the preliminary demonstration of a continuum AD time-delay-luminosity relation for low-luminosity AGNs (\citealt{2005ApJ...622..129S}). In this case, the time delays are attributed to the light travel time across different regions of the disc (\citealt{1998ApJ...500..162C}). 
Consequently, AD time delay can indicate, to a first approximation, the physical size of the continuum-emitting region. 
Unfortunately, this relation has not yet been fully explored, mainly due to observational limitations, as the time delays of AD are about a factor of 10 smaller than the BLR sizes. Therefore, light curves with very high cadence are required to map AD with sufficient accuracy. 
The demonstration of the relation also for high redshift quasars could offer new possibilities for the use of quasars as efficient standard cosmological candles up to high $z$ (see the review by \citealt{2022arXiv220906563C}).
In this context, the main advantage over measuring the BLR response is that the AD sizes could be determined on shorter timescales (of only a few years) for larger samples of objects. 
For example, the standard theory of AGN accretion discs (\citealt{1969Natur.223..690L}; \citealt{1972A&A....21....1P}; \citealt{1973A&A....24..337S}; \citealt{1973blho.conf..343N}) predicts a temperature profile $T\propto R^{-3/4}$ where the delays between the bands of the UV/optical continuum $\tau_{c}$ correspond to $R(\lambda)\propto \tau_{c}(\lambda)\propto \lambda^{4/3}$ (\citealt{1998ApJ...500..162C}; \citealt{2005ApJ...622..129S}; \citealt{2007MNRAS.380..669C}). 
The predicted AD spectrum follows $f_{\nu}\propto\tau_{c}(\lambda)^2\cos{i}/\lambda^{3}D^{2}$, where $D$ is the distance to the AGN and $i$ is the disc inclination with respect to the observer. 
Thus, the redshift-independent distance to the AGN is $D\propto\tau_{c} f_{\nu}^{1/2}\cos{i}^{1/2}/\lambda^{3/2}$ and the Hubble constant $H_{0}\propto 1/D$ (\citealt{1992ARA&A..30..499C}).

\begin{figure*}
  \centering
  \includegraphics[width=15cm,clip=true]{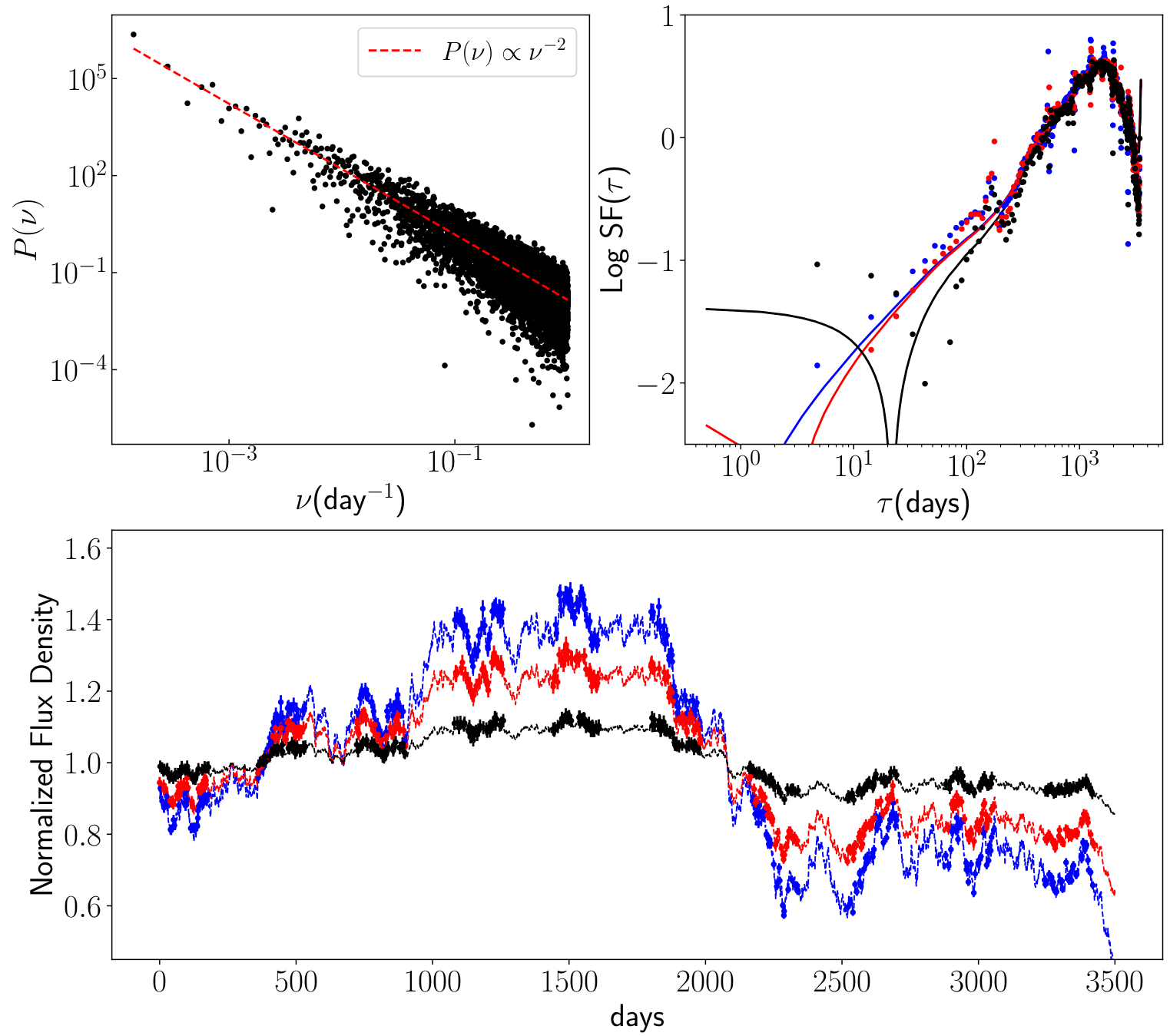}
  \caption{Examples of simulated light curves using a CAR(1) random walk process with a power spectrum of the form $P(\nu)\propto \nu^{-2}$ (top left). The light curves are characterized by a SF with log\,$SF(\infty)\sim0.3$, and timescale $\tau_{\rm char} \sim 200$ days (top right). The blue and red dotted lines correspond to a light curve for a quasar located at $z=0.5$, and observed through the $u$ and $y$ filters respectively. For illustration, we show how the variability amplitude drastically changes if the quasar is located at $z=3.0$ (black dotted line) in the $u$-band. The light curves have been randomly sampled (dots) considering an average time sampling of 5 days and including a seasonal gap of 180 days. This particular example shows the light curves with a photometric signal-to-noise ratio of S/N $=60$, which correspond to a measurement uncertainties at the $\sim2$\% level. The S/N calculations have been performed using the total source counts, sky background contamination, instrumental noise and gain. More information about the LSST limiting magnitudes and S/N can be found in \url{https://smtn-002.lsst.io/}.}
\label{lc_xrw}
\end{figure*}

However, there are some issues that need further investigation before this relationship can be used in practise. For example, various techniques involving microlensing, spectroscopic and photometric RM have measured AD sizes which are a factor $\sim3$ larger than those expected by SS73 theory (e.g. \citealt{2006ApJ...648...67P}; \citealt{2015ApJ...806..129E}; \citealt{2016ApJ...821...56F}; \citealt{2016AN....337..356C}; \citealt{2018ApJ...857...53C}; \citealt{2020MNRAS.494.1165L}). 
The contribution of emission lines to the broadband PRM data, together with the diffuse continuum emission from the BLR (\citealt{2001ApJ...553..695K}; \citealt{2019NatAs...3..251C}; \citealt{2019MNRAS.489.5284K}; \citealt{2022MNRAS.509.2637N}), has been suggested to be the main source of contamination that could bias the time lags to larger values. 
Other models, such as slim discs (\citealt{1988ApJ...332..646A}; \citealt{2019Univ....5..131C}), clumpy discs (\citealt{2011ApJ...727L..24D}) and non-blackbody discs (\citealt{2018ApJ...854...93H}) have been explored to reconcile AD sizes with the larger observed values. 
These models have different predictions for the growth rate of SMBHs and the influence of AD radiation on their environment (e.g. due to winds), and yet they have difficulty explaining the observed increase in time delay from the UV to the optical wavelength range. As an alternative, \cite{2017MNRAS.467..226G} suggested that a correction for internal exctinction would lead to an increase in AGN luminosity by a factor of 4 and 10 in the optical and UV, respectively, which would explain the discrepancy between the measured AD sizes and the prediction of the standard theory. This was recently confirmed by \cite{2022arXiv220811437G} for the well-studied AGN NGC5548, where the AD is greatly underestimated by a factor of 2.6 without taking reddening into account.
Models of the X-ray illumination of AD could also explain the observed larger delays for certain cases where the corona is located at a distance of more than $\sim40$ gravitational radii above the black hole (\citealt{2022A&A...666A..11P}).

New techniques using narrowband PRM have been introduced to measure the time delay in the optical continuum, which is less affected by the BLR contribution (\citealt{2017PASP..129i4101P,2018rnls.confE..56P}). 
For instance, the PRM results suggest that an underestimation of the black hole masses caused by the unknown BLR geometry scaling factor could easily compensate for the difference between the observations and the AD theory (e.g., for Mrk509; \citealt{2019MNRAS.490.3936P}). 
PRM is therefore an efficient and powerful method for deriving $\tau_{c}$ for a large number of objects. 
It can be applied in the next generation of large-data photometric surveys, especially in view of the upcoming public 8.4-meter optical/near-infrared LSST (e.g., \citealt{2020A&A...636A..52C}; \citealt{2021MNRAS.505.5012K,2022ApJS..262...49K})

With a field of view of 9.6 deg$^{2}$, the future LSST will cover about 20000 deg$^{2}$ of sky with about 200 visits for each $ugrizy$ photometric band and will provide a full sample of light curves for about 16 million AGNs with $i_{mag} \leq 26.25$ and redshift $z < 6.5$ (\citealt{2009arXiv0912.0201L}). 
The LSST observation sequence allows the study of long-term variability up to 10 years, the total duration of the survey (\citealt{2019ApJ...873..111I}). 
Using the luminosity function of \cite{2009AJ....138..305J} and the density distribution of quasars of \cite{2006AJ....132..117F}, LSST predicts the identification of up to 10000 quasars at very high redshift ($6.5 < z < 7.5)$ in the y-band. 
Quasars at redshift $z\sim7.5$ will be detected as z-band dropouts and followed up by ground-based optical spectroscopy (\citealt{2009arXiv0912.0201L}).

Due to the large amount of photometric data that LSST will provide, it is essential to investigate the effectiveness of the PRM method. 
In this work, we focus our attention on how to derive the size of quasar accretion discs using PRM time delay measurements.

\section{QUASAR LIGHT CURVE SIMULATIONS}

In this section, we simulate the AD continuum light curves by considering the characteristics of the LSST survey and the intrinsic properties of quasars at different redshifts. 
The LSST properties include (i) the total duration of the survey, (ii) the time sampling, and (iii) the data quality. 
The intrinsic properties of quasars include (i) variability and (ii) the dependence of luminosity on redshift. In the course of this work we assume a concordance cosmology with ${H_{0}=70\ \mathrm{km\ s^{-1}\ Mpc^{-1}}}$, $\Omega_{\Lambda}=0.73$ and $\Omega_{m}=0.27$.
   
\subsection{The driving AD variability}\label{ranwalksec}

According to the AD reprocessing model (e.g. \citealt{2002apa..book.....F}; \citealt{2007MNRAS.380..669C}; \citealt{2018MNRAS.480.2881M}), the UV/optical variations are driven by an X-ray emitting corona located at about 10$r_{g}$ (e.g., \citealt{2012ApJ...756...52M}; \citealt{2016MNRAS.458..200W}), where $r_{g}$ is the gravitational radius of the black hole $r_{g} = GM_{\rm BH}/c^2$. 
This has been confirmed by observations of time delays between X-rays\footnote{It remains unclear whether soft/hard X-ray emission or far-UV emission is responsible for driving the variability observed in the innermost part of the AD (e.g., \citealt{2012ApJ...760...73L}; \citealt{2017MNRAS.470.3591G}). It has been shown that the X-rays do not correlate with the UV-optical variations and that the extreme UV has enough power to drive the variations at longer wavelengths (see Figure 2 in \citealt{2008RMxAC..32....1G} and discussion in \citealt{2003A&AT...22..661G}). However, for the purposes of our simulations, the effect of this assumption is negligible.} to optical wavelengths in multiple objects (e.g., \citealt{2014MNRAS.444.1469M}; \citealt{2019ApJ...870..123E}).
The variable X-ray emission has a stochastic or chaotic nature, which could be related to the presence of turbulent magnetic fields that produce fluctuations in the AD radiation. 
This stochastic input mechanism leads to optical light curves of quasars that are often well described by a random walk process with a power spectral density (PSD) $P(\nu)\propto \nu^{-\alpha}$ and $\alpha \sim2$ (\citealt{1967AJ.....72.1341K}; \citealt{1999MNRAS.306..637G}; \citealt{2001ApJ...555..775C}; \citealt{2007A&A...462..581H}; \citealt{2017ApJ...834..111C}) corresponding to a red noise spectrum. 

Such lightcurves can be modelled working in frequency domain or in time domain. The use of the frequency domain as a starting point is useful since we have observational determination of the PDS for a number of sources, in X-ray and in the optical band, and the power spectra have complex shapes. A method to construct the lightcurve in the time domain from the power spectrum has been introduced by \citet{1995A&A...300..707T} (hereafter TK95). It can be used to generate synthetic light curves with any PSD model. 
According to TK95, the discrete Fourier transform of a time series $x(t)$ can be expressed as a complex Gaussian random variable,

\begin{align}
  \hat{x}(\nu) = \mathscr{F}{(x(t))} &= \frac{1}{\sqrt{N}} \left[ \sum_{t} x(t)\cos{(2\pi\nu t)} + i  \sum_{t} x(t)\sin{(2\pi\nu t)}  \right] \nonumber \\ 
  &= \sqrt{\frac{P(\nu)}{2}}\left[\mathscr{N}(0, 1) + i\mathscr{N}(0, 1)\right]
\label{eq:tk95}
\end{align}
where $P(\nu)$ is the scaling PSD model. We assume here a power law $\nu^{-\alpha}$ with $\alpha = 2.0$, which corresponds to a random walk process. 
The time series $x(t)$ is obtained from the inverse Fourier transform of Eq~[\ref{eq:tk95}], $x(t) = \mathscr{F}^{-1}{(\hat{x}(\nu))}$, where the Fourier components for negative frequencies were chosen to be complex conjugates of the positive ones, $\hat{x}(-\nu) = \hat{x}(\nu)^{\ast}$, and $x(t)$ is normalised to a mean flux and variance (see section~\ref{advar}) that depend on the bolometric luminosity of the quasar ($L_{\rm bol}$) and the redshift ($z$).

An alternative approach to modelling quasar variability is to work directly in the time domain. The method might seem simpler but confronting such models with the actual data and their statistical properties is not simple, apart from the simplest cases. Such a simple case - the first-order stochastic continuous-time autoregressive process (CAR (1))  was proposed by \cite{2009ApJ...698..895K}. This process is a special case of a more general class of continuous-time autoregressive moving average (CARMA) models (\citealt{brockwell02}; \citealt{2014ApJ...788...33K}). 
The process CAR (1) is consistent with a PSD of the form $P(\nu)\propto \nu^{-2.0}$ and has the advantage of including characteristic time scales $\tau_{\rm{char}}$ directly related to the physical processes in the AD\footnote{An exponential decay time scale has been used, for example, by \cite{1987ApJS...65....1G} to model AGN light curves and quantify the accuracy of cross-correlation estimates.}. In the TK95 the corresponding parameter is the frequency break, i.e. $\tau_{\rm{char}}^{-1}$.
Following \cite{2009ApJ...698..895K}, a CAR (1) process is defined as the solution of the stochastic differential equation

\begin{align}
    dx(t) = -\frac{1}{\tau_{\rm{char}}} x(t) dt + \sigma \sqrt{dt} \epsilon(t) + b\ dt
    \label{eq-car1}
\end{align}
where the time series $\epsilon(t)$ correspond to a white noise process\footnote{Similarly to the TK95 method (equation~[\ref{eq:tk95}]), the white noise process is assumed to be Gaussian $\sim \mathscr{N}(0, 1)$.} with a mean of zero and a variance of 1. 
In this context, $x(t)$ is the light curve of the quasar characterised by the time scale or "relaxation time" $\tau_{\rm{char}}$ with a mean $b\tau_{\rm{char}}$ and a variance $\tau_{\rm{char}}\sigma^{2}/2$. 
The value of $\tau_{\rm{char}}$ is often constrained by the power law section of the light curve structure function (SF; \citealt{1992ApJ...396..469H}). 
The SF measures the mean variability amplitude of $x(t)$ within measurements separated by a time interval $\tau = t_{j} - t{i}$

\begin{align}
  \mathcal{S(\tau)} = \frac{1}{N(\tau)} \sum_{i<j} \left[ x(t_j) -
      x(t_{i})\right]^2
\label{equ:sf}
\end{align}
for which correlated variations are characterized by the power law section between the interval $\tau_{min} \le \tau \le \tau_{max}$ with $\tau_{\rm{char}} \sim \tau_{max}$. 
It has been observed that the variability timescale $\tau_{\rm{char}}$ correlates with the mass of the black hole and luminosity (\citealt{2009ApJ...698..895K}), as expected from theoretical models of orbital or accretion disk thermal timescales (e.g. \citealt{1998ApJ...504..671K}; \citealt{1999ApJ...514..682E}).

In this work we quantify $\tau_{\rm{char}}$ based on the procedure used by \cite{2001ApJ...555..775C} which uses theoretical SF derived from symmetric triangular flares profiles (\citealt{2000ApJ...544..123C}).
Examples of simulated time series $x(t)$ with a PSD corresponding to a random-walk process are shown in Figure~\ref{lc_xrw}.

\subsection{AD spectrum and variability luminosity}\label{advar}

The total energy $\mathcal{E}_{T}$ radiated by an optically thick and geometrically thin AD, is given by the corresponding blackbody radial temperature profile $T(r)$ across the disc $T(r)^{4}=\mathcal{E}_{T}(r)/\sigma = [\mathcal{E_{V}}(r) + \mathcal{E_{I}}(r)]/\sigma$ (\citealt{1998ApJ...500..162C}; \citealt{2002apa..book.....F}; \citealt{2007MNRAS.380..669C}). 
The mechanism of energy production is achieved by a viscous heating process with an energy of $\mathcal{E_{V}}(r)=3GM_{\rm BH}\dot M/8\pi r^3$, and due to irradiation by the external UV/X rays emitting corona $\mathcal{E_{I}}(r)=L_{*}(1 - a)/4\pi r^3 H_{*} \cos{\theta}$ with bolometric luminosity $L_{*}$. 
In this model, the corona is at a height $H_{*}$ along the rotation axis of the black hole, where $a$ is the albedo and $\theta$ is the angle between the surface normal of the disc and the incident radiation (\citealt{2005ApJ...622..129S}; \citealt{2013peag.book.....N}). 
The emitted spectrum $F_{\nu}$ can be calculated by integration from the radius of the innermost stable circular orbit $r_{\rm in}\sim 6R_{g}$ to an outer radius $r_{\rm out}$, and considering the radial temperature profile $T(r)\propto (M_{\rm BH} \dot M)^{1/4}r^{-3/4}$ (Figure~\ref{model}), $F_{\nu} = \int_{r_\mathrm{in}}^{r_\mathrm{out}} B_\nu(T(r)) \mathrm{d}\Omega$, where $B_\nu(T(r))$ is the Planck function of a black body at temperature $T(r)$, $\mathrm{d}\Omega = 2 \pi r \mathrm{d}r \cos{i}/ D^{2}$ is the solid angle, subtended by a ring between radii $r$ and $r + \mathrm{d}r$, and $i$ is the inclination of the disc seen by an observer at distance $D$ (see \citealt{2002apa..book.....F} Sect.5.5 for a complete derivation).

The variability amplitude of a quasar depends on the luminosity of the quasar, the redshift, the wavelength and the time span of the observing programme (e.g. \citealt{1996MNRAS.282.1191C}; \citealt{2009ApJ...698..895K}; \citealt{2014ApJ...784...92M}; \citealt{2017ApJ...834..111C}). 
Light curves for more luminous high-$z$ objects have smaller standard deviations ($\sigma$) compared to less luminous Seyfert galaxies\footnote{Such an anti-correlation with luminosity should also be expected for black hole masses, but the relationship remains unclear (see \citealt{2017ApJ...834..111C} and references therein).}. 
We account for this effect in our simulations using the quasar variability- luminosity relation of \cite{2014ApJ...784...92M} obtained with data from the Panoramic Survey Telescope and Rapid Response System (Panstarrs) and the Sloan Digital Sky Survey (SDSS)\footnote{This equation does not account for the effect of decreasing variability with increasing Eddington ratio (i.e. in narrow-line Seyfert 1s) as found by \cite{2004ApJ...609...69K}.}

\begin{eqnarray}
\sigma(L,\lambda,z,t)&=&0.079\left(1+z\right)^{0.15}\left(\frac{L}{10^{46}\,\rm{erg}\,\rm{s}^{-1}}\right)^{-0.2}\nonumber\\
& \times &\left(\frac{\lambda}{1000\,\rm{nm}}\right)^{-0.44} \left(\frac{t}{1\,\rm{year}}\right)^{0.246}
\end{eqnarray}
with $z$ the redshift, $L$ the bolometric luminosity, $\lambda$ the central wavelength of the filter used and $t$ the time span of the observations. 
We take a bolometric luminosity correction $L_{\rm Bol} = 10\lambda L_{\lambda}(5100$\AA) (\citealt{2000ApJ...533..631K}; \citealt{2004MNRAS.352.1390M}) and account for contamination by a host galaxy in the photometric aperture using the flux variation gradient method (\citealt{1981AcA....31..293C}; \citealt{1992MNRAS.257..659W}; \citealt{2014A&A...568A..36P}; \citealt{2022A&A...657A.126G}). 
The contamination from the host galaxy contributes up to $\sim50$\% and $\sim20$\% of the total luminosity for local and high-redshift quasars\footnote{We note that the contamination of BLR emission can affect the linear relation between the fluxes as shown in \cite{2018A&A...620A.137R}. For our simulations we have assumed that BLR emission in the bandpass is negligible or that it can be accounted for when applying the flux variation gradient method (see \citealt{2023MNRAS.518..418H}).}, respectively.
The nuclear extinction caused by the internal reddening of the AGN (\citealt{2004ApJ...616..147G}; \citealt{2007arXiv0711.1013G}) is taken into account by assuming the reddening curve of \cite{2007arXiv0711.1013G}. 

\begin{figure}
  \centering
  \includegraphics[width=\columnwidth]{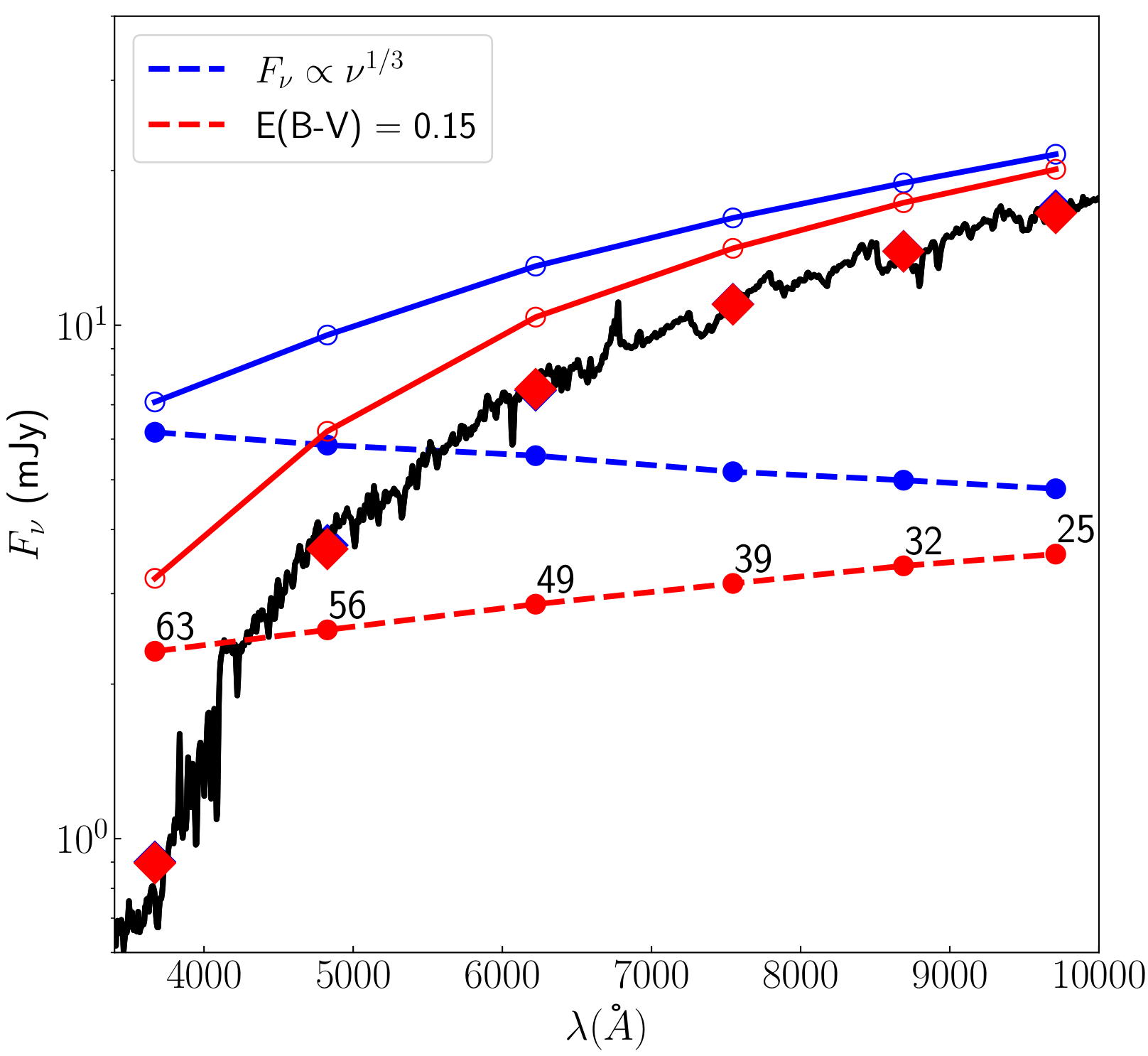}
\caption{Simulated host galaxy and extinction correction using the FVG method. The total fluxes in the photometric aperture are shown with open circles for an unreddened (blue) and reddened (red) AGN spectrum. The black solid line is the spectrum of the Sa-type host galaxy (\citealt{1996ApJ...467...38K}) used for the simulations. The diamonds show the determined mean contribution of the host galaxy for each filter combination. The blue and red filled circles show the unreddened and reddened AGN spectra, respectively, after subtracting the galaxy contribution. The numbers above the red circles indicate the percentage of underestimation of the flux when reddening is not taken into account. In this particular example, a reddening of $E(B-V) = 0.15$ was used. The FVG diagrams and light curves for a S/N 100 and a sampling of 3 days can be found in Figure \ref{fvgext} in the Appendix.}
\label{fvghost}
\end{figure}

An example of our flux decomposition procedure is shown in Figure~\ref{fvghost}.

\subsection{AD time delays}\label{adtimedel}

\begin{figure}
  \centering
  \includegraphics[width=\columnwidth]{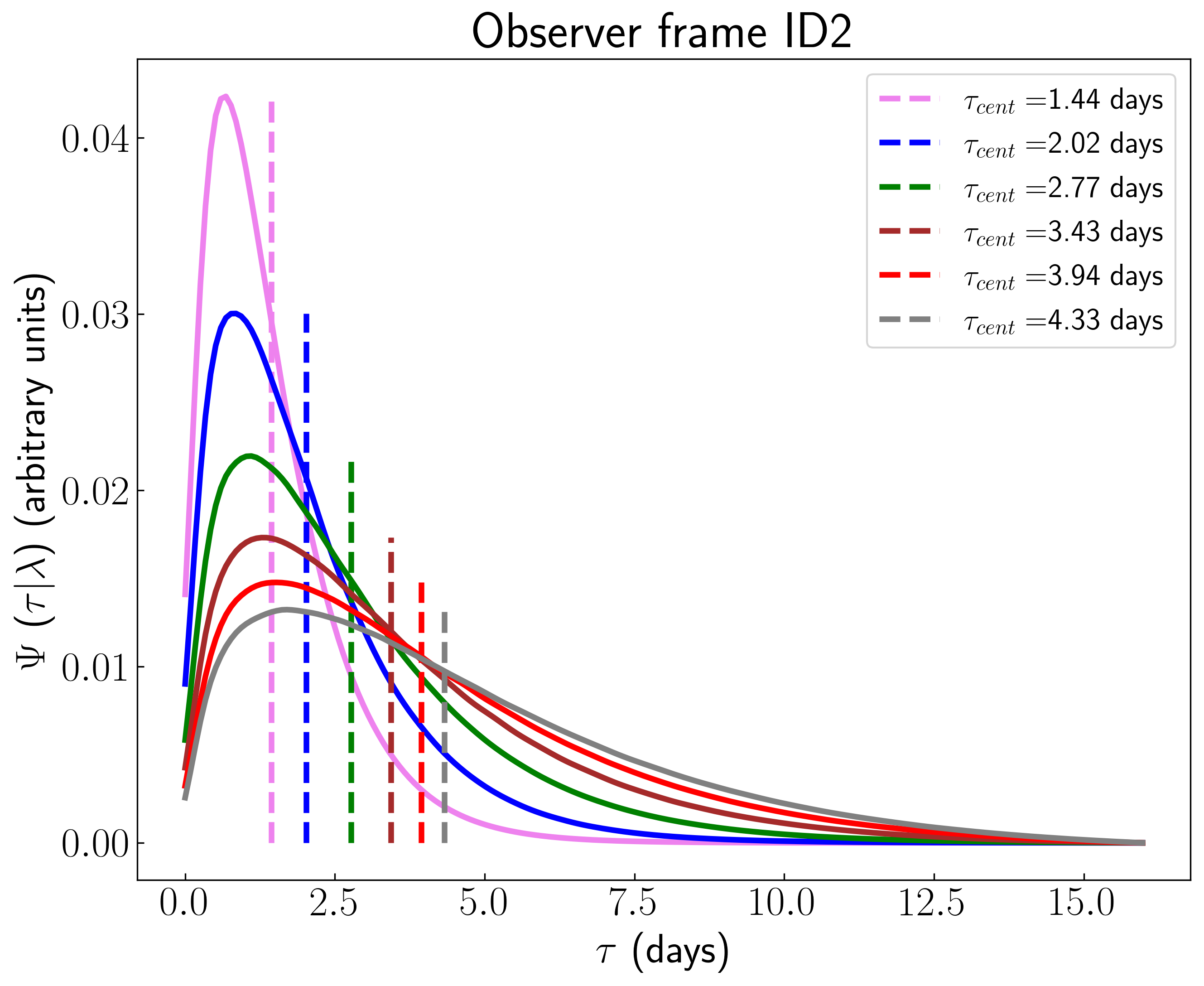}
  \includegraphics[width=\columnwidth]{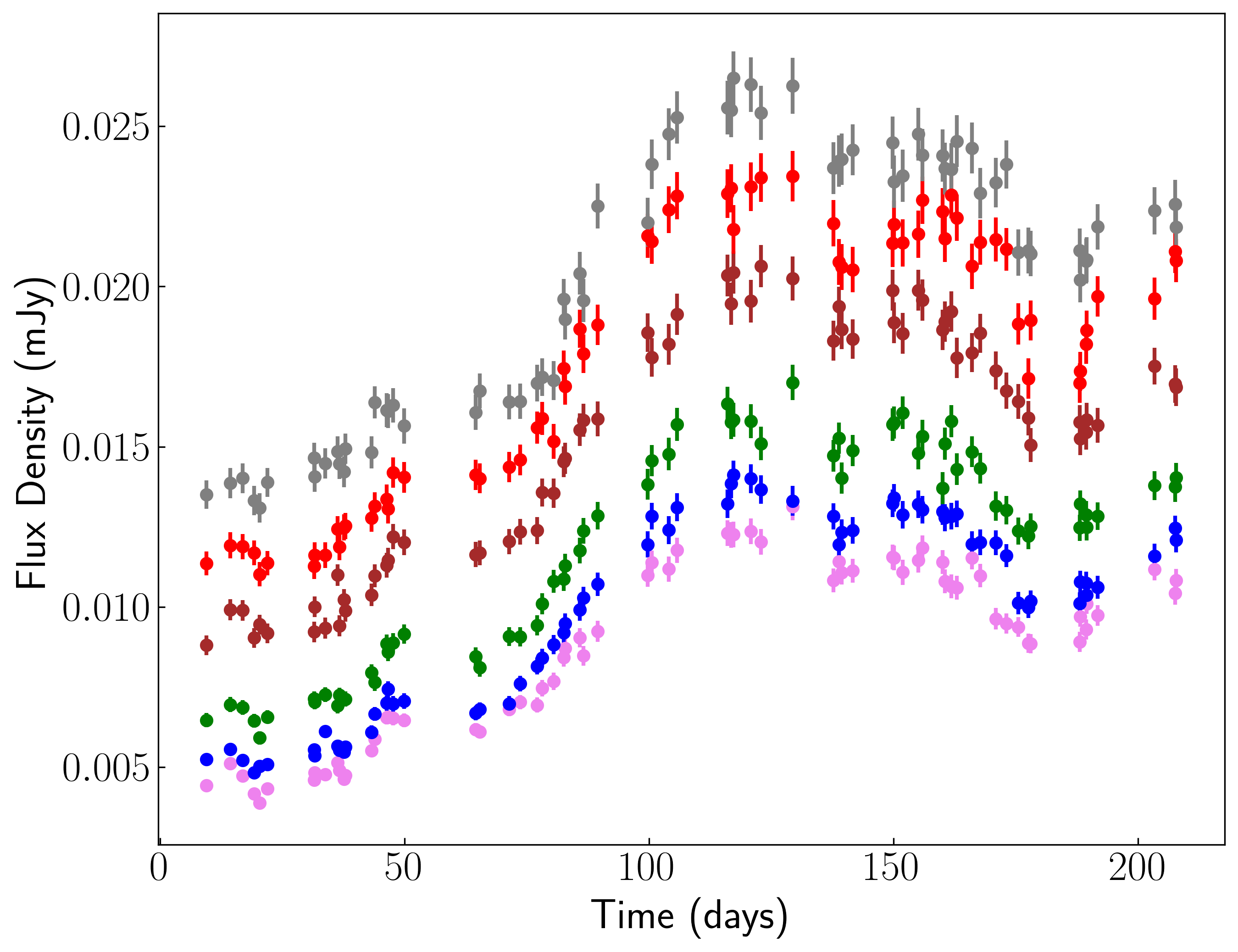}
  \caption{AD response functions (top) and convolved light curves (bottom) for the LSST filters $ugriyz$ (magenta,blue,green,brown,red,grey). The centroid of the response function is marked with vertical dotted lines and is given in the observer frame. The resampled and convolved light curves include the host galaxy and nuclear reddening contribution (see text).}
\label{adresponse}
\end{figure}

The reprocessed UV/optical continuum emission $F_{c}(t)$ from the AD is the result of the convolution of the X-ray driving light curve $F_{x}(t)$, with a response function $\Psi(\tau|\lambda)$

\begin{align}
F_{c}(\lambda,t) = \int_{0}^{\infty} \Psi(\tau|\lambda) F_{x}(t-\tau) \mathrm{d}\tau
\label{eq:tf}
\end{align}
where $\tau$ is the time delay function for a Keplerian ring/disc structure
\begin{equation}
    \tau(r,\phi,i) = \frac {r} {c} (1 + \sin i \cos \phi)
\end{equation}
with $r$ the radius of the ring, $i$ the inclination ($0 \leq i
\leq 90^\circ $) of the axis of the disk with respect to the observer's line of sight ($0$=face-on, $90$=edge-on) 
and $\phi$ is the azimuthal angle between a point on the disk and the projection of the line of sight onto the disk (\citealt{1991ApJ...379..586W}). If the UV/optical variations are driven by the X-ray corona, which is located at a distance $h$ above the disk (\citealt{2005ApJ...622..129S}), the time delay takes the form
\begin{equation}
    \tau(r,\phi,i) =\frac {1} {c} \left[\sqrt{r^2+h^2} +  r \sin i \cos \phi + h \cos i \right]
\end{equation}

In other words, the response function $\Psi(\tau|\lambda)$ characterizes the propagation and temporal behavior of the wavelength-dependent luminosity $L_{\lambda}$ across the disk with respect to a continuum pulse $L_{*}$ emitted by the X-ray corona

\begin{align}
\Psi(\tau|\lambda) = \frac{\partial L_{\lambda}(\tau)}{\partial L_{*}(t-\tau)}
\label{eq:tfex}
\end{align}
since the observed flux is modeled as black-body emission $F_{c}(\lambda,t) = \int B_\nu(\lambda,T(t-\tau)) \mathrm{d}\Omega$, the transfer function can be expressed as
\begin{equation}
    \Psi(\tau|\lambda) = \int_{r_\mathrm{in}}^{r_\mathrm{out}} \frac{ \partial B_{\nu}}{ \partial T} \frac{ \partial T}{ \partial L_{*}} \delta\left(\tau - \tau(r,\phi,i)\right) \mathrm{d}\Omega
\end{equation}
where the delta function ensures that only the radii $\tau$ that produce the particular time delay $\tau(r,\phi,i)$ contribute to the transfer function (\citealt{2007MNRAS.380..669C}; \citealt{2019MNRAS.488.2780M}). Figure~\ref{adresponse} shows an example of the response functions and simulated light curves for a quasar at $z = 0.7$ with a black hole mass $M_{\rm BH} = 4 \times 10^{7} M_{\odot}$, a mass accretion rate $\dot M = 0.2 M_{\odot} yr^{-1}$, an inclination $i = 0.0^{\circ}$, and a bolometric luminosity $L_{\rm Bol} = (1.2 \pm 0.10)\times 10^{45}{\mathrm{erg\ s^{-1}}}$.

Assuming the AD reprocessing scenario, the wavelength dependent time delay is given as

\begin{align}
  \tau_{jk}= \gamma \left[\lambda_{k}^{4/3} - \lambda_{j}^{4/3} \right] \left[ \frac{3GM\dot M}{8\pi \sigma} + \frac{L_{*}(1 - a)}{4\pi \sigma} H_{*} \cos{\theta} \right]^{1/3}
\label{equ:delayfunction}
\end{align}
with $\lambda_{k} > \lambda_{j}$ the central wavelengths for two different continuum light curves and $\gamma = c^{-1}(xk/hc)^{4/3}$ (e.g. \citealt{2019MNRAS.490.3936P}). Here $x = 2.49$, which is obtained by assuming a flux-weighted mean radius $\langle R \rangle = {\int_{R_0}^{\infty} B(T(R))R^2\,dR}/{\int_{R_0}^{\infty} B(T(R))R\,dR}$, where $T(R)$ is the combined temperature profile of the disc (see Figure~\ref{model}) as predicted by the standard theory of AGN accretion discs.

\begin{figure}
  \centering
  \includegraphics[width=\columnwidth]{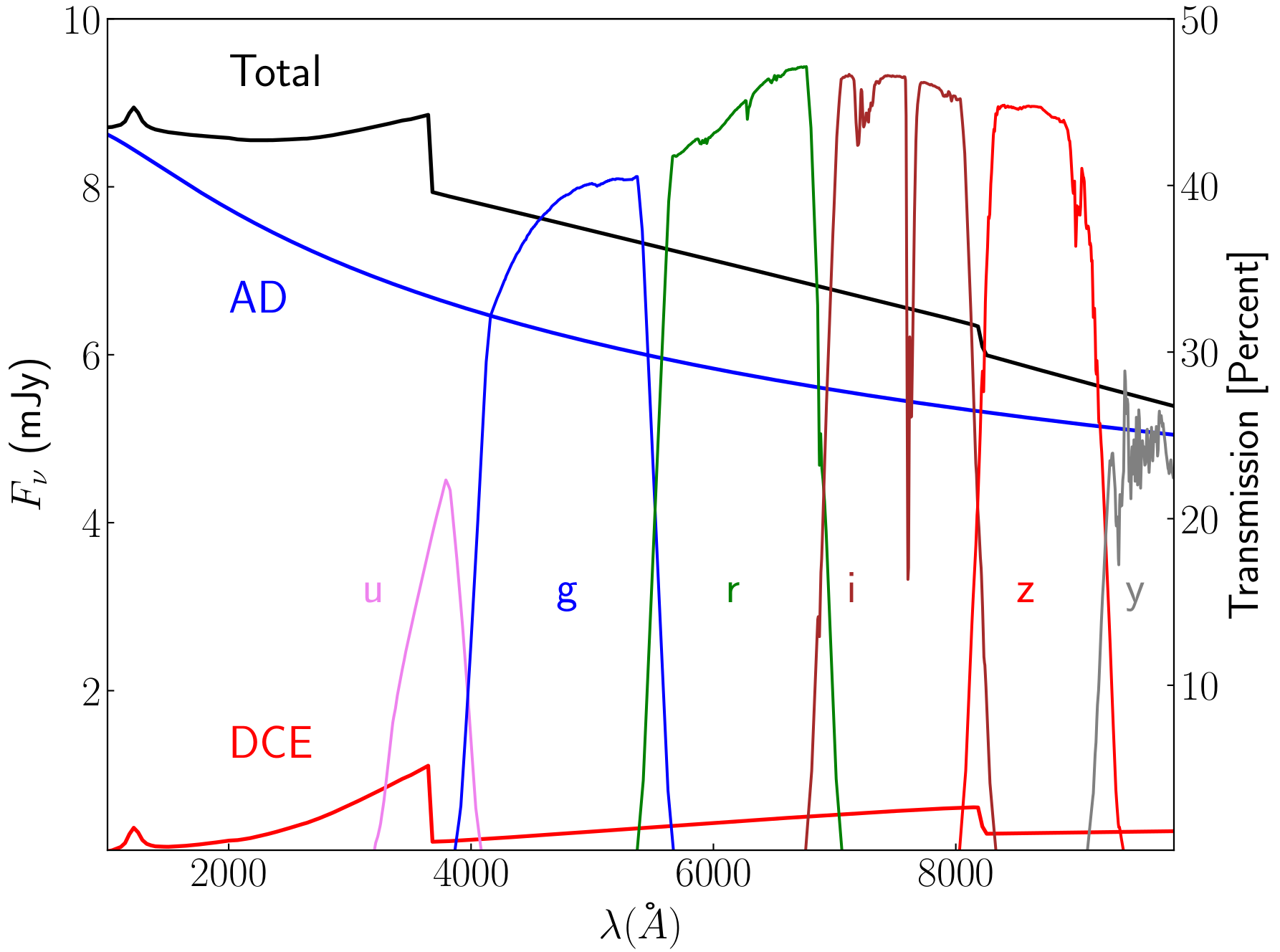}
\caption{DCE and AD models for an arbitrary quasar obtained with the same model parameters as shown in Figure~\ref{model}. The LSST transmission curves ($ugrizy$) are convolved with the quantum efficiency of the CCD camera and denoted by colored solid lines.}
\label{dceex}
\end{figure}

\begingroup
\setlength{\tabcolsep}{10pt} 
\renewcommand{\arraystretch}{1.5}
\begin{table*}
\begin{center}
\caption{Simulation parameters}
\label{table1}
\begin{tabular}{@{}ccccccc}
\hline\hline

Case & Redshift & Mass & Luminosity & Accretion Rate & $R_{\rm {min}}$ & $R_{\rm {max}}$ \\
 & $z$     & $M_{\rm {BH}}$ (M$_{\odot}$) &  $L$ (erg s$^{-1}$) &  ($M_{\odot}$ yr$^{-1}$) & Light-days & Light-days \\
\hline
(a) & $0.01 - 0.5$ & $2 \times 10^{7}$ & $5 \times 10^{44}$ & $0.1$ &  $15$ & $30$ \\
(b) & $0.51 - 1.0$ & $5 \times 10^{7}$ & $1 \times 10^{45}$ & $0.2$ &  $34$ & $46$ \\
(c) & $1.01 - 1.49$ & $2 \times 10^{8}$ & $5 \times 10^{45}$ & $0.8$ & $48$ & $64$ \\
(d) & $1.5 - 2.0$ & $8 \times 10^{8}$ & $2 \times 10^{46}$ & $3.0$ & $65$ & $87$ \\
\hline
\end{tabular}
\end{center}
\noindent
\end{table*}
\endgroup

\subsection{Diffuse continuum emission}\label{dce}

Recently, it has been shown that the optical continuum time delay can be significantly overestimated due to contamination from the diffuse continuum emission (DCE) of BLR clouds (\citealt{2019NatAs...3..251C}; \citealt{2019MNRAS.489.5284K}). The DCE composition is mostly associated with hydrogen/helium free-bound continua, including free-free emission, electron and Ly$\alpha$ Rayleigh scattering (\citealt{2001ApJ...553..695K}). This contamination was observed as an excess in the Lag spectrum, with more pronounced effects at shorter wavelengths, near the Balmer ($3646$\,\AA) limit (e.g $U$- band; \citealt{2016ApJ...821...56F}; \citealt{2019ApJ...870..123E}), and at longer wavelengths near the onset of the Paschen ($8204$\,\AA) lines. However, its fractional flux contribution is expected to be extended across a wider spectral range: $1000$\,\AA -$10000$\,\AA \, (\citealt{2019MNRAS.489.5284K}).

Under the assumption of the Local Optimally-emitting Cloud (LCO; \citealt{1995ApJ...455L.119B}) model for a spherical BLR geometry, and through the use of photoionisation calculations, \cite{2019MNRAS.489.5284K} modeled the DCE contribution with respect to the total continuum emission. The authors quantified the DCE lag spectrum for the case of the well studied AGN NGC\,5548, and provided a detailed methodology to estimate the DCE for other sources. In order to create even more realistic time-delay simulations that will allow us to quantify potential biases during the AD time delay recovery process (Section~\ref{sec3}), we follow the procedure implemented in \cite{2019MNRAS.489.5284K} by scaling the DCE fractional contribution and lag spectrum obtained for NGC\,5548 using the BLR radius-luminosity relation $R_{\rm BLR}\propto L_{\rm{AD}}^{\alpha}$ ($\alpha = 0.533^{+0.035}_{-0.033}$; \citealt{2013ApJ...767..149B}). Figure~\ref{dceex} shows an example of the DCE and AD models for an arbitrary quasar. We discuss the impact of the DCE on the lag spectrum in Section~\ref{bhmsec}.

\section{AD time delay recovery}\label{sec3}

With the present simulations, we aim to quantify the accuracy with which time delays between different continuum bands can be recovered during the next LSST survey.
To this end, we used three different methods to recover the time delays: the traditional interpolated cross-correlation function (ICCF, \citealt{1987ApJS...65....1G}; \citealt{2000ApJ...533..631K}; \citealt{2004ApJ...613..682P}), the discrete correlation function (DCF, \citealt{1988ApJ...333..646E}), including the Z-transformed DCF (\citealt{1997ASSL..218..163A}), and the von Neumann statistical estimator (VN; \citealt{2017ApJ...844..146C}), which is not based on the interpolation and binning of the light curves but on the degree of randomness of the data. We note that the methodology and application of the cross-correlation functions follow essentially the same way as in previous PRM studies (see \citealt{2019MNRAS.490.3936P} and references therein). Here, we only briefly characterize them.

For the three methods, we used a common time delay search interval $[\tau_{min},\tau_{max}$] = $[-100,100]$ days, and we estimated delays relative to the $u$- band. At this stage, we do not consider the effects of DCE (section~\ref{bhmsec}). For the ICCF analysis, we used the search interval at 0.1-day intervals, which corresponds to the minimum sampling used for linear interpolation of the fluxes. Unlike the ICCF, the DCF does not interpolate the data, but performs discrete binning of the time difference pairs and reports the mean correlation coefficient for each bin. We evaluate the DCF using a bin size equal to the median sampling of the light curves. The use of equal population binning instead of equal time difference pairs is one of the main advantages of ZDCF over the traditional DCF method. The ZDCF applies Fishers' z-transform to estimate the confidence level of the DCF correlation measurements. We estimated the time lag using the centroid $\tau_{\rm cen}$ of the cross-correlation function $R(\tau)$ calculated above the correlation level at $R \geq 0.7R_{max}$. The time lags derived from the centroid of the cross-correlation function are then compared to the centroids of the response functions obtained in Section~\ref{adtimedel}.

The statistical VN estimator is an alternative approach to the traditional cross-correlation analysis and has recently been introduced in the field of quasars RM. The VN estimator for the randomness of the combined light curve is defined as the mean square of the successive differences,

\begin{align}
  \mathcal{V_{N}(\tau)} = \frac{1}{N-1} \sum_{i=1}^{N-1} \frac{\left[ F(t_i) -
      F(t_{i+1})\right]^2}{W_{i,i+1}}
\label{equ:vn}
\end{align}
where $W_{i,i+1} = 1/[\sigma_{lc}^2(t_i) + \sigma_{lc}^2(t_{i+1})]$ is a weighting factor introduced by \cite{1994A&A...286..775P}, which takes into account the flux uncertainty ($\sigma_{lc}$) from the light curves (see \cite{2017ApJ...844..146C} for a slightly modified version of this factor). The goal is to find a time delay $\tau_{0}$ from a predefined search interval $[\tau_{min},\tau_{max}$] that minimizes the VN estimator such that $V_{N}(\tau_{0}) \equiv min[V_{N}(\tau)]$.

We begin first with a test simulating an ideal scenario for the case of a noise-free light curve with a sampling of $\Delta t = 0.1$ days and a zero contribution from the emission lines. 
AD mock light curves were generated for an arbitrary source with a black hole mass $M_{\rm BH} = 2 \times 10^{7}\, M_{\odot}$ and an accretion rate $\dot M = 0.1\, M_{\odot} yr^{-1}$. We have assumed that the source accretes at 10\% Eddington. The results of the cross-correlation functions and the VN estimator for this test are shown in the Appendix in Figure~\ref{ccfexamp} and Figure~\ref{vnex}, respectively. 
The three methods are able to recover the delay with an accuracy of $1$\%, as expected for such an ideal case scenario.

In the next test, we consider the time sampling during LSST monitoring, which is expected to be between 2 and 5 days.
We set $\rm{S/N}=30$ corresponding to the highest expected photometric noise, especially for bands $u$ and $y$ ($\sim3$\%). 
The recovered distributions of time delays ($\tau^{*}$) are shown in the Appendix in Figure~\ref{ccfsampling}. 
For a time sampling of $\Delta t = 2.0$ days, the ICCF and ZDFC methods show similar performance and the delays are recovered with an accuracy of $36.2$\% and $35.3$\%, respectively.
The accuracy obtained with the ICCF decreases up to a value of $\sim50$\% at $\Delta t = 5.0$ days, while the ZDCF remains more stable ($\sim45$\%). 
The ICCF shows a larger dispersion around the true value at a time sampling of more than 2 days. 
This is an artefact caused by the linear interpolation of the data. 
If the interpolation resolution is set to a constant value much lower than the time sampling, the recovered distributions are smoothed by adding artificial centroids as a result of the interpolation. 
Therefore, it is safer to use a time resolution that is close to the time sampling of the light curves. 

The VN estimator shows similar performance compared to the cross-correlation analysis with a recovery accuracy of $35.8$\% and $49.3$\% at $\Delta t = 2.0$ and $\Delta t = 5.0$, respectively. Finally, the delay cannot be recovered if the time sampling exceeds 8 days, so the distribution is skewed towards zero delay. Since all three methods perform similarly on both low and high quality data, we show only the ICCF results in the following analysis, as this is the most commonly used method to recover the delay in RM studies.

\subsection{The influence of BLR emission lines}\label{emissionlinesinf}

We consider the impact of BLR emission line contamination in the calculation of the AD continuum time delays.
For simplicity, we consider two filters $f_{\lambda_{1}}$ and $f_{\lambda_{2}}$. Each filter contains the emission from two components; the AD continuum, $ f_{c,\lambda}$, and the line emission $f_{l,\lambda}$. We define the relative strength of the emission line fluxes as
\begin{align}
  \alpha = f_{l,\lambda_{1}}/\left[f_{l,\lambda_{1}}+f_{c,\lambda_{1}}\right] \\
  \beta = f_{l,\lambda_{2}}/\left[f_{l,\lambda_{2}}+f_{c,\lambda_{2}}\right]
\label{equ:contributionlines}
\end{align}
with $\alpha$ and $\beta$ the average contribution of the line to the total flux in filters $f_{\lambda_{1}}$ and $f_{\lambda_{2}}$, respectively. The total flux observed by each filter is therefore,

\begin{align}
  f_{\lambda_{1}}(\alpha,t) = \alpha F_{L}(t) + (1-\alpha) F_{\rm c}(t) \\
  f_{\lambda_{2}}(\beta,t) = \beta F_{L}(t) + (1-\beta) F_{\rm c}(t)
\label{equ:contributionlines2}
\end{align}
where $F_{\rm L}(t)$ and $F_{\rm c}(t)$ are the result of the convolution between the driving X-ray light curve (see Section~\ref{ranwalksec}) with the BLR and AD transfer functions, respectively.
We model the BLR transfer function assuming clouds with Keplerian orbits distributed in a disc-shaped geometry with inclination $i=25^{\circ}$. 
We set the BLR extent to $R_{\rm {max}}/R_{\rm {min}} = 2.0$, where $R_{\rm {min}}$ is fixed at 10 times the average size of the AD.

We have selected a few test cases based on the redshift of the source, where the relative contribution of the emission lines changes in different filters. Figure~\ref{speclines} illustrates this selection using the composite spectrum of \cite{2001AJ....122..549V} and \cite{2006ApJ...640..579G}, as well as the transmission curves of the photometric LSST bands. 
The physical properties used in the simulations are listed in Table~\ref{table1}. 
An example of the transfer functions calculated with the parameters corresponding to $0.01 < z < 0.5$ (see Table~\ref{table1}) for the filters $u$ and $z$ is given in Figure~\ref{tfunc}.

Light curves with flux uncertainties of $\sim1$\% are expected to be routinely recorded by the LSST.
Therefore, we assume here an S/N = $100$ for the following analysis.

Since the BLR has a larger delay compared to the AD, we expect an overestimation of the delay due to a higher fraction of line emission, i.e. higher $\beta$ values, whereas a higher value of $\alpha$ should lead to an underestimation of the delay. Most filters configurations at the chosen redshift contain a small contribution from emission lines, i.e. H$\gamma$ and H$\beta$ (+narrow) $\sim5$\% in the g-band (case a) or HeI, H$\epsilon$, H$\delta$ and H$\gamma$ $\sim6$\% in total (case b) or the blue wing of MgII $\sim3$\% (case c) and the red wing of MgII $\sim4$\% (case d), with the exception of H$\alpha$ and CIV, where the contribution reaches up to $\sim10$\%.  
Figure~\ref{fig:Apx6} shows the recovered delays ($\tau^{*}$) for a fixed S/N = $100$ and an emission line contribution fraction of up to $10$\%.
The distributions of $\tau^{*}$ obtained from 1000 mock light curve simulations and for selected emission line contributions are shown in Figure~\ref{fig:Apx7}. 
We describe the results for each redshift case in detail below.

\subsubsection{Case (a) -- $0.01 < z < 0.5$}

In this redshift range we expect AD with time delays of less than 1 day. The centroid of the transfer function ($\tau_{cen,\lambda}$) at bands $u$ and $z$ is $0.59$ and $1.67$ days, respectively. As can be seen in Figure~\ref{fig:Apx6}, the delay is recovered with an accuracy of $\sim20$\% when the line contribution is $\leq2$\% and $\Delta t = 2.0$ days. In the case of $\Delta t = 5.0$ days, the delay is biassed towards lower values by up to $\sim50$\%. 

We also experimented with different configurations. For example, if the time sampling can be reduced to $\Delta t = 0.5$ days, i.e. two observations per day, and the contribution of the lines in both filters is $\leq4$\%, the delay can be recovered with an accuracy of $5$\% (the LSST survey will not yield observations with a sampling of less than 2 days, hence not shown). Increasing the S/N to higher values, e.g. $1000$ (flux uncertainties of $\sim0.1$\%), does not improve the results. 

In order to estimate in advance the accuracy with which the time delay can be measured in different redshift ranges for a given light curve quality, we estimated a heuristic function using the recovered distributions of the delays for different emission line contributions, S/N and sampling (hereafter referred to as delay maps). We applied multi-variable linear regression analysis to reconstruct the delay maps using the high-Performance Deep Learning Library PyTorch (\citealt{Paszke2019}; \citealt{2019ASPC..523...63N}). An example of the delay map reconstruction and the heuristic equation can be found in the Appendix and in Figure \ref{delmap}.

\subsubsection{Case (b) -- $0.51 < z < 1.0$}

In this redshift range we expect AD with time delays of $\sim2$ days. The centroids are $\tau_{cen,u} = 0.95$ days and $\tau_{cen,z} = 2.77$ days. The delay is recovered with an accuracy of $\sim15$\% for $\alpha \leq3$\% and $\beta \leq1$\%. When the contributions of $\alpha$ and $\beta$ are $\leq1$\%, it is possible to achieve an accuracy of $10$\% for $\Delta t = 2.0$ days. The better performance compared to case (a) is expected because the relative delay ($\Delta \tau = \tau_{cen,u} - \tau_{cen,z} = 1.87$ days) is closer to $\Delta t = 2.0$ days. In the case of $\Delta t = 5.0$ days, the lag is again underestimated even if the contribution of the lines is only $1$\%. Similar to case (a), the delay can be recovered with an accuracy of $5$\% when $\Delta t = 0.5$ days and the contribution of the line is below $2$\%. A higher S/N does not improve the results.

\subsubsection{Case (c) -- $1.01 < z < 1.49$}

In this redshift range we expect AD with time delays of $\sim4$ days. The centroids are $\tau_{cen,u} = 1.78$ days and $\tau_{cen,z} = 5.19$ days. The LSST sampling rate $\Delta t = 2.0$ days satisfies the Nyquist criterion, since $\Delta \tau = 3.4$ days. This explains the higher accuracy of $\sim8$\% compared to cases (a) and (b), where the contribution of the lines is $\leq1$\%. An accuracy of up to $10$\% can be expected when the contribution of the lines is $\sim4$\%. When the sampling is $\Delta t = 5.0$ days, the delay is recovered with an accuracy of $\sim20$\% for $\alpha \leq3$\% and $\beta \leq1$\%.

\subsubsection{Case (d) -- $1.5 < z < 2.0$}

In this redshift range we expect AD with time delays of $\sim6$ days. The centroids are $\tau_{cen,u} = 3.47$ days and $\tau_{cen,z} = 8.91$ days. In this case both time sampling of $\Delta t = 2.0$ and $5.0$ days satisfies the Nyquist theorem. For a sampling of $\Delta t = 2.0$ days, the delay is recovered within an accuracy of $\sim5$\% if the line contribution in both filters is $\leq3$\%. If the sampling is $\Delta t = 5.0$ days and the line contribution is $\leq2$\%, the delay is recovered at $10$\% precision.

\begin{figure}
  \centering
  \includegraphics[width=\columnwidth]{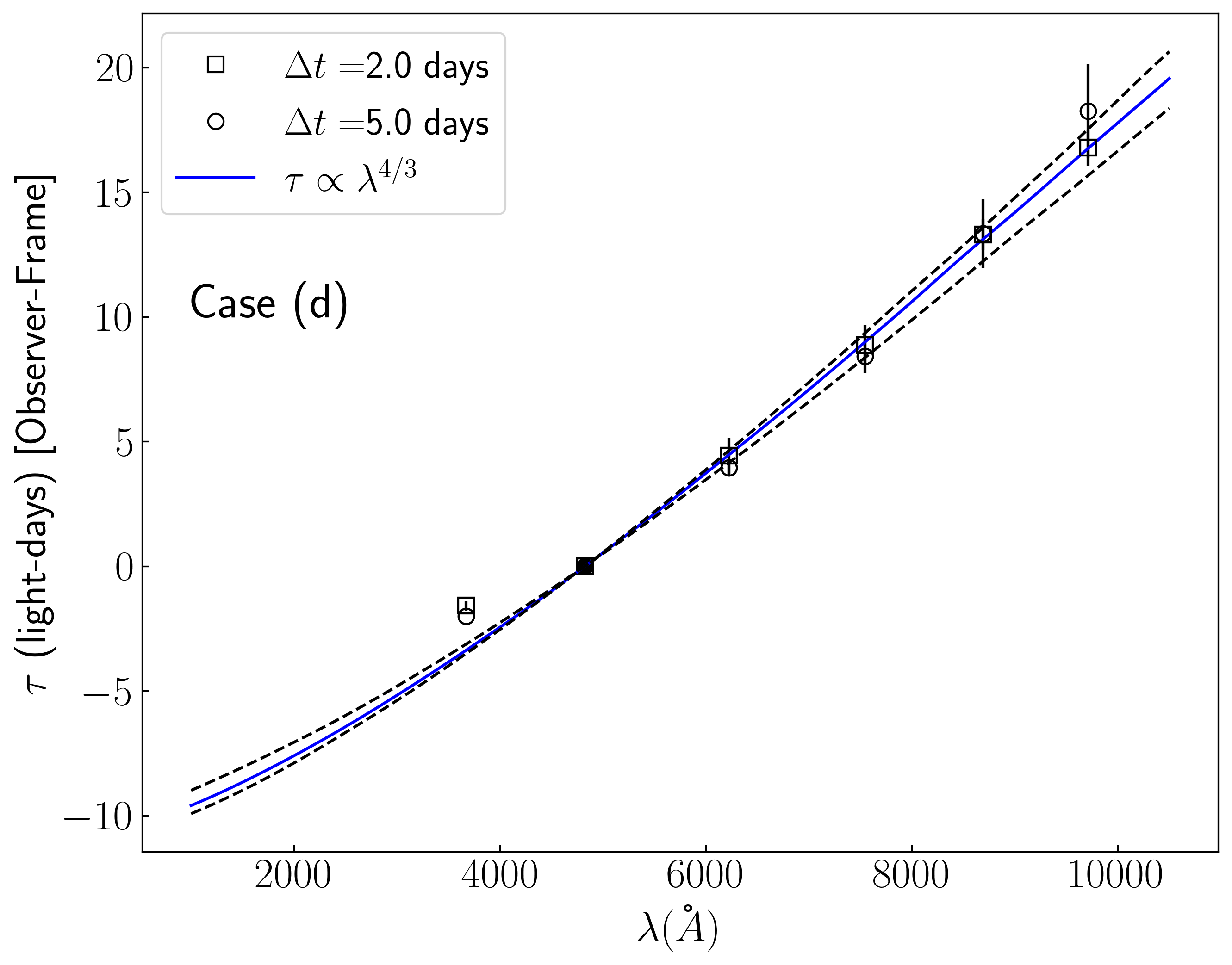}
  \caption{AD time delay spectrum $\tau_{c}(\lambda)\propto \lambda^{4/3}$ (blue line) as predicted from the response functions. Filled squares and open circles denote the recovered delays at sampling 2 and 5 days respectively. The dotted lines show the delay spectrum obtained for a black hole mass with $30$\% uncertainty.}
\label{delayavgres}
\end{figure}

\section{Discussion}\label{bhmsec}
As we have shown in Section~\ref{emissionlinesinf}, the accuracy of the recovered delay depends on the relative contribution of the emission lines and on how well the light curves are sampled with respect to the sought delay.
Figure~\ref{delayavgres} shows the expected time delays for an optically thick and geometrically thin accretion disc model together with the delays determined for case (d) in section~\ref{emissionlinesinf}. This case concerns quasars with a redshift of $1.5 < z < 2.0$, for which the delay was determined with an accuracy between $5$ and $10$\%. The results for other cases are shown in the Appendix in Figure~\ref{fig:Apx8}.

We find that a sampling $\Delta t = 2$ yields a time-delay spectrum which is consistent with an assumed black hole mass with uncertainties of the order of $\sim30$\%. This black hole mass uncertainty is typically recovered with reverberation mapping of the BLR using single epoch spectra measurements (e.g. \citealt{2009ApJ...692..246D}). Of course, any assumption for the black hole mass in Equation~\ref{equ:delayfunction} (Section~\ref{adtimedel}) must also hold for the mass accretion rate, which in turn depends on the bolometric luminosity $\dot M = L_{\rm Bol}/\eta c^2$. Neglecting nuclear extinction and host galaxy corrections, for example, leads to incorrect estimates of the AGN luminosity and mass accretion rate and thus to an under- or overestimation of the delay spectrum\footnote{Here we assume a radiation efficiency $\eta = 0.1$, as commonly used for co-rotating discs. However, we note that $\eta$ can vary between 0.038 and 0.42 depending on the spin of the black hole (\citealt{2011ApJ...728...98D}).}. We note that this effect is more pronounced for low $z$ objects, where the host galaxy contributes to about $\sim50$\% of the total luminosity.

\begin{figure}
  \centering
  \includegraphics[width=\columnwidth]{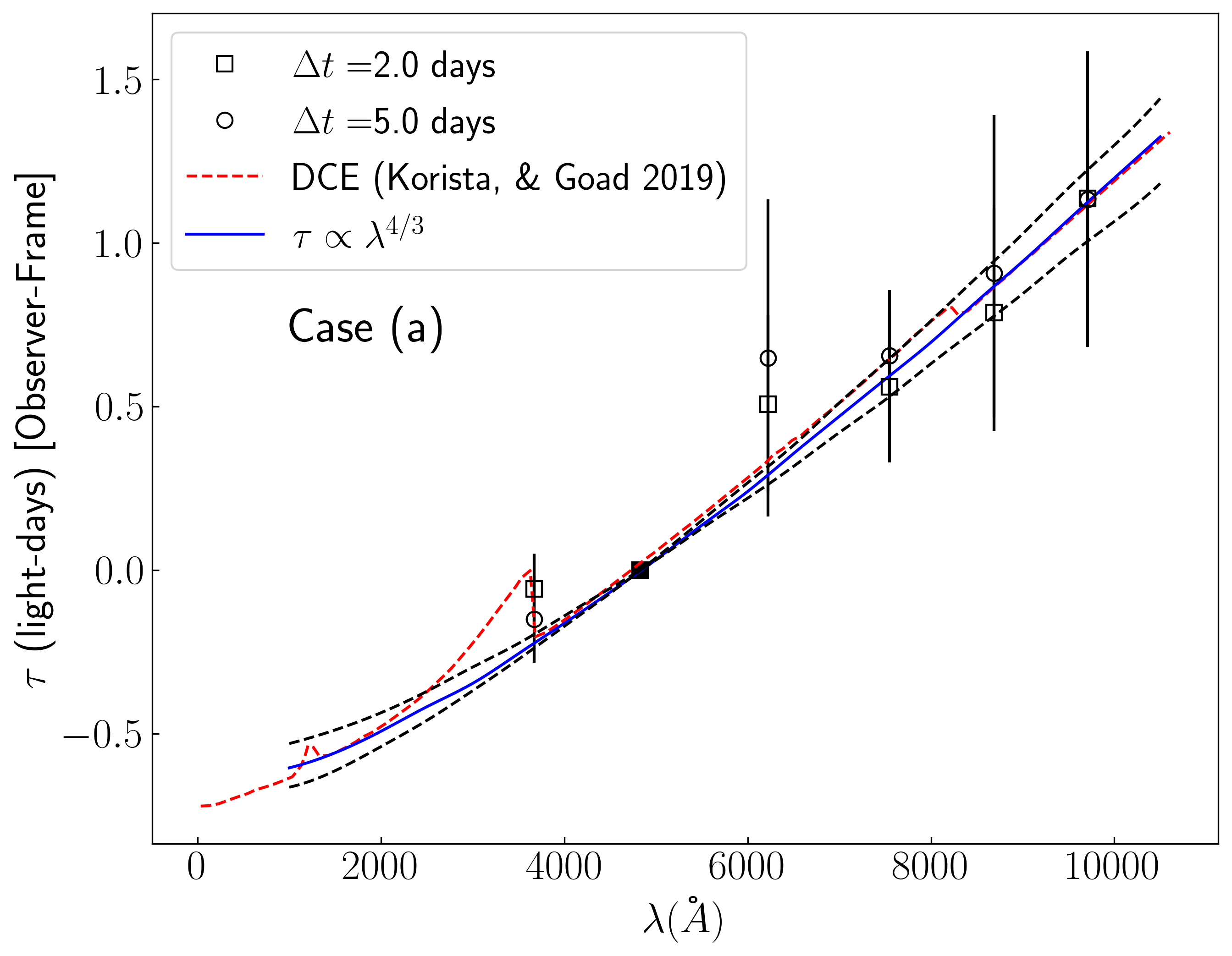}
  \caption{Same as Figure \ref{delayavgres}, but for Case(a) and an AD time delay spectrum recovered from observations with variable DCE from the BLR (red dotted line). The results are shown for a time sampling $\Delta t = 2$ days (filled squares).}
\label{dcespecdel}
\end{figure}

For the case of variable DCE contamination from the BLR (Section~\ref{dce}). We have performed the same analysis to determine the time delay as in Section~\ref{sec3}, but this time we have simulated a mixture of continuum emission from AD including the theoretically expected variable DCE and emission line contamination. We estimated the delays using the $g$- band as a reference. The results are shown in Figure~\ref{dcespecdel}. DCE contamination leads to time delays that are larger by a factor of $\sim1.5$. It is clear that the effect of DCE is particularly strong in the $u$ band, as expected for the case of low-$z$ quasars (see Figure~\ref{speclines}).

Other effects such as inclination, albedo, limbdarkening or partial blocking by optically thick dusty material could also lead to a bias in the time delay spectrum. These effects are directly related to the actual "AD size problem" and are closely linked to a correct isolation of the AGN luminosity. As a result, there are some challenges to overcome in this direction, for example to use AGNs as standard candles through a AD -size-luminosity relation. In principle, such a relationship should be straightforward as it relies on the same temperature structure of the BLR. This suggests that the temperature at the onset of the BLR is universal, as suggested by the model of \cite{2011A&A...525L...8C}, in which this temperature is identified as the dust sublimation temperature. However, the challenge lies in the actual measurements and their interpretation. In addition, there is the contamination of the disc emission by the surrounding medium. For example, \cite{2022MNRAS.509.2637N} have recently shown a new relationship between delay and luminosity for certain objects dominated only by the diffuse continuum emission of the BLR. These delays do not originate in the AD itself. Therefore, the origin of such a relationship is still debated and it seems to be very complex to determine it through observations. For this reason, the time delay of AD, measured for a large number of objects, will provide a method to decipher some of these effects.

\section{Conclusions}

We have investigated, by means of extensive simulations, the ability of the next LSST survey to recover AD time delays and black hole masses. Our results can be summarized as follows:

\begin{enumerate}
      \item A minimum signal-to-noise ratio (S/N) of 100 with a BLR emission line contribution of less than 10\% in the bandpasses can lead to recovery of the time delays with $5$ and $10$\% accuracy for a time sampling of 2 and 5 days, respectively, and for quasars at $1.5 < z < 2.0$. An accuracy of 10 to 20\% can be achieved for quasars at $z < 1.5$ only if the contribution of the BLR emission lines is less than 5\%. Increasing the S/N does not improve the results significantly. Reducing time sampling and BLR emission line contamination is the only solution to improve time delay accuracy.
      
      \item By assuming an optically thick and geometrically thin AD model, the recovered time-delay spectrum is consistent with black hole masses inferred with an accuracy of $30$\%. Under the current observational conditions of the LSST survey, this means covering a range of intermediate black hole masses between $\sim10^{8} - 10^{9}$ M$_{\odot}$. For special DDF fields of quasars, which may have a time sampling of 1 to 2 days, the range could be pushed to a lower limit of $\sim5\times10^{7}$ M$_{\odot}$.
      
      In order to achieve the precision obtained in this work, we recommend the following procedure:
      
      \begin{enumerate}
          \item First, the fluxes in each photometric band should be corrected for contamination by the host galaxy, including internal reddening. This can be done using the flux variation gradient method and the AGN reddening curve of \cite{2007arXiv0711.1013G}.
          \item The recovered delay spectrum should be corrected for the variable diffuse continuum emission (DCE) of the BLR. This can be achieved by scaling the DCE lag spectrum modeled by \cite{2019MNRAS.489.5284K}. The scaling should be performed using the BLR size-luminosity relation to account for differences in luminosity.
      \end{enumerate}
      \item Finally, it is important to note that our analysis is not limited to the next LSST survey and can be applied to any future photometric reverberation mapping survey of the AD. Due to cosmological time dilation, continuum- emission line RM studies of luminous quasars at $z\sim2$ require about 15 years of observations (e.g., using CIV lines). The AD time delays are about 10 times shorter than BLR time delays, and therefore have the potential to better constrain a AD size-luminosity relation, which could provide new opportunities for using quasars as efficient standard cosmological candles up to high-$z$. However, the time sampling is a crucial factor and special surveys are needed to study specific quasar populations. In this context, we are planning a AD PRM campaign of quasars up to redshift $z\le1.5$ using the optical and near-infrared telescopes of the Cerro Armazones Observatory (OCA) in Chile. More details on this new survey will be given in an upcoming paper.
\end{enumerate}

\section*{Acknowledgements}

The authors are very grateful to the referee, Martin Gaskell, for raising important points that helped significantly to clarify the aim of the paper and the methodology.
Authors F.P, S.D, and K.L.P gratefully acknowledge the generous and invaluable support of the Klaus Tschira Foundation.
This project has received funding from the European Research Council (ERC) under the European Union's Horizon 2020 research and innovation programme (grant agreement No 951549). We also acknowledge support from the DIR/WK/2018/09 grant of the Polish Ministry of Science and Higher Education.
S.P acknowledges the financial supports from the Conselho Nacional de Desenvolvimento Cientifico e Tecnologico (CNPq) Fellowship (164753/2020-6).
This research has made use of the NASA/IPAC Extragalactic Database (NED) which is operated by the Jet Propulsion Laboratory, California Institute of Technology, under contract with the National Aeronautics and Space Administration. This research has made use of the SIMBAD database, operated at CDS, Strasbourg, France.

\section*{Data Availability}

The simulated data underlying this article will be shared on reasonable request to the corresponding author.





\bibliographystyle{mnras}
\bibliography{mnras_fpozo}



\appendix

\section{Simulations}
\label{app:simul}


\begin{figure*}
  \centering
  \includegraphics[width=18cm,clip=true]{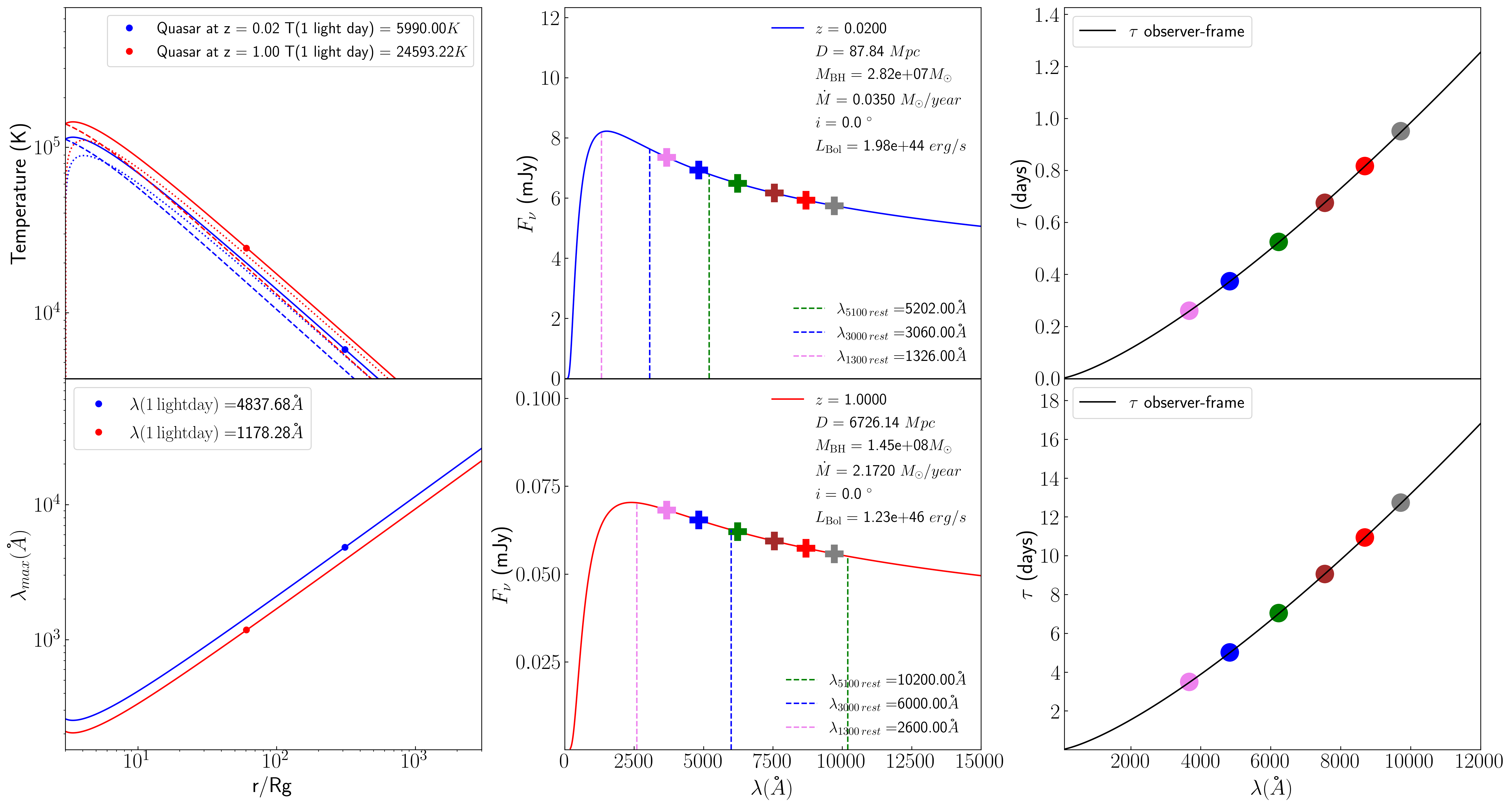}
  \caption{Characteristics of the simulations for two quasars. The temperature profile for the AD (top left panel). The Wien's law for a temperature at 1-light day distance from the innermost stable circular orbit (bottom left panel). The simulated spectrum (middle panels) with the central wavelengths for the LSST filters (ugrizy) indicated by color crosses. The dotted magenta, blue and green lines mark the rest-frame wavelengths typically used to estimate AGN bolometric luminosities. The right panels show our time-delay predictions in the observers and rest-frame, respectively.}
\label{model}
\end{figure*}

\begin{figure}
  \centering
  \includegraphics[width=\columnwidth]{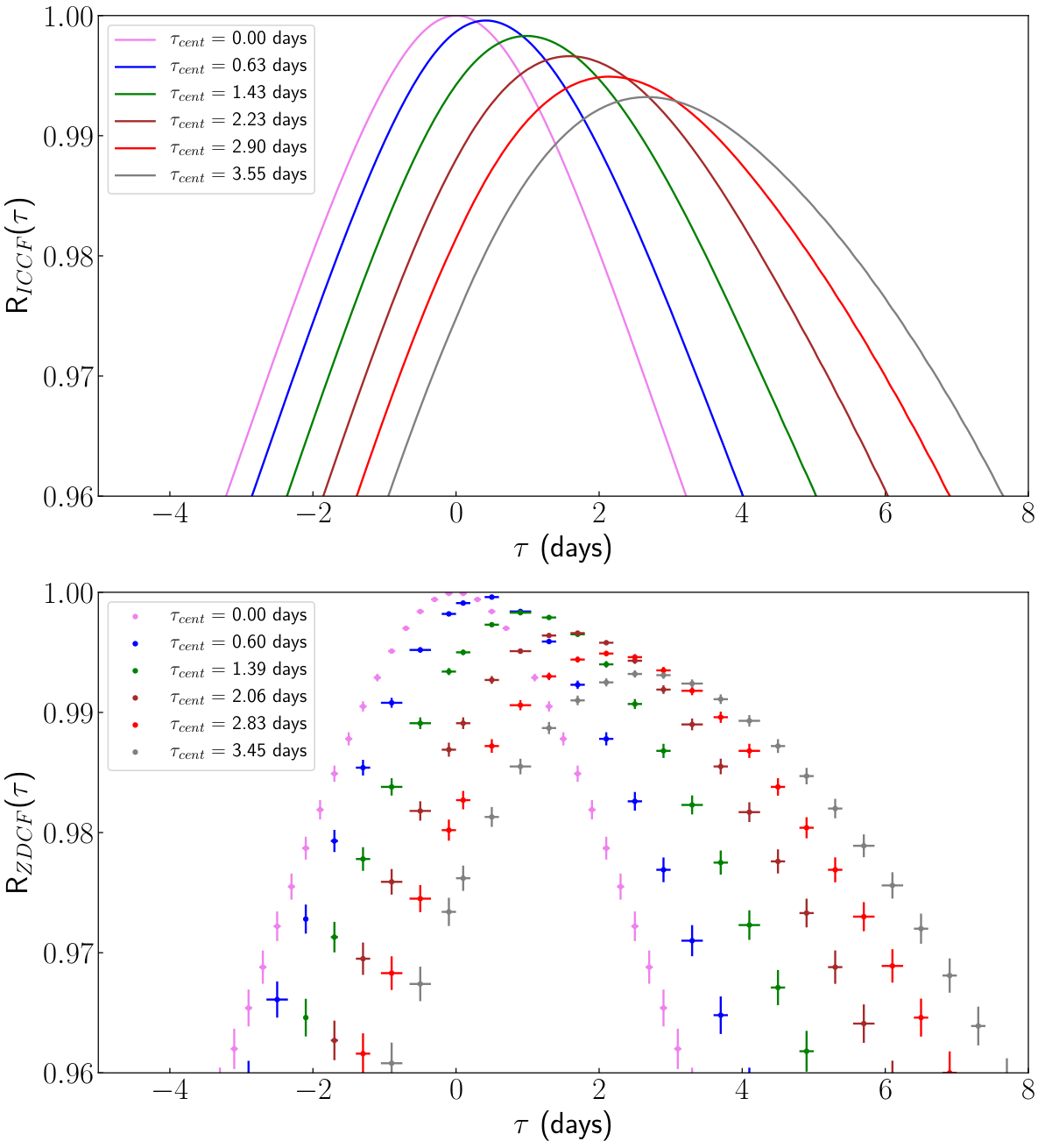}
\caption{Cross-correlation functions with respect to the $u$-band obtained with the ICCF (top) and ZDCF (bottom) methods for an ideal time sampling of 0.1 days.}
\label{ccfexamp}
\end{figure}

\begin{figure}
  \centering
  \includegraphics[width=\columnwidth]{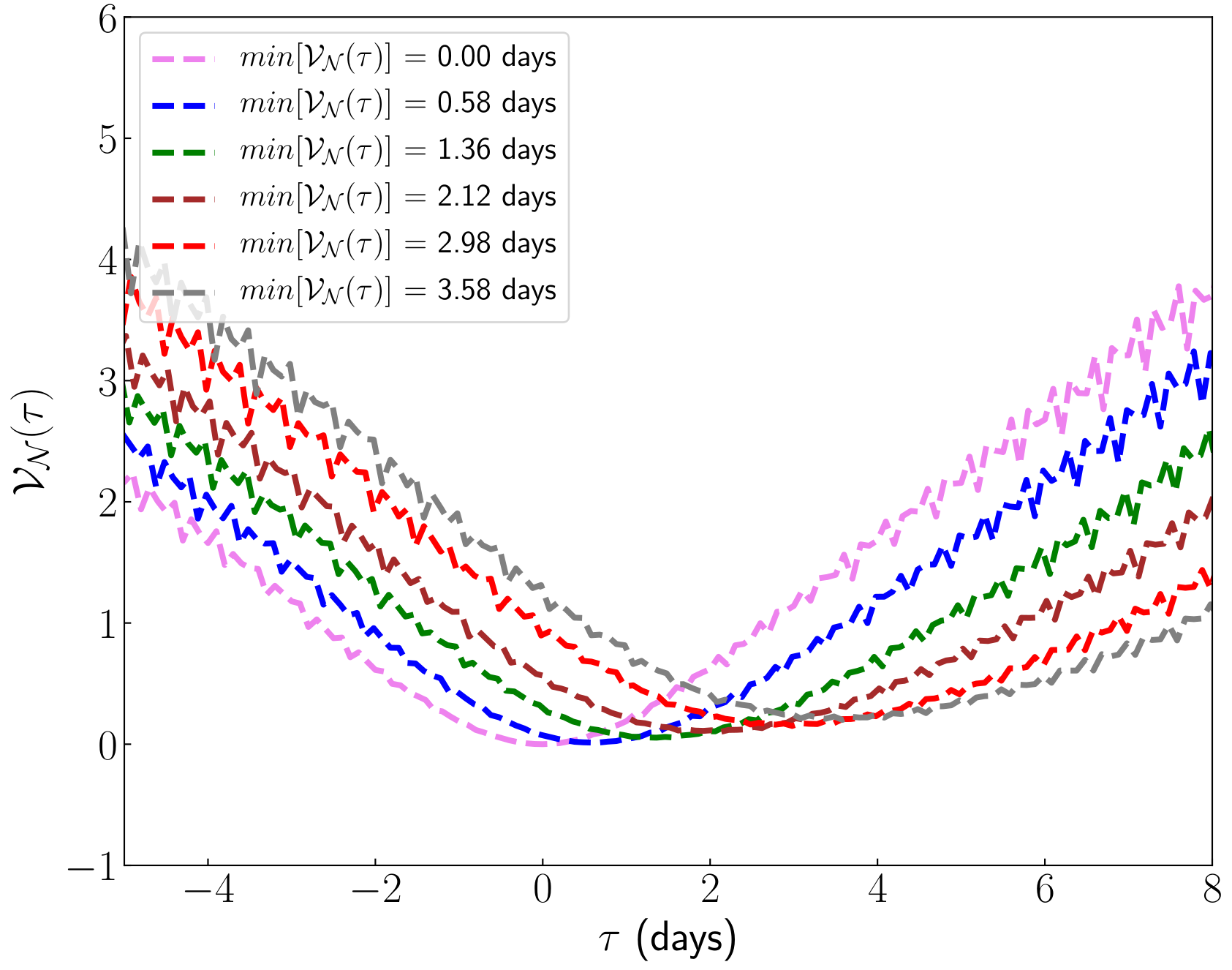}
\caption{The von Neumann estimator for an ideal time sampling of 0.1 days.}
\label{vnex}
\end{figure}

\begin{figure}
  \centering
  \includegraphics[width=\columnwidth]{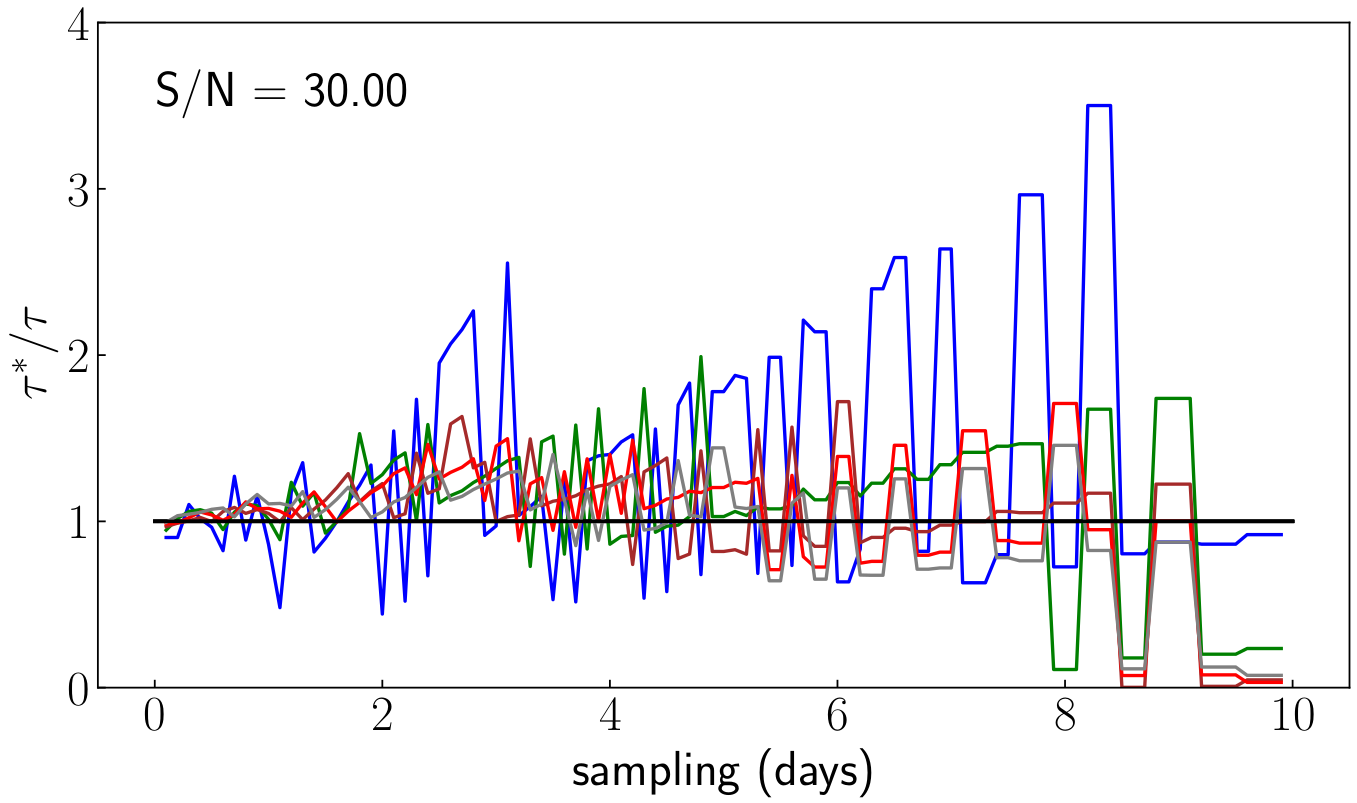}
  \includegraphics[width=\columnwidth]{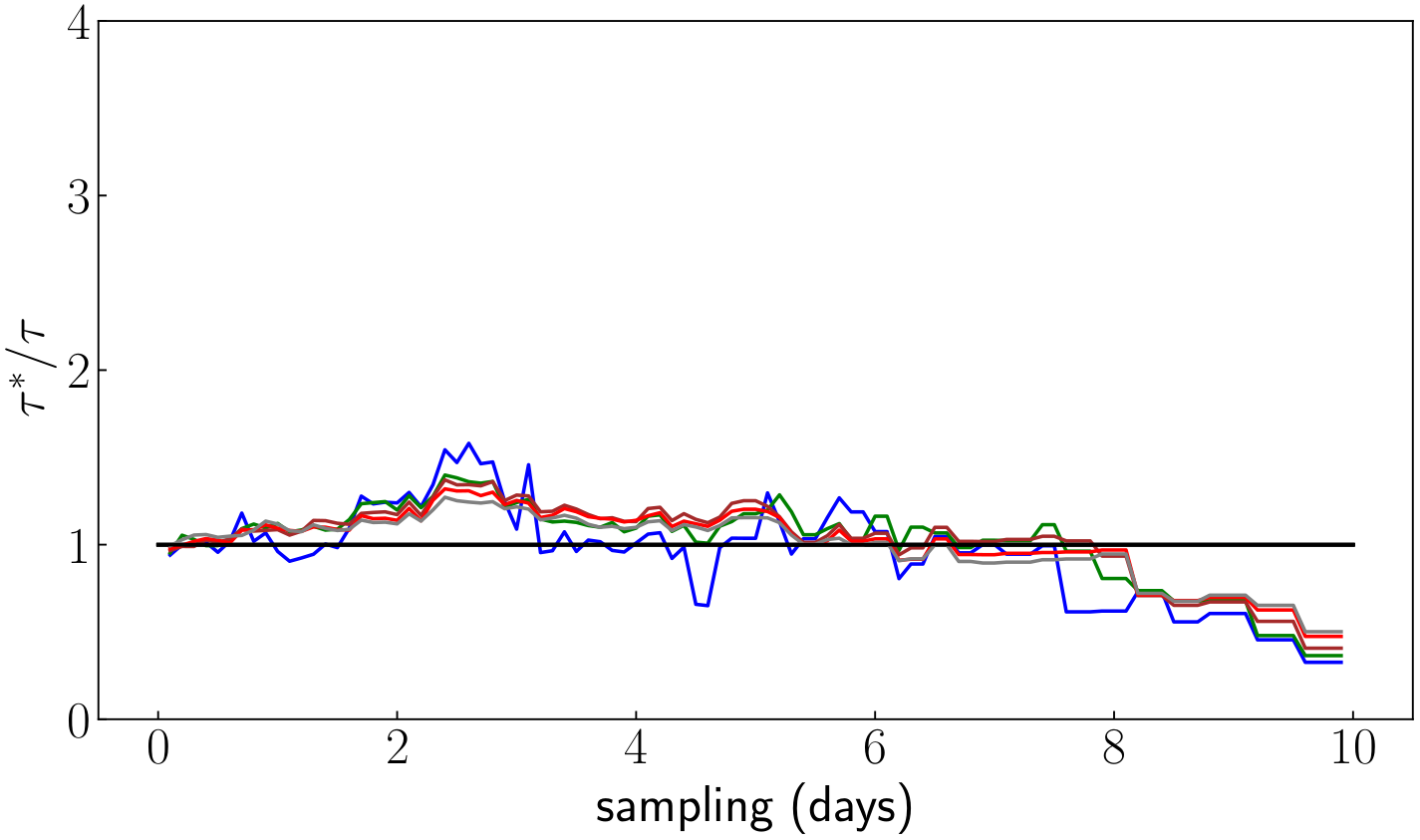}
  \includegraphics[width=\columnwidth]{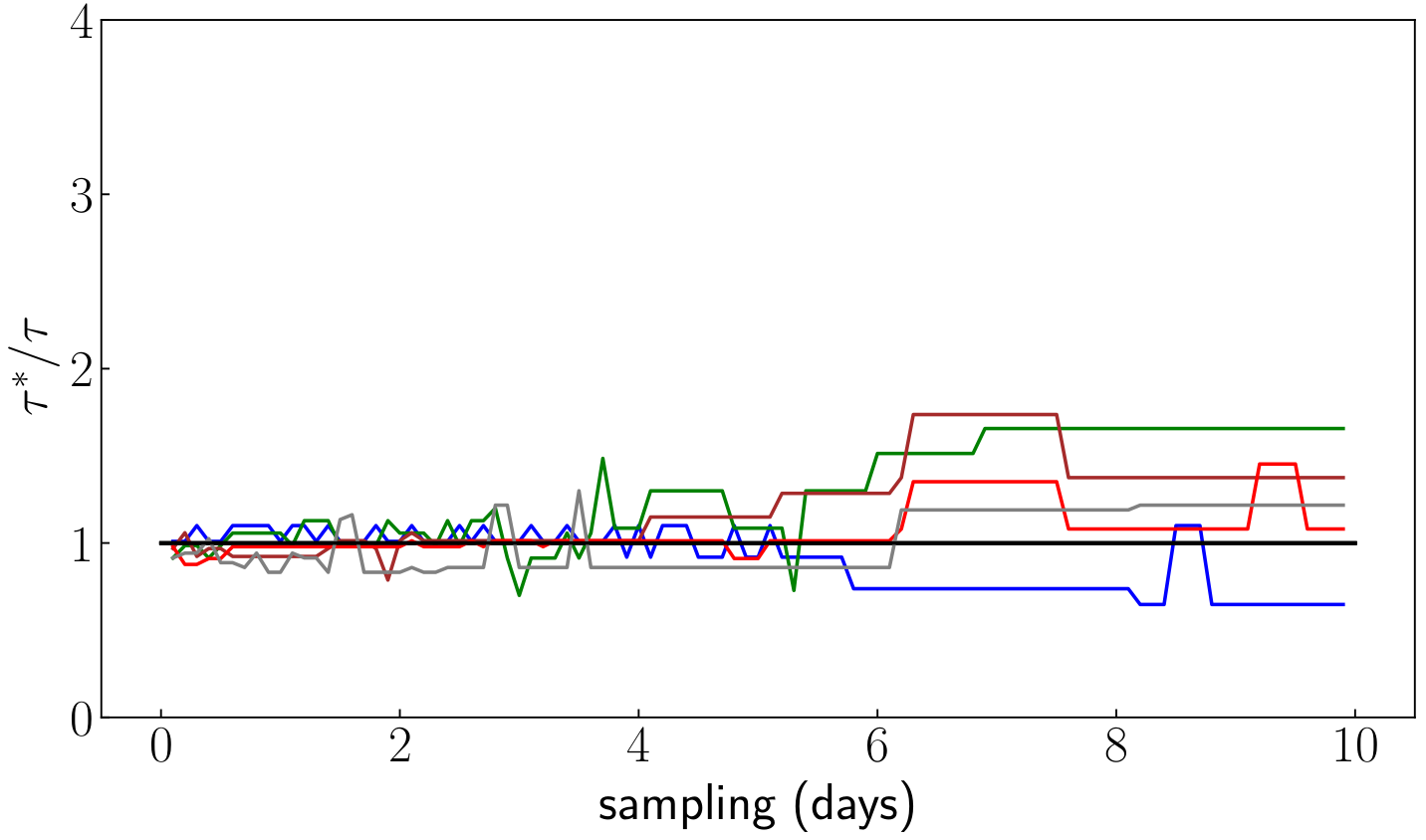}
\caption{Recovered ICCF (top), ZDCF (middle), and VN (bottom) distributions of time delays ($\tau^{*}$) for various sampling and S/N $= 30$ corresponding to measurement uncertainties at the $\sim3$\% level.}
\label{ccfsampling}
\end{figure}

\begin{figure}
  \centering
  \includegraphics[width=\columnwidth]{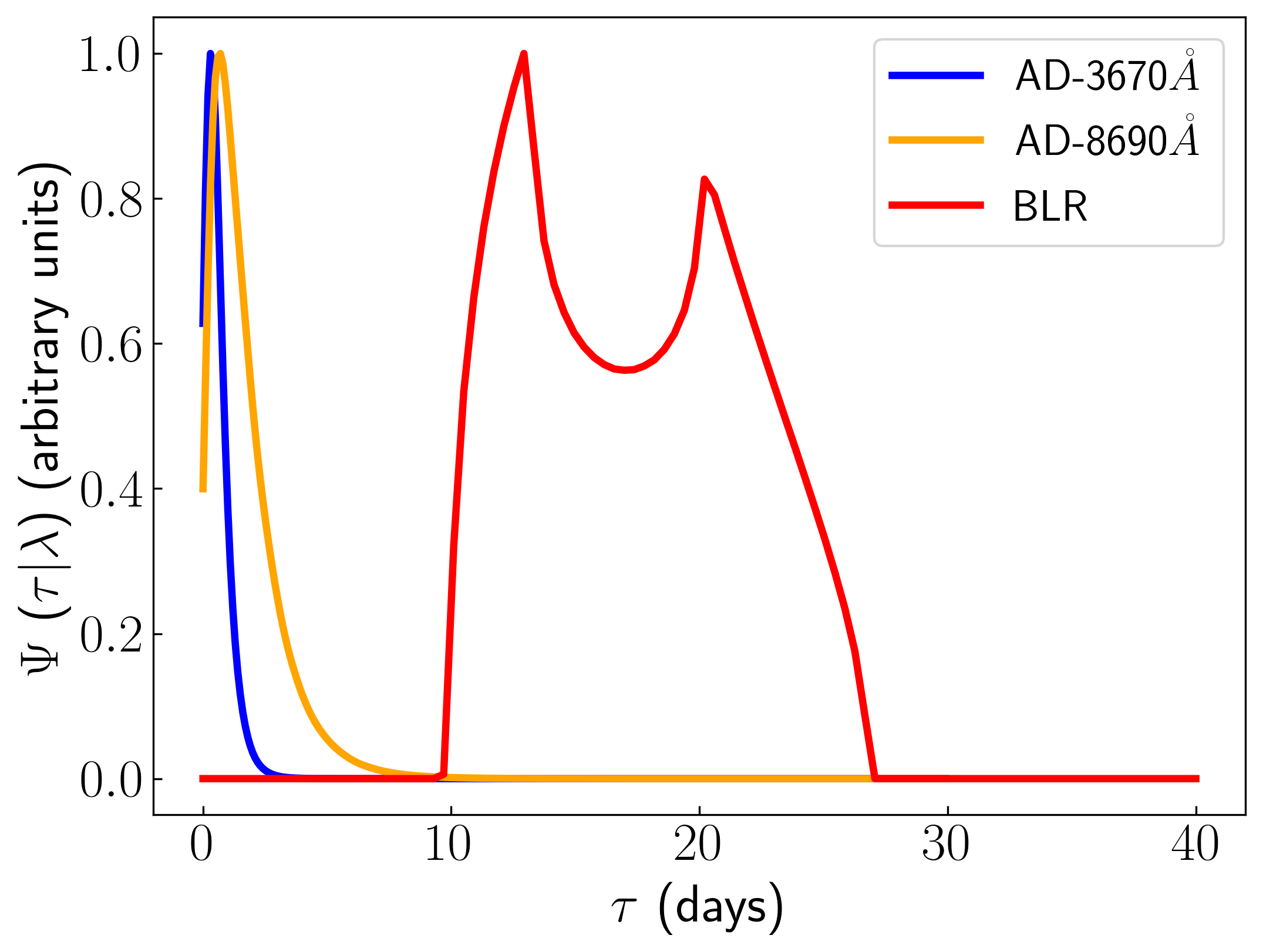}
\caption{AD and BLR transfer functions.}
\label{tfunc}
\end{figure}

\begin{figure}
  \centering
  \includegraphics[width=90mm]{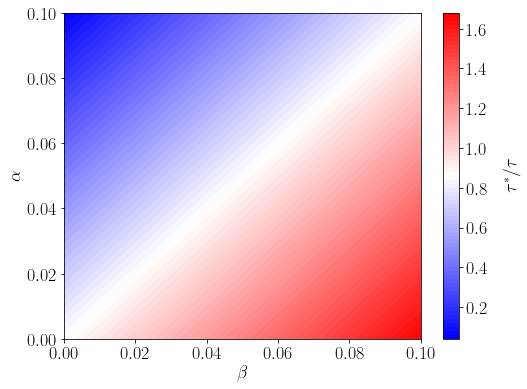}
\caption{Time delay map for case (a) $0.01 < z < 0.5$. For this particular example, we obtain the heuristic equation $\tau_{\rm {PR}}(\alpha,\beta,S/N,\Delta t) = 0.973 - 5.445\ \alpha + 6.662\ \beta + 1.387\times10^{-6}\ S/N + 1.160\times10^{-2}\ \Delta t$.}
\label{delmap}
\end{figure}

\section{Prediction for time delay accuracy}
\label{app:tda}

Using the delay maps, we predicted the accuracy of the time delay measurements based on different redshift ranges (see Table 1).
To obtain the precision in percentage we use $\mid \tau_{\rm {PR}} - 1 \mid \times 100\%$.

\paragraph{Case (a)}

\begin{align}
    \tau_{\rm {PR}}(\alpha,\beta,S/N,\Delta t) =& 
    0.973 - 5.445\ \alpha + 6.662\ \beta \notag\\
    & + 1.387\times10^{-6}\ S/N + 1.160\times10^{-2}\ \Delta t
\
\label{equ:contributionlines}
\end{align}

\paragraph{Case (b)}
\begin{align}
    \tau_{\rm {PR}}(\alpha,\beta,S/N,\Delta t) =& 
    0.984 - 4.625\ \alpha + 5.605\ \beta \notag\\
    & + 2.076\times10^{-7}\ S/N - 4.710\times10^{-3}\ \Delta t
\
\label{equ:contributionlines}
\end{align}

\paragraph{Case (c)}
\begin{align}
    \tau_{\rm {PR}}(\alpha,\beta,S/N,\Delta t) =& 
    0.982 - 2.350\ \alpha + 3.247\ \beta \notag\\
    & + 4.214\times10^{-7}\ S/N - 2.273\times10^{-3}\ \Delta t
\
\label{equ:contributionlines}
\end{align}

\paragraph{Case (d)}
\begin{align}
    \tau_{\rm {PR}}(\alpha,\beta,S/N,\Delta t) =& 
    0.992 - 5.346\ \alpha + 6.472\ \beta \notag\\
    & + 1.216\times10^{-6}\ S/N + 3.130\times10^{-3}\ \Delta t
\
\label{equ:contributionlines}
\end{align}

\begin{figure*}
\begin{tabular}{cc}
  \includegraphics[width=70mm]{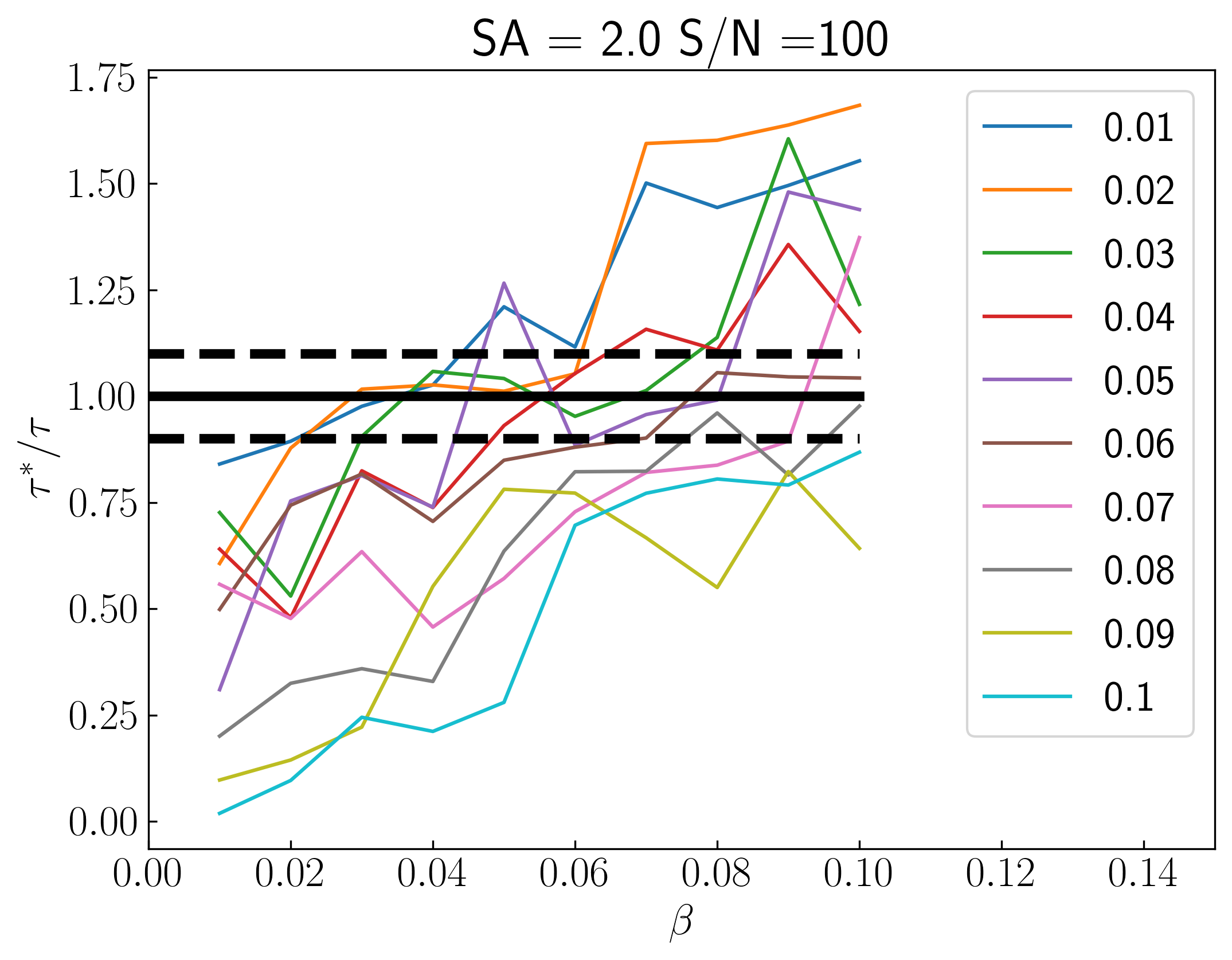} &   \includegraphics[width=70mm]{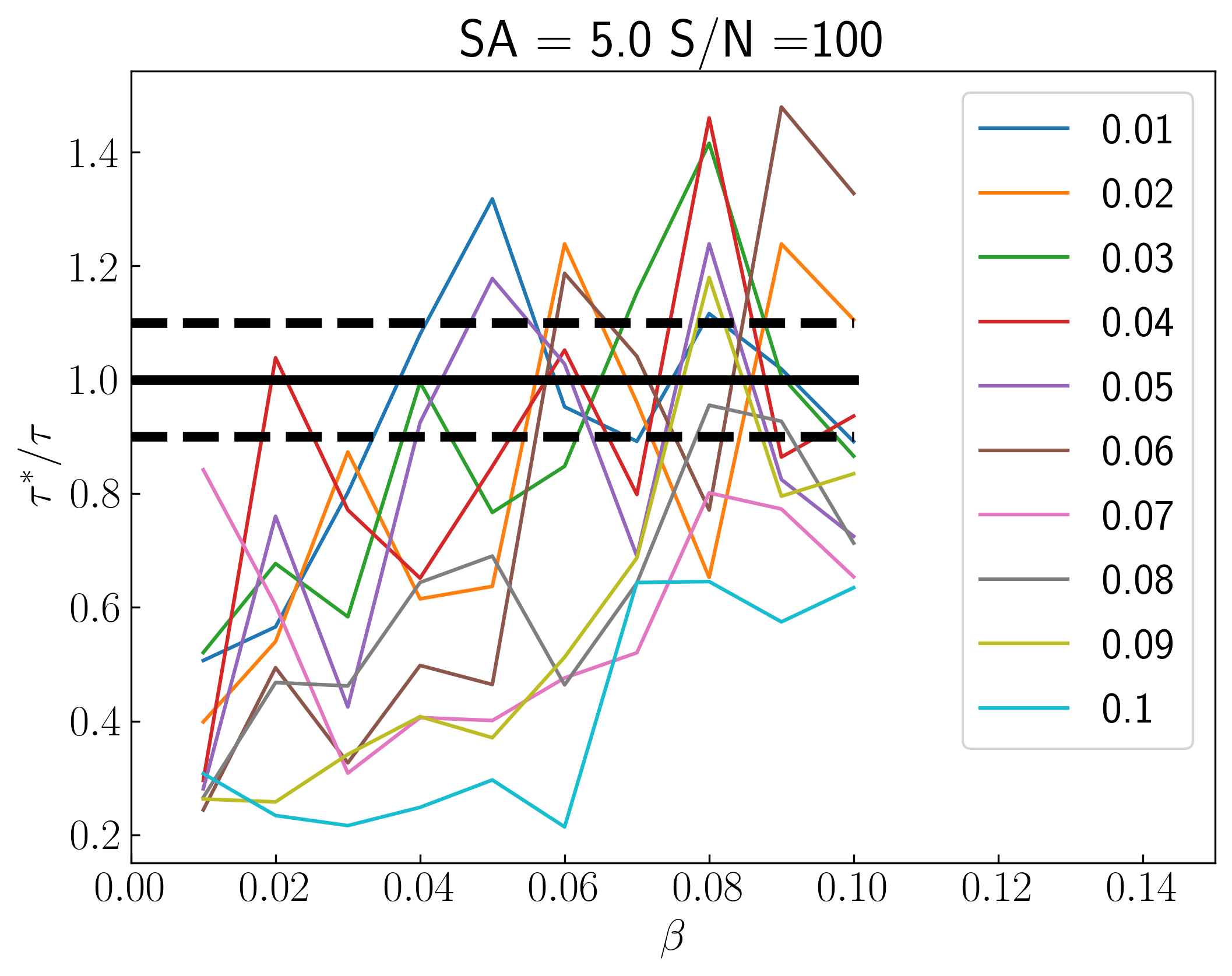} \\
 \includegraphics[width=70mm]{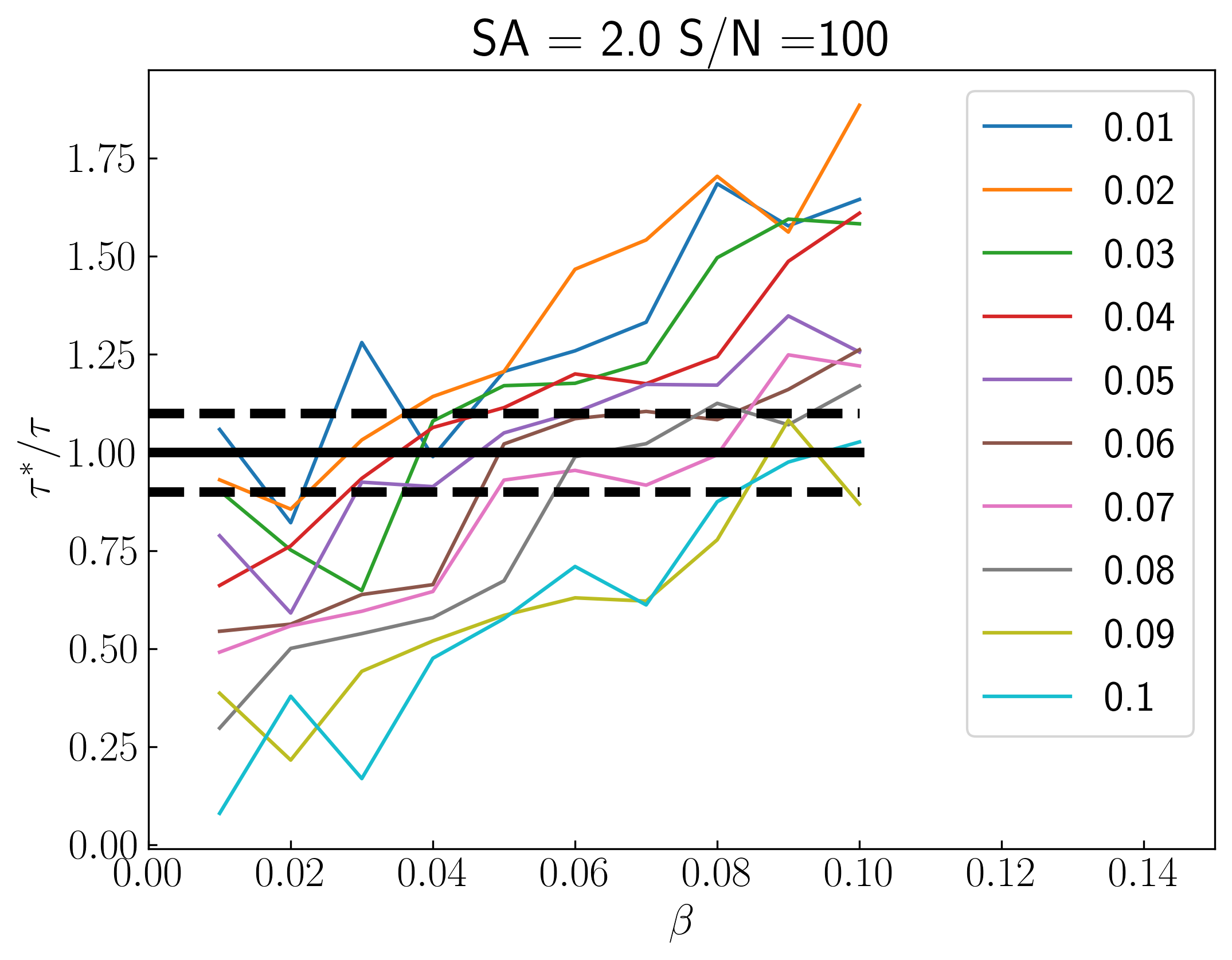} &   \includegraphics[width=70mm]{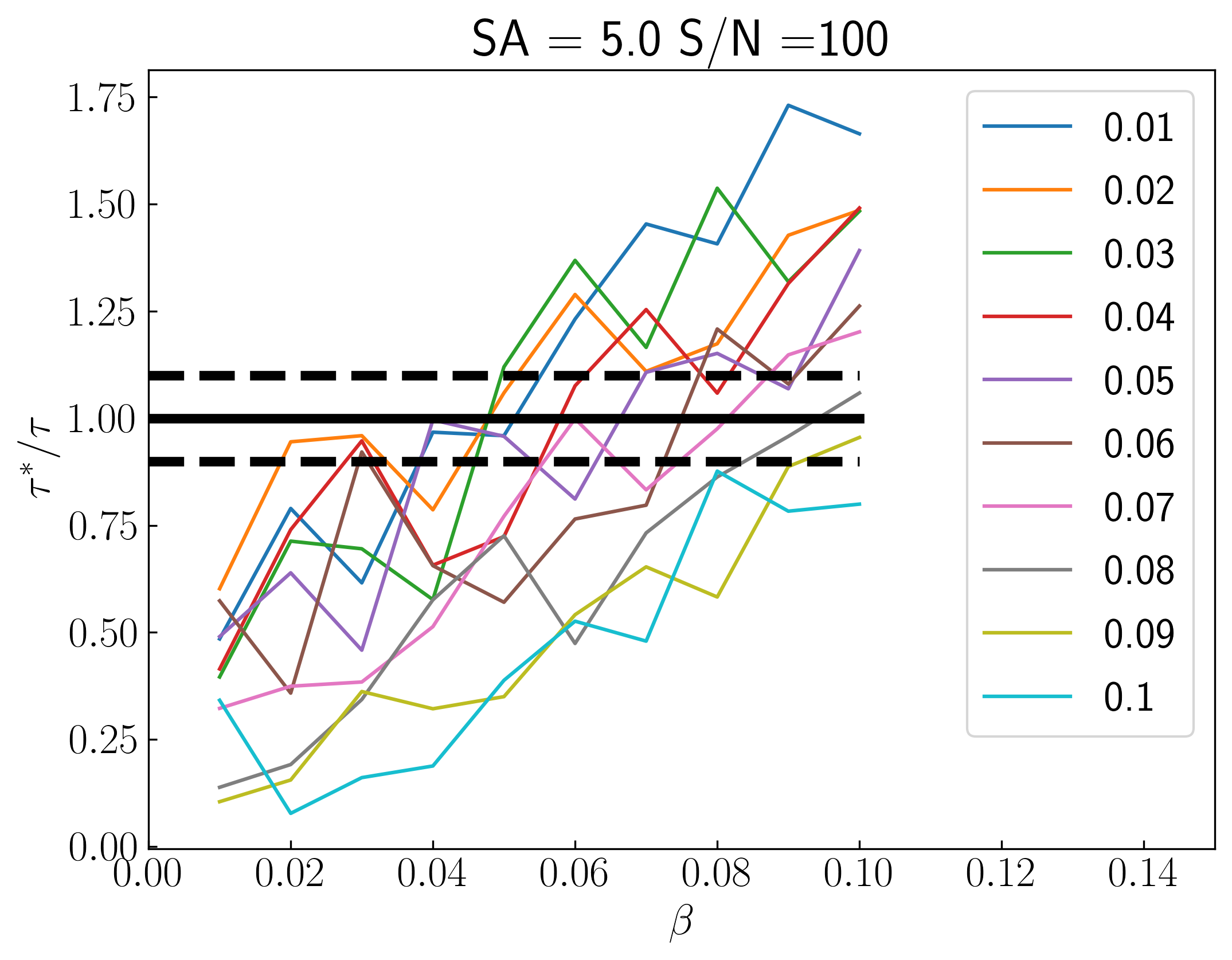} \\
\includegraphics[width=70mm]{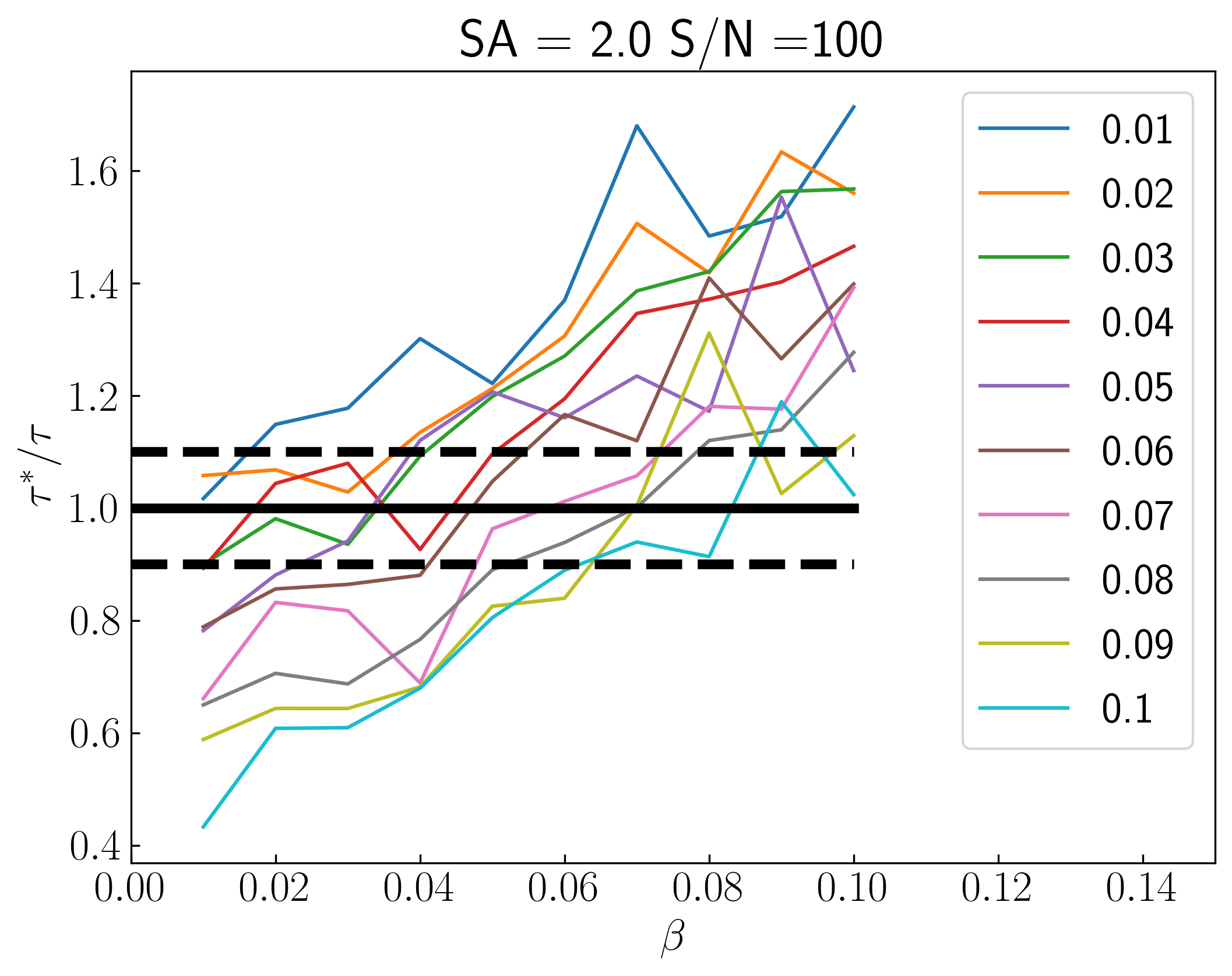} &   \includegraphics[width=70mm]{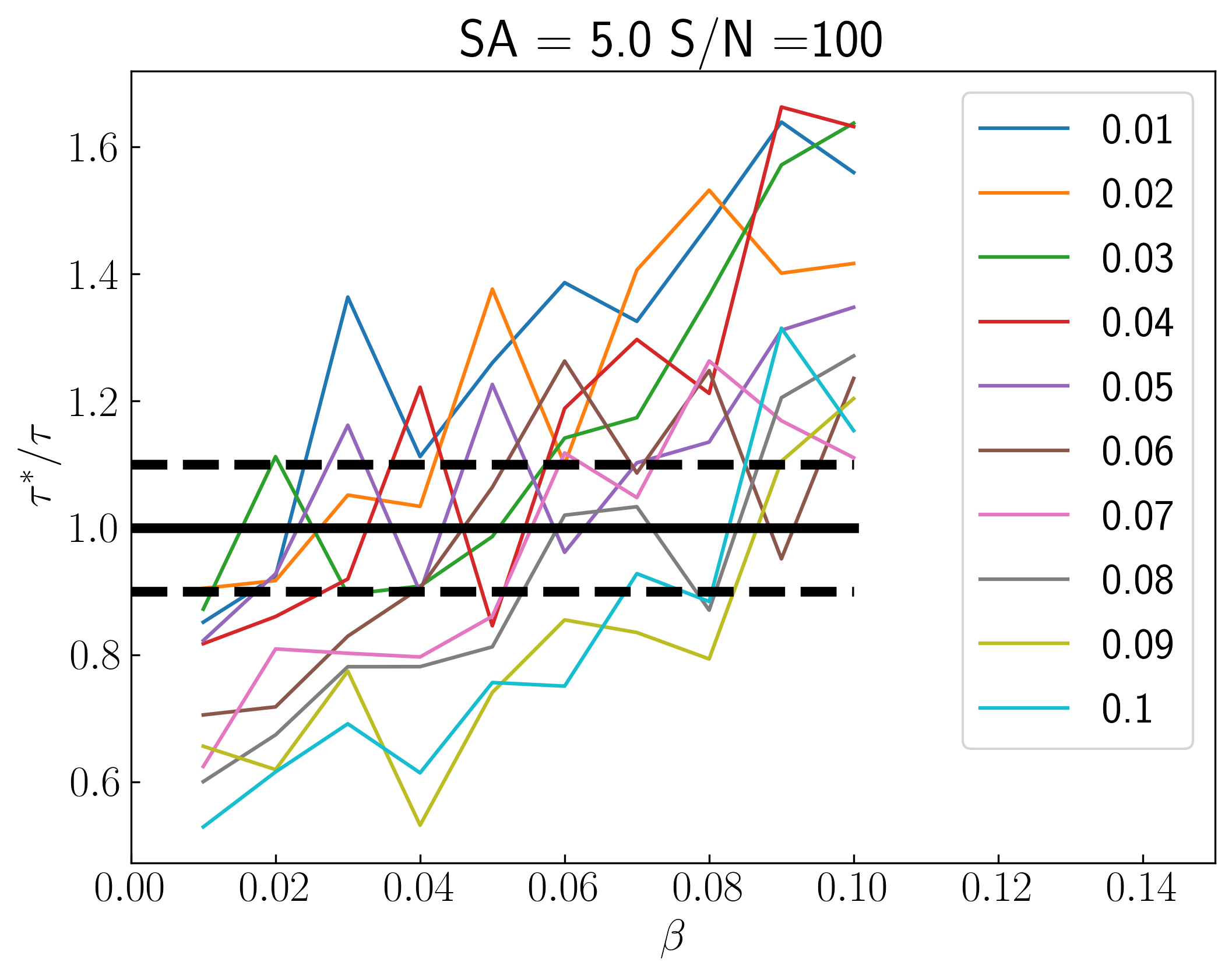} \\
\includegraphics[width=70mm]{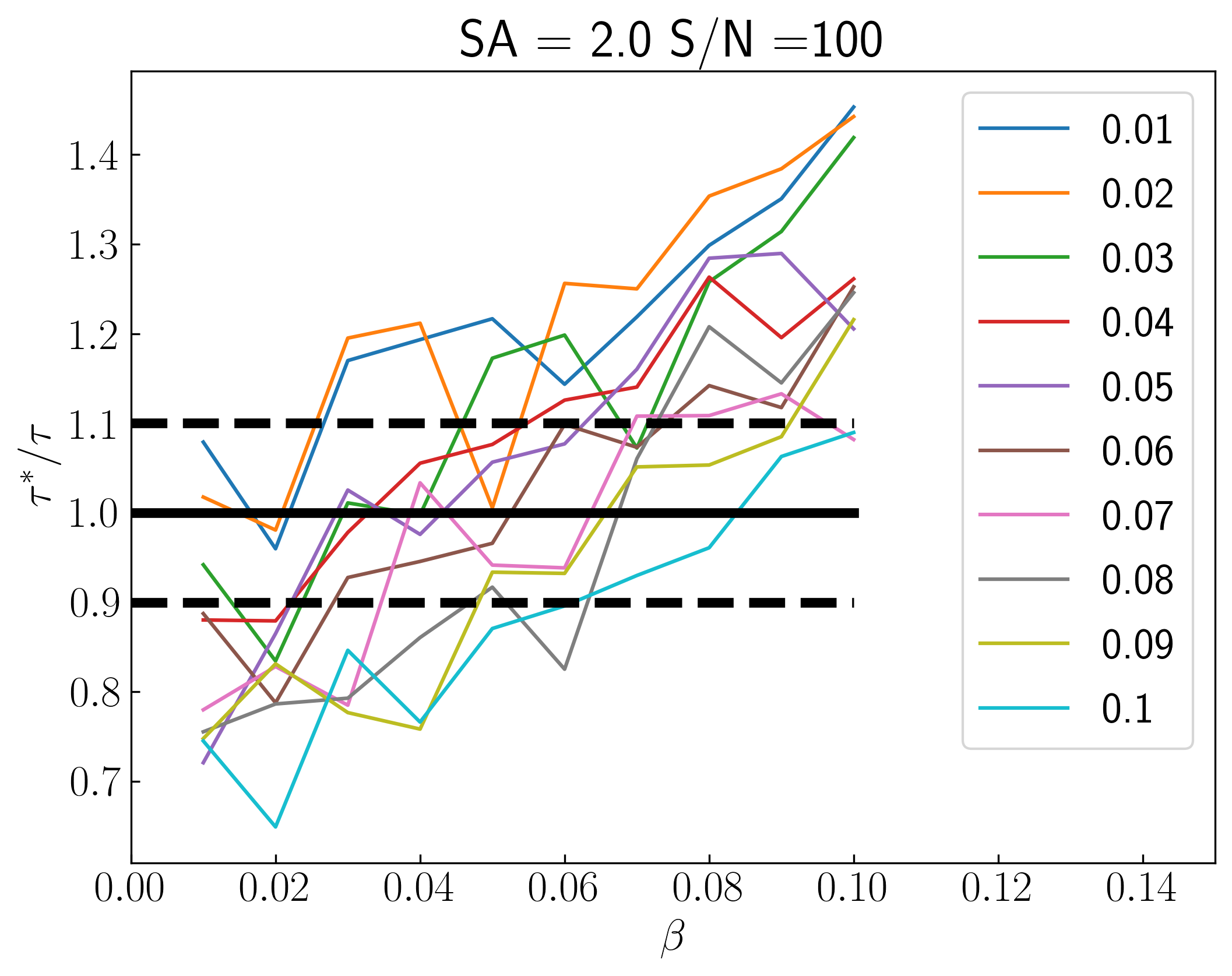} &   \includegraphics[width=70mm]{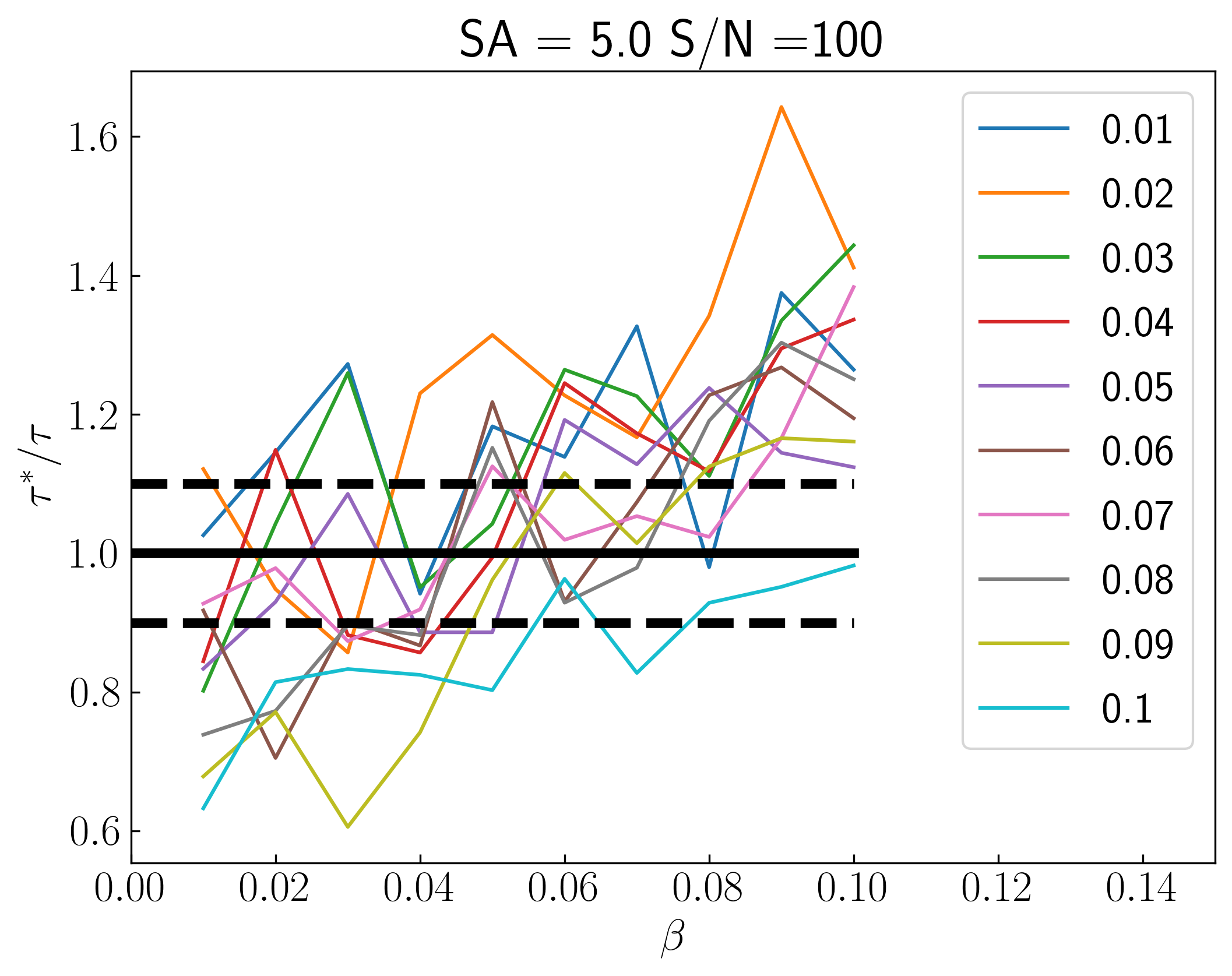} \\
\end{tabular}
\caption{Recovered distributions of delays ($\tau^{*}$) for different redshift cases. From top to bottom: Cases a, b, c and d for a fixed S/N = $100$ with $\alpha$ (coloured lines) and $\beta$ between 1\% and 10\%. The left and right panels show $\tau^{*}$ for $\Delta t = 2.0$ and 5 days, respectively.}
\label{fig:Apx6}
\end{figure*}

\begin{figure*}
\begin{tabular}{cc}
  \includegraphics[width=70mm]{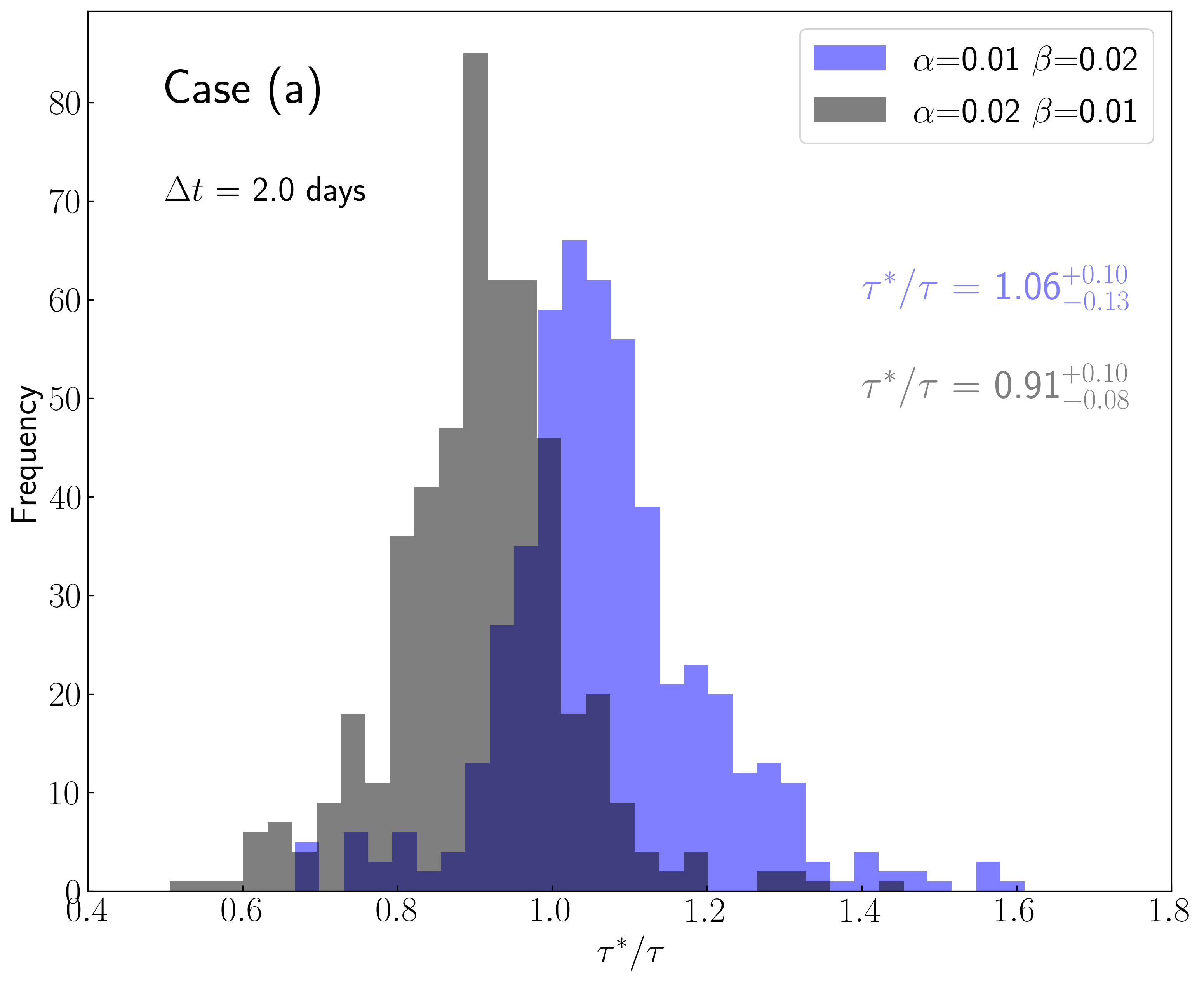} & \includegraphics[width=70mm]{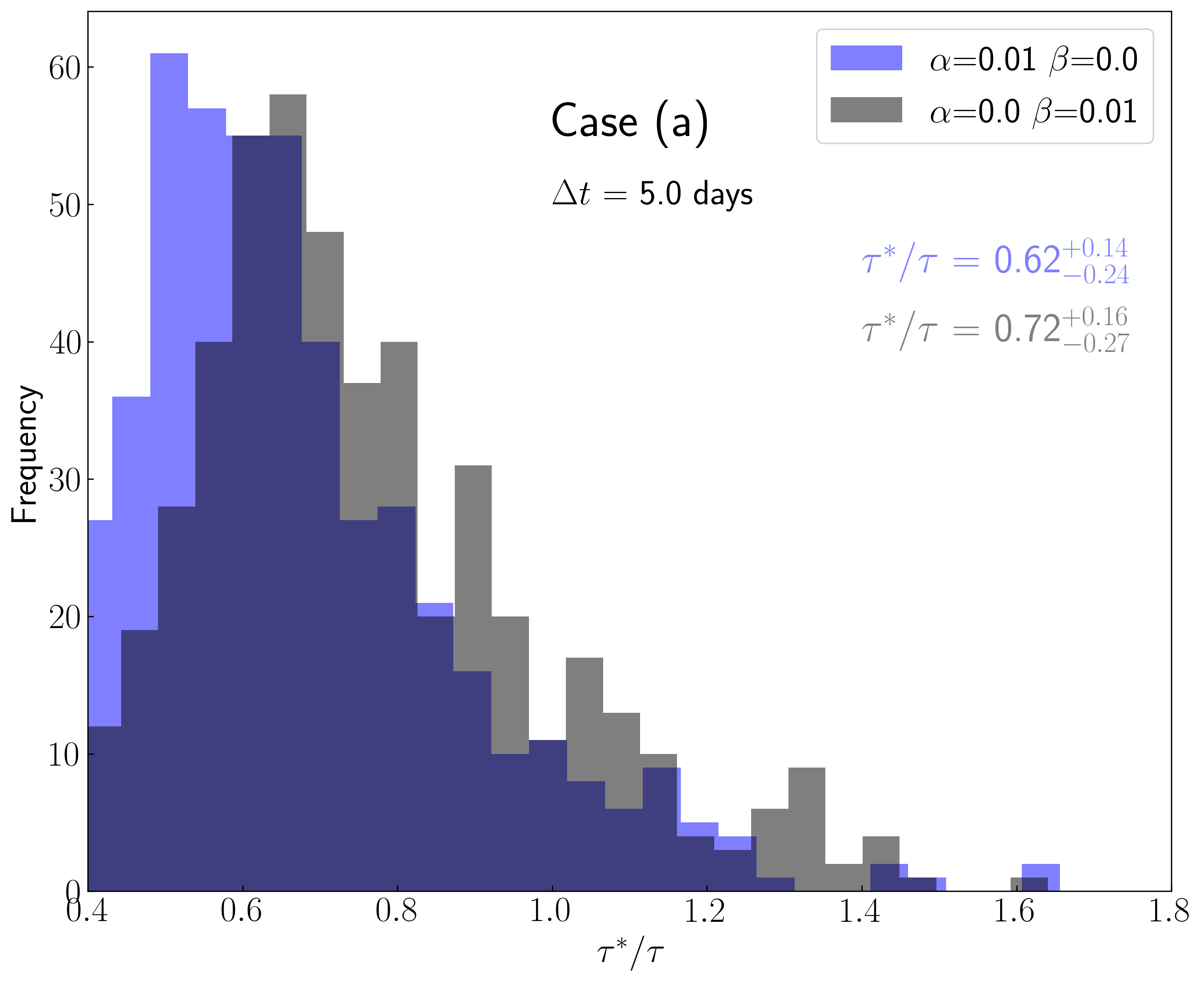} \\
 \includegraphics[width=70mm]{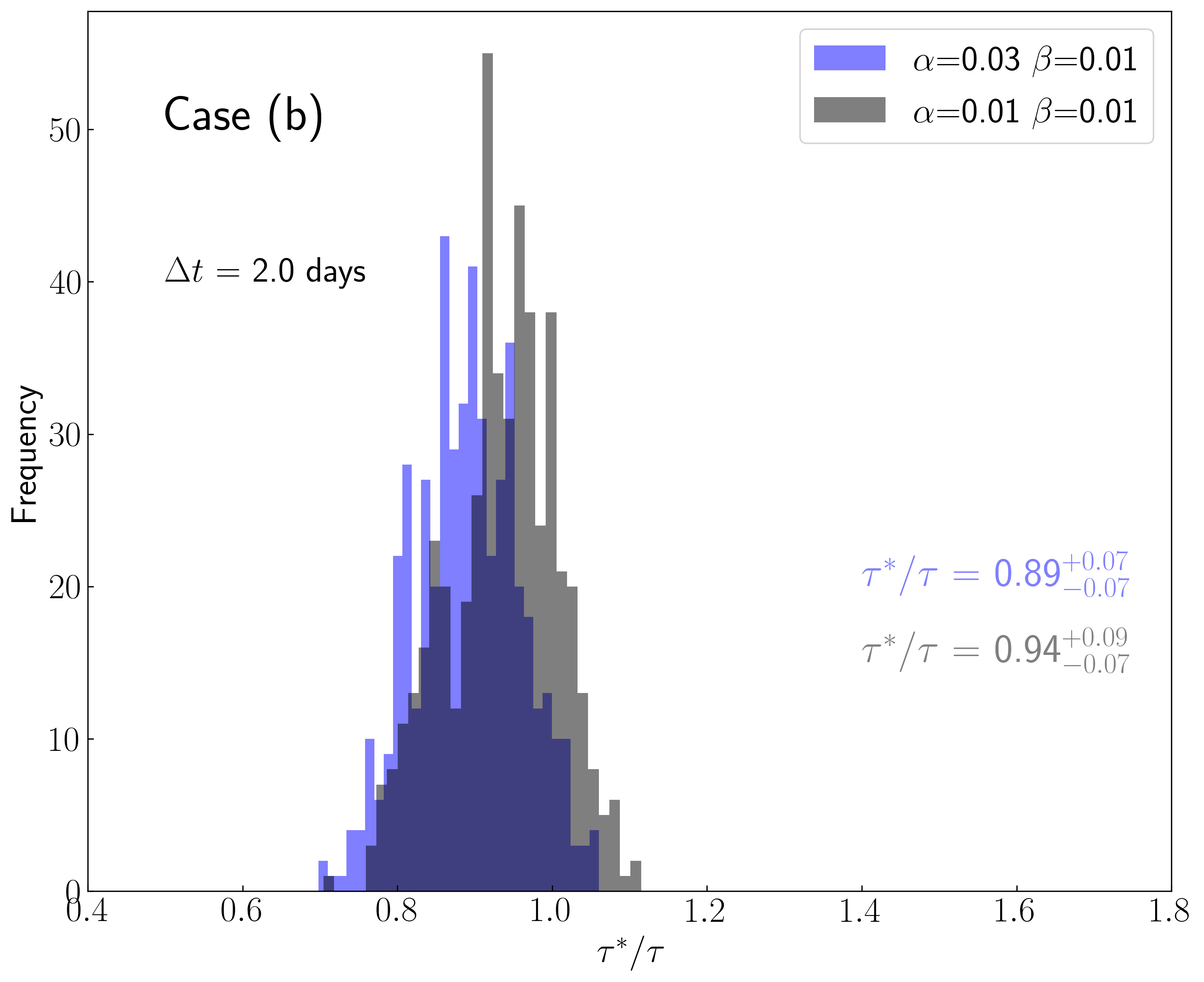} & \includegraphics[width=70mm]{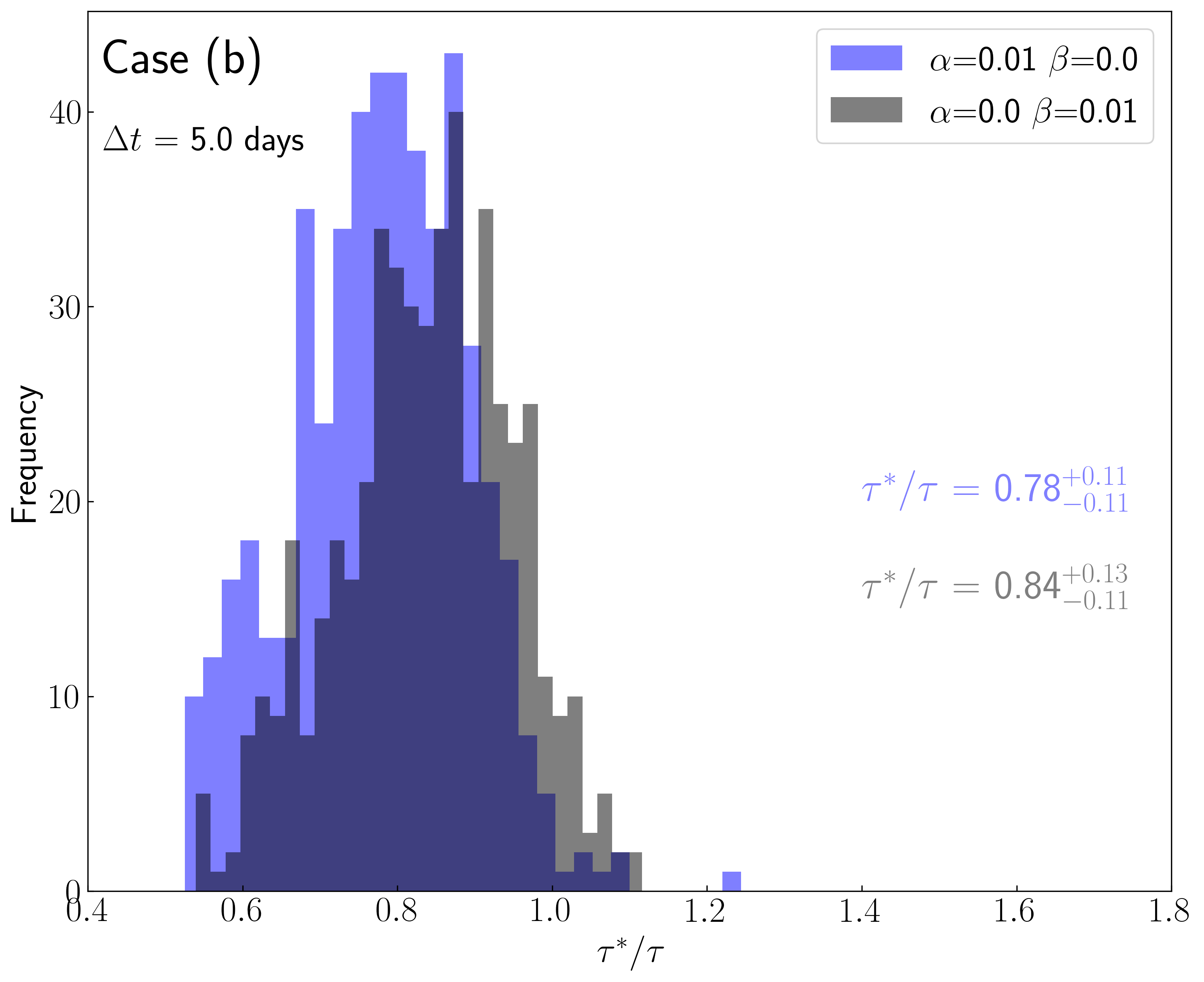} \\
\includegraphics[width=70mm]{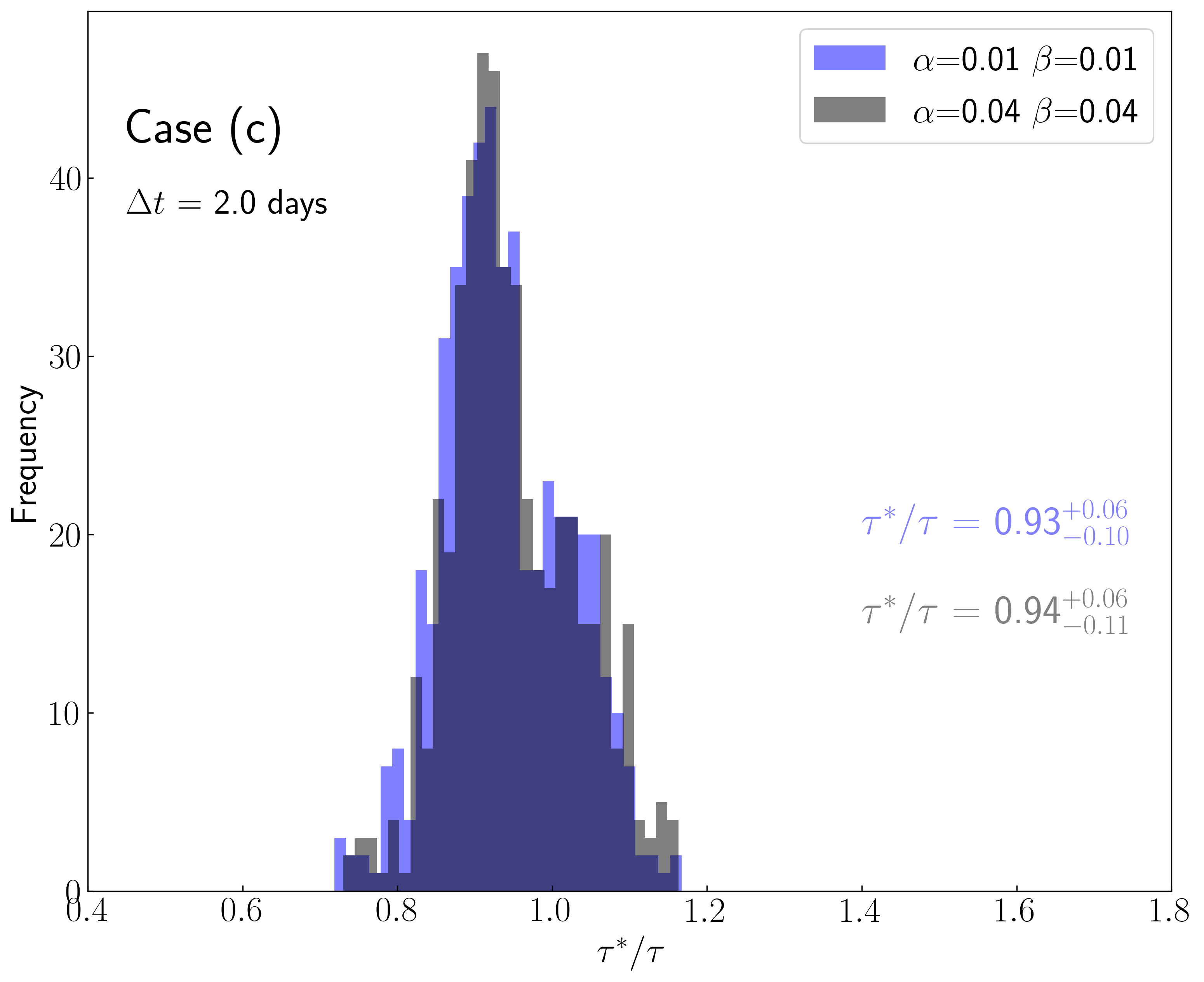} &   \includegraphics[width=70mm]{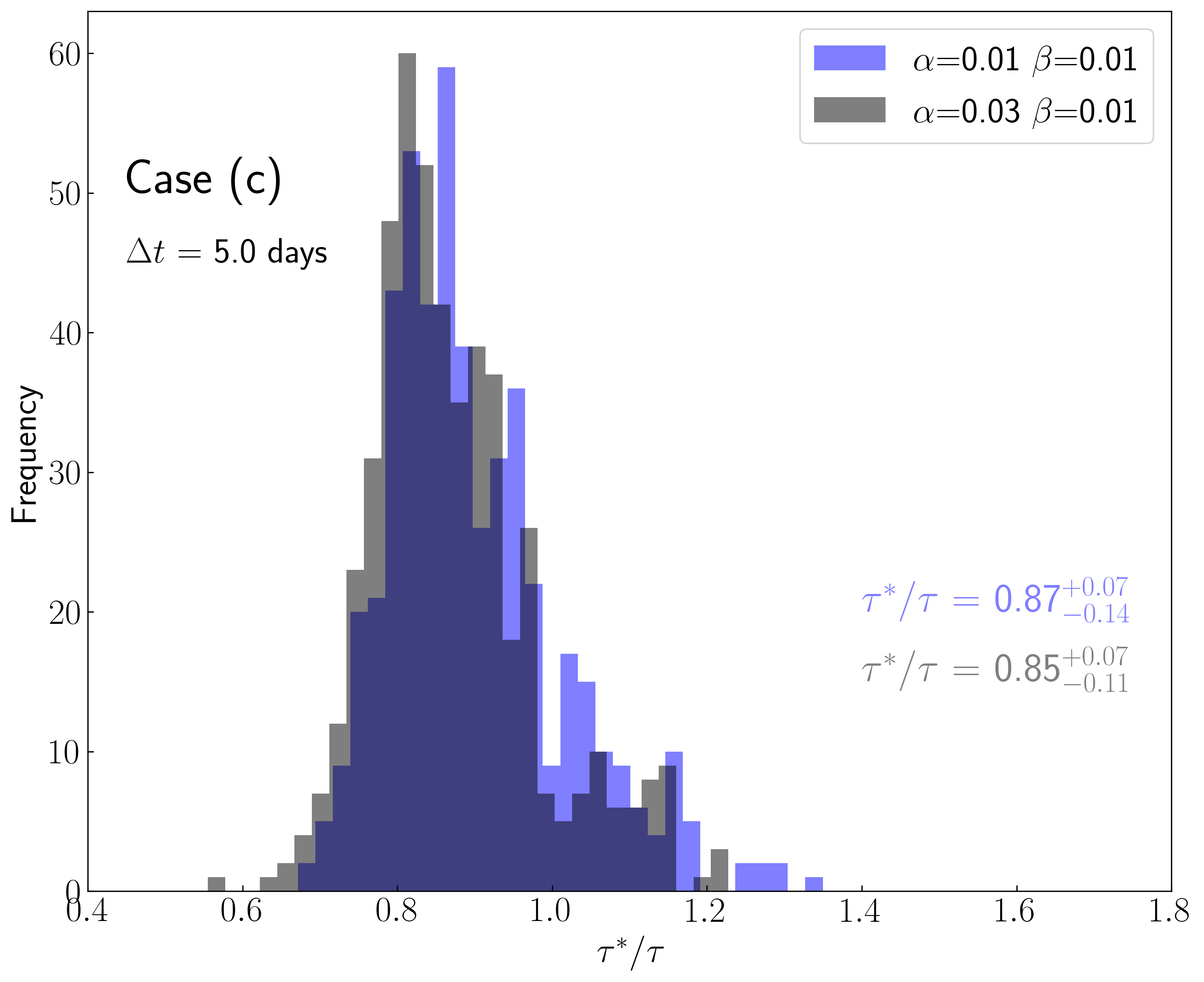} \\
\includegraphics[width=70mm]{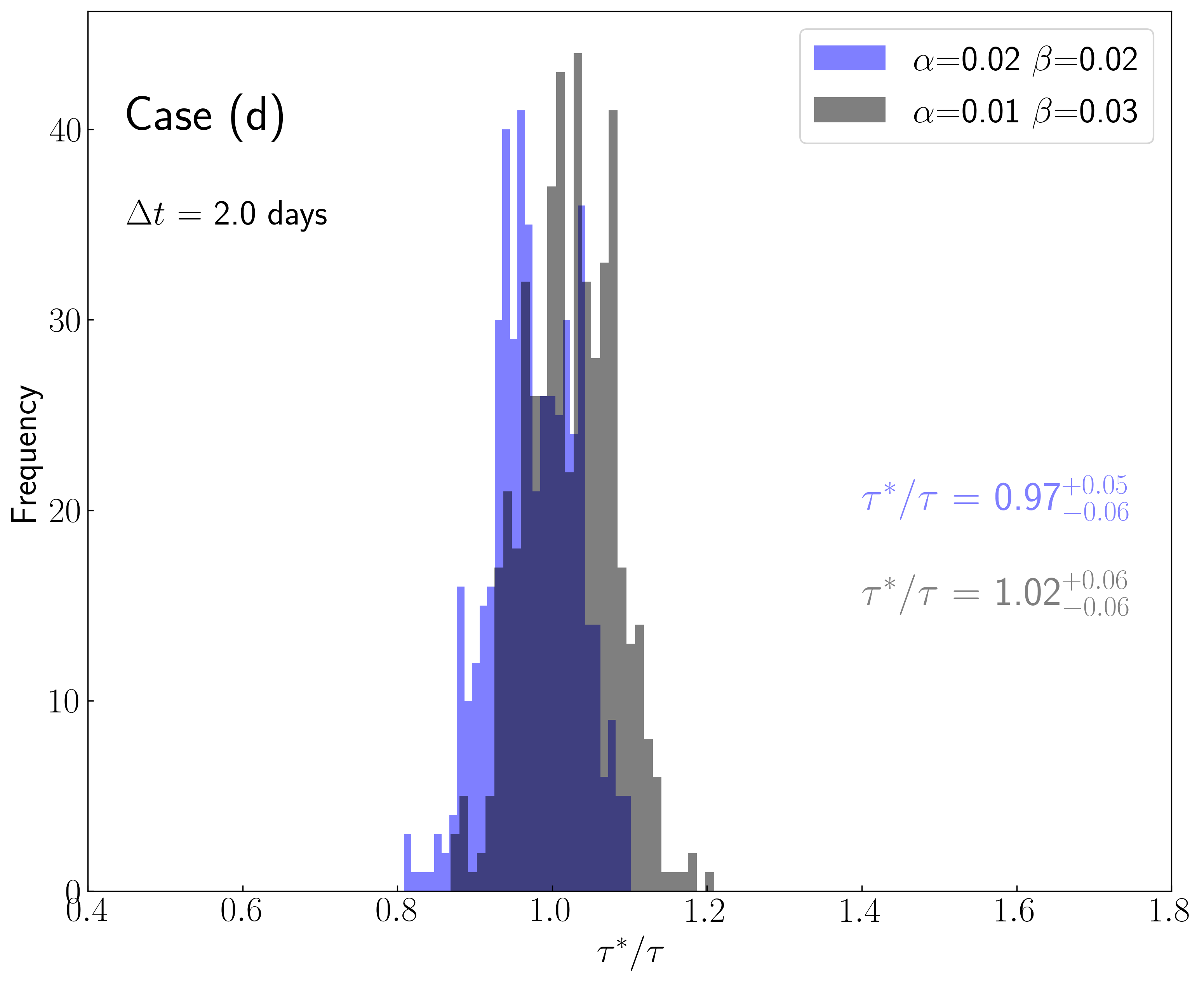} &   \includegraphics[width=70mm]{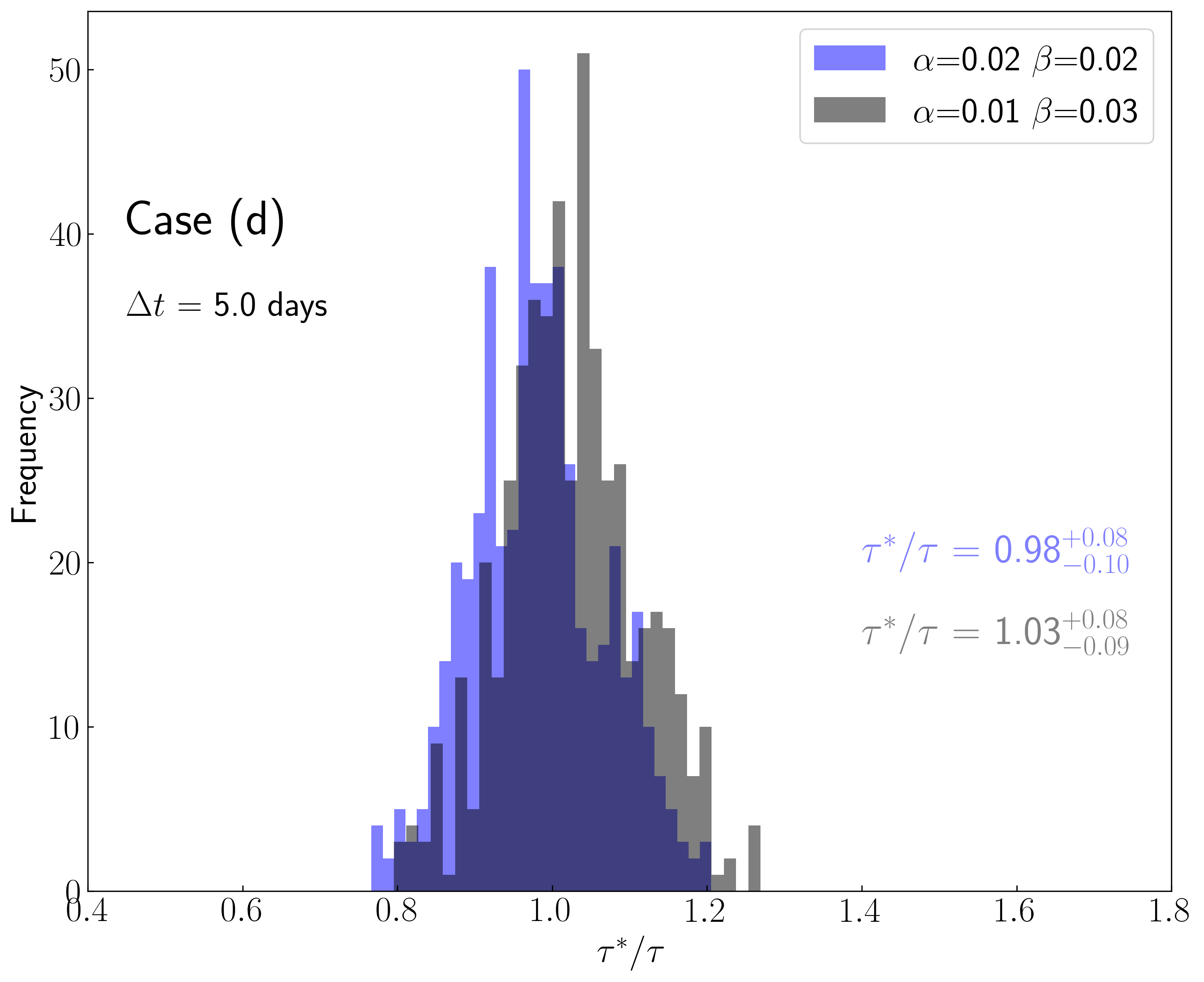} \\
\end{tabular}
\caption{Recovered distributions of delays ($\tau^{*}$) obtained for 1000 mock light curves and specific $\alpha$ and $\beta$ contributions. The numbers indicate the median and the central 68\% confidence (1$\sigma$) intervals of the distributions.}
\label{fig:Apx7}
\end{figure*}

\begin{figure*}
\begin{tabular}{cc}
  \includegraphics[width=72mm]{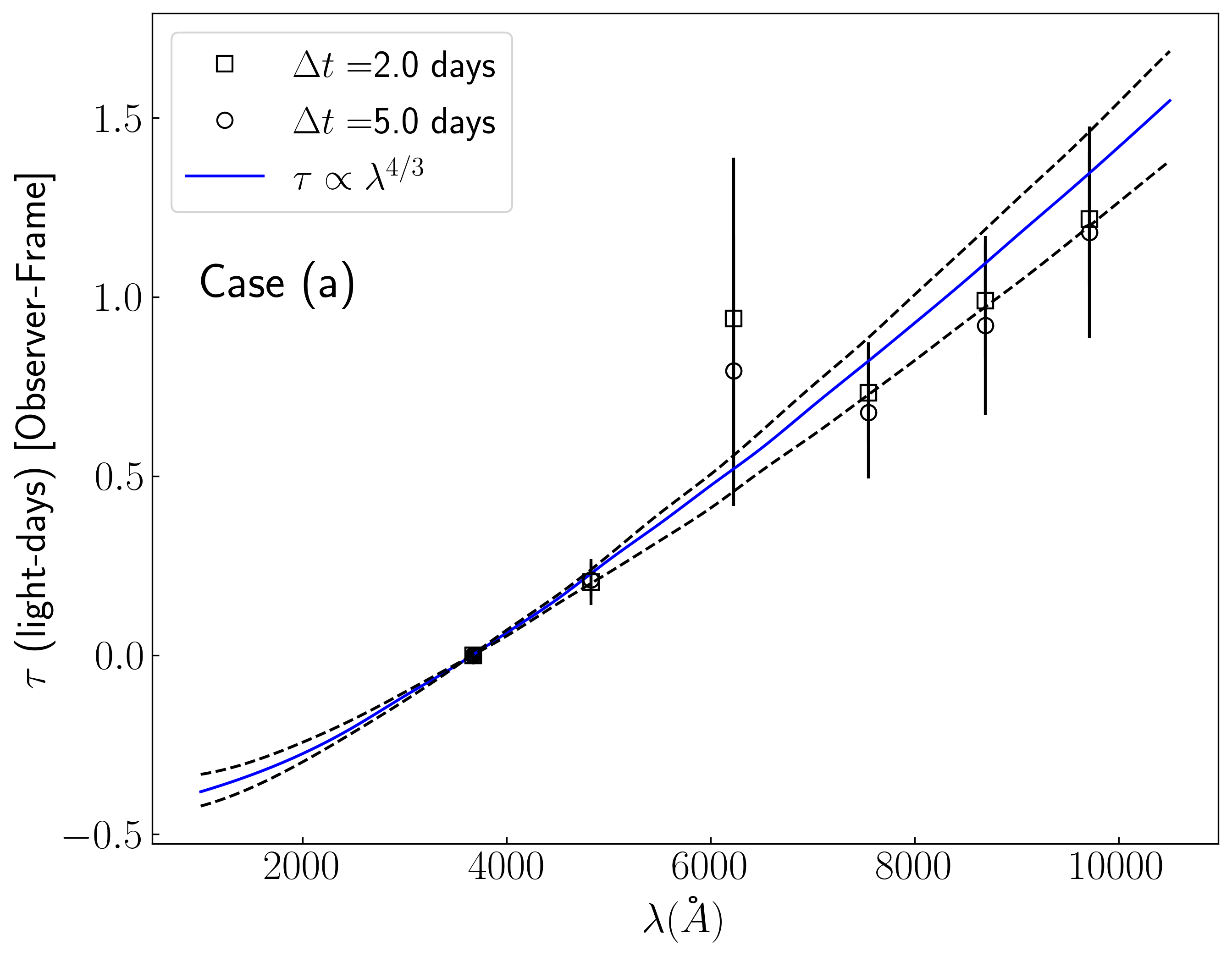} &   \includegraphics[width=70mm]{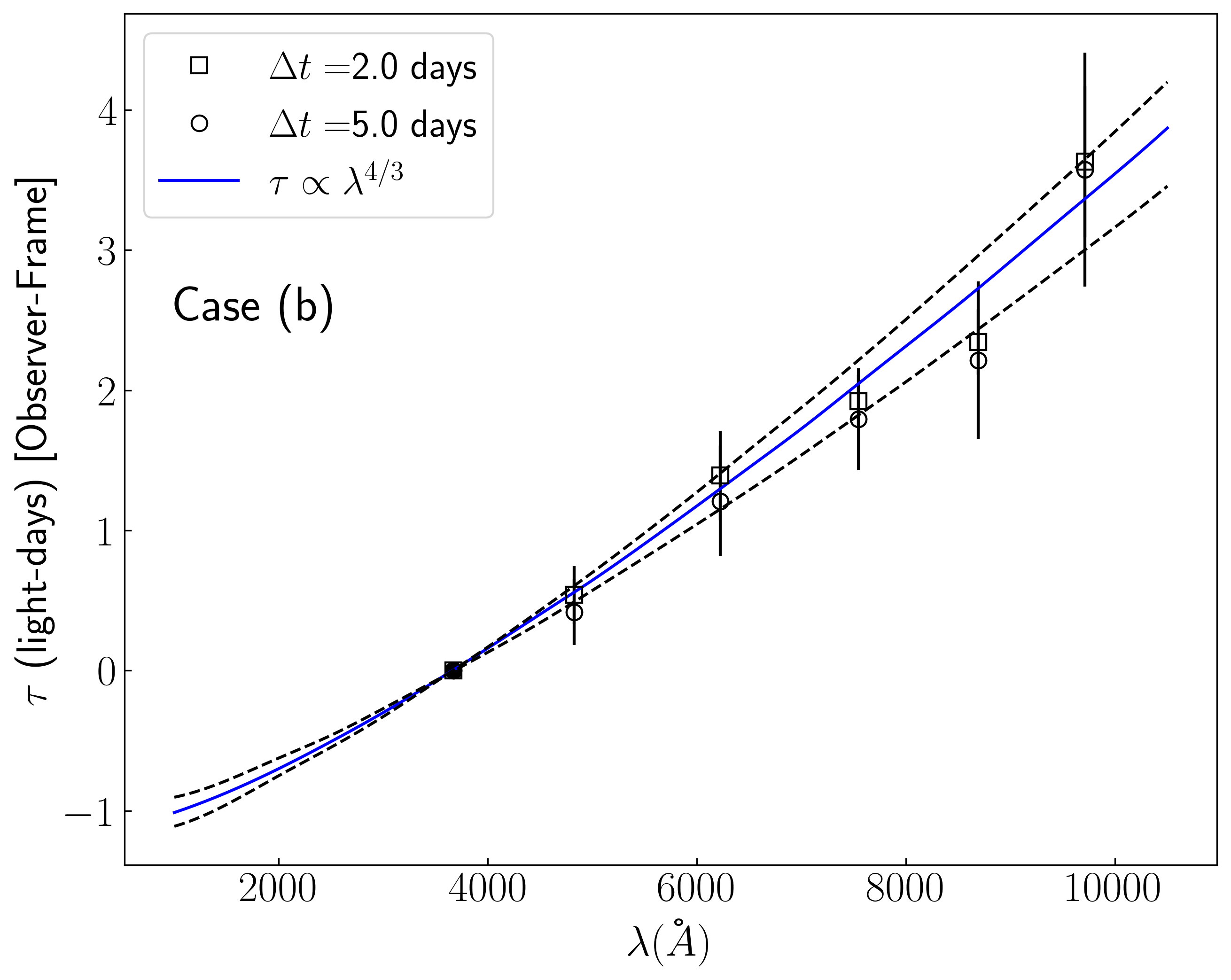} \\
 \includegraphics[width=70mm]{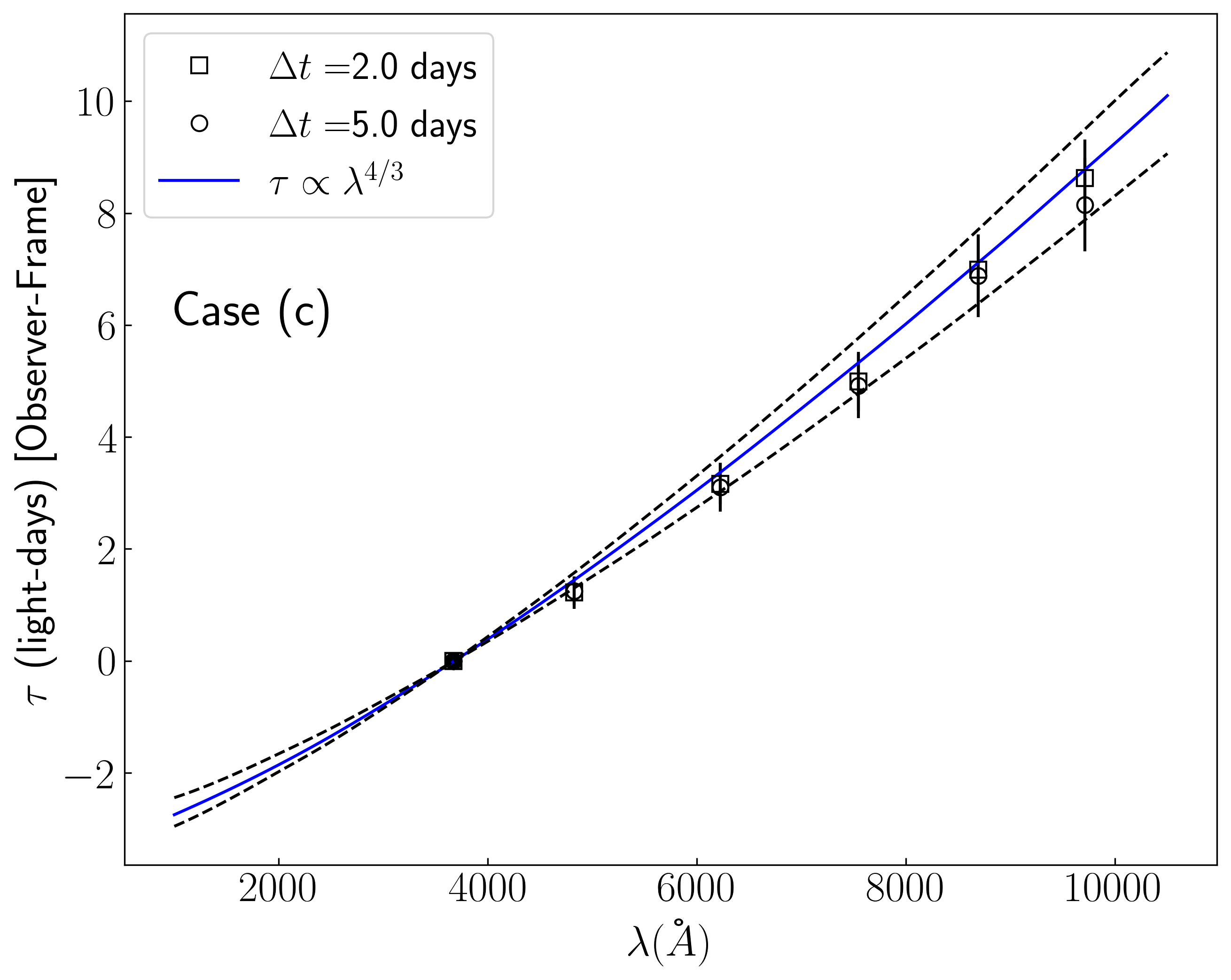} &   \includegraphics[width=71mm]{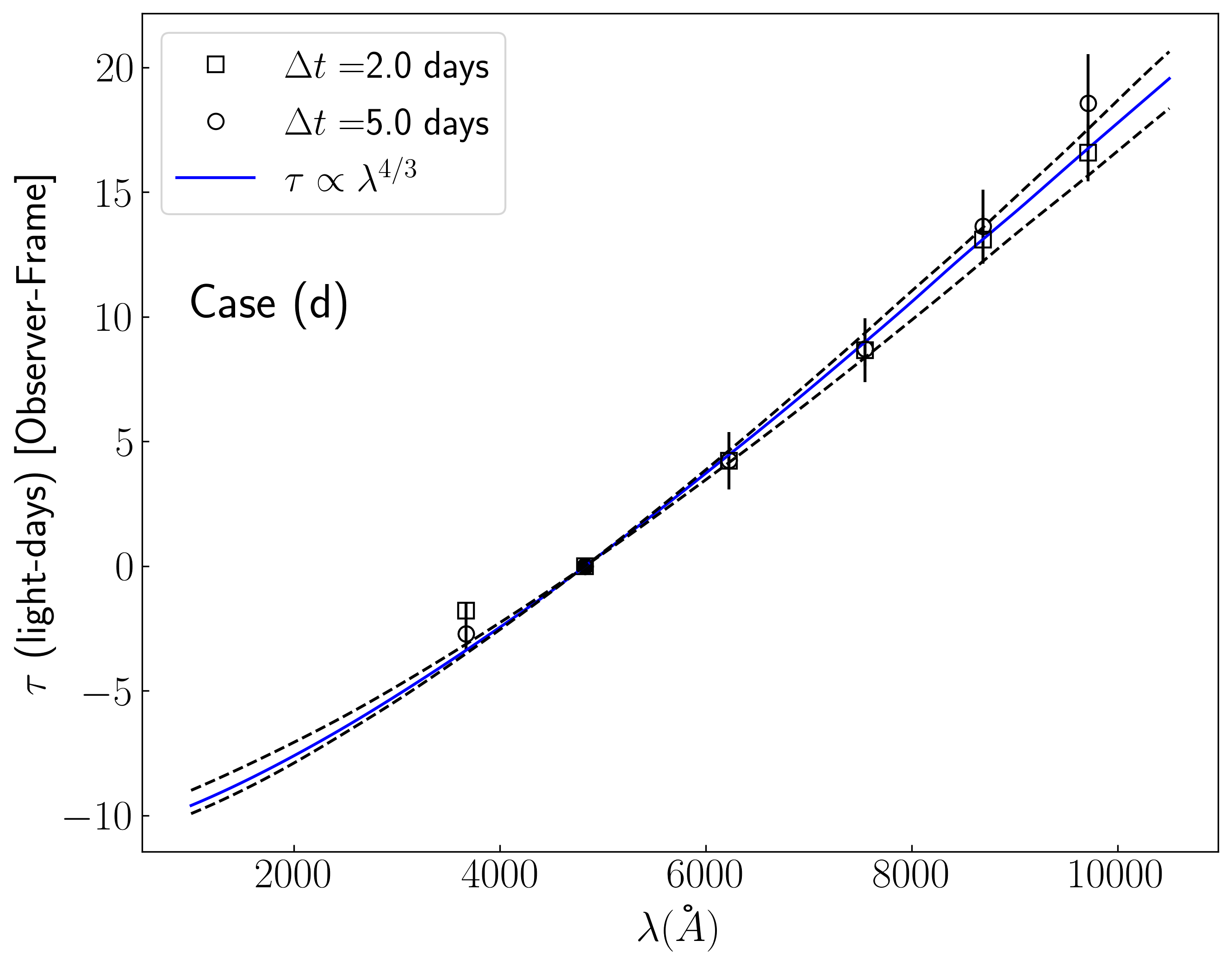} \\
 \end{tabular}
\caption{Same as Figure \ref{delayavgres}, but for different cases as listed in Table~\ref{table1}.}
\label{fig:Apx8}
\end{figure*}

\begin{figure*}
   \includegraphics[width=0.67\columnwidth]{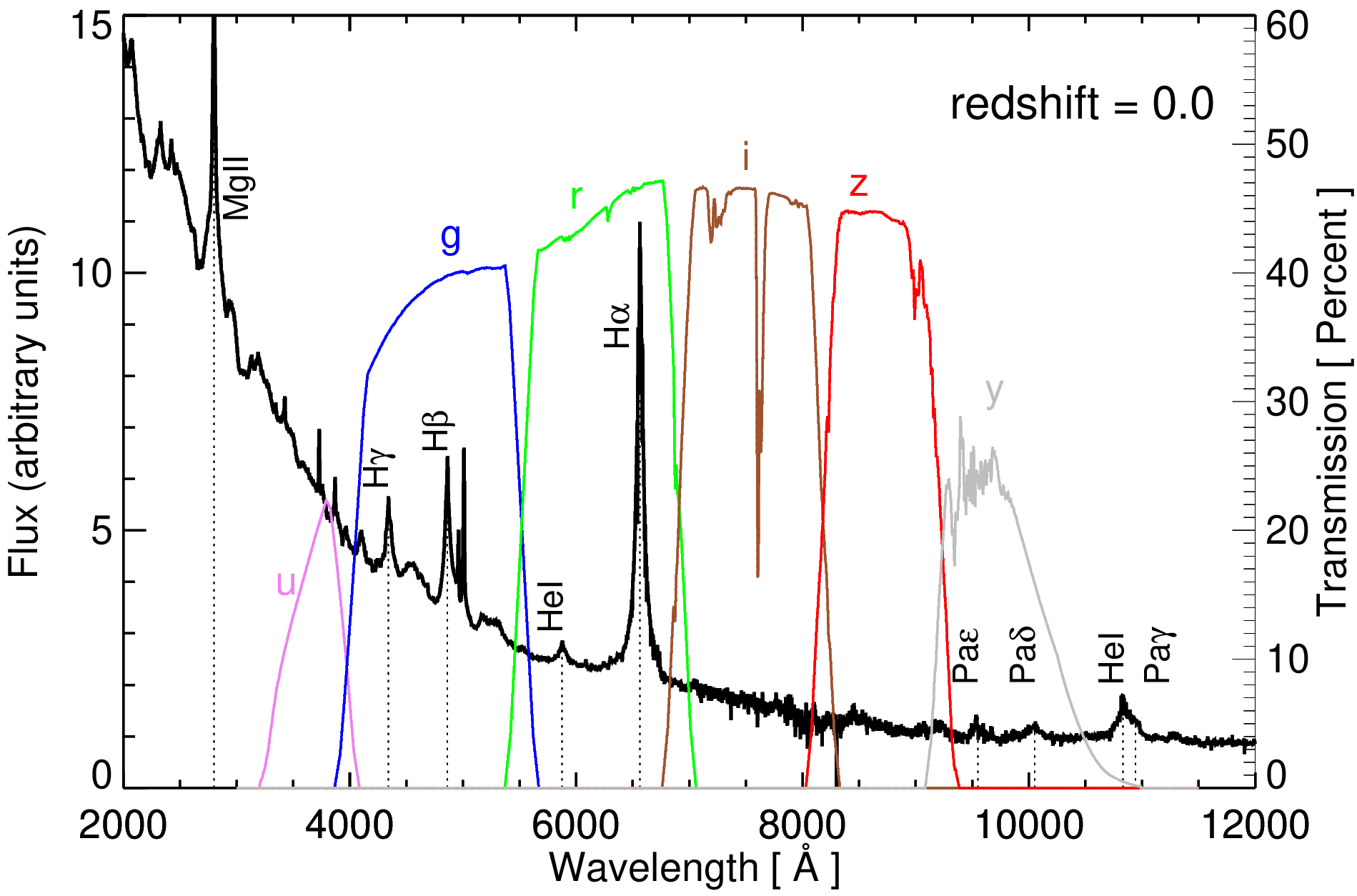} 
   \includegraphics[width=0.67\columnwidth]{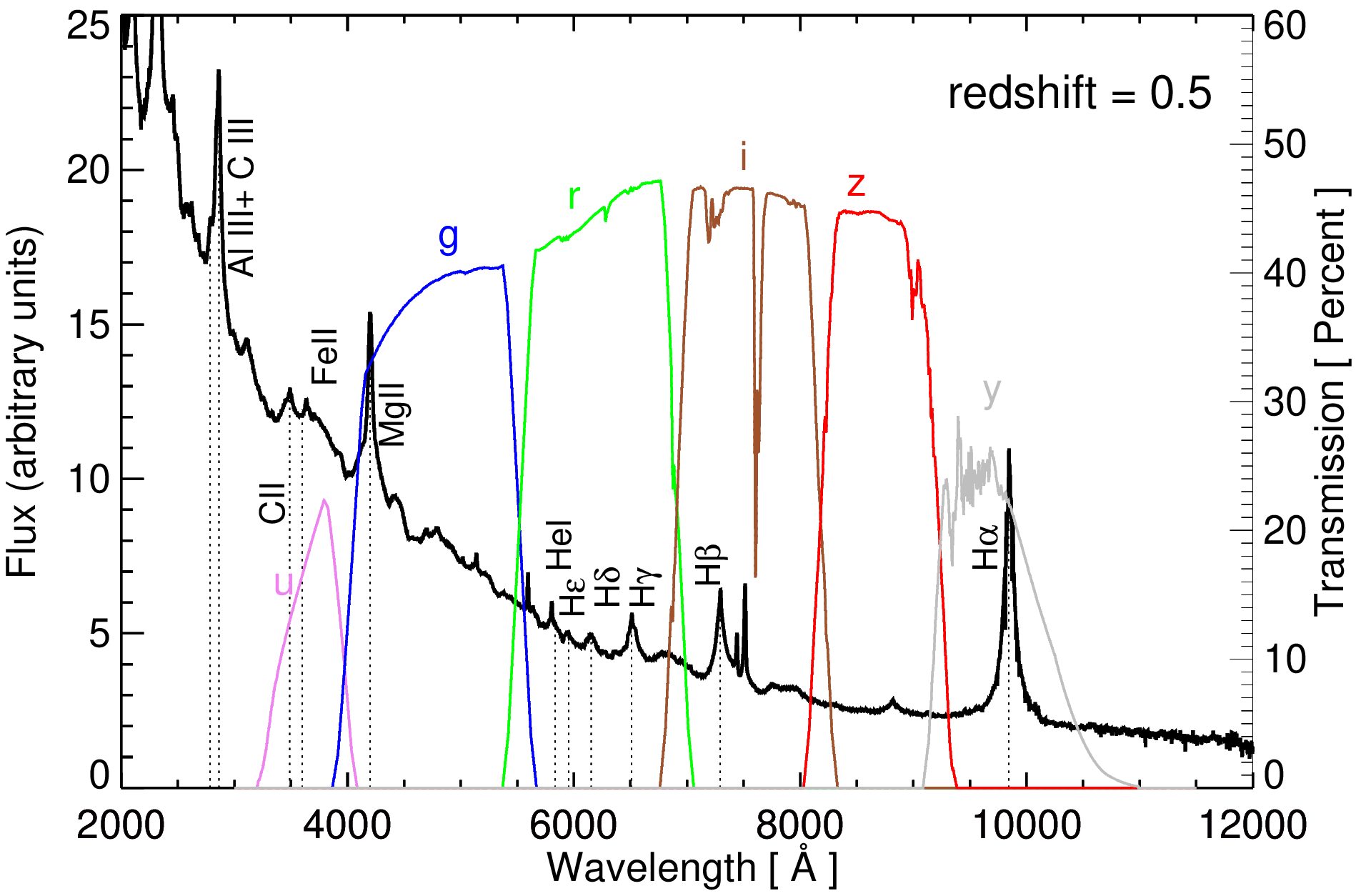} 
   \includegraphics[width=0.67\columnwidth]{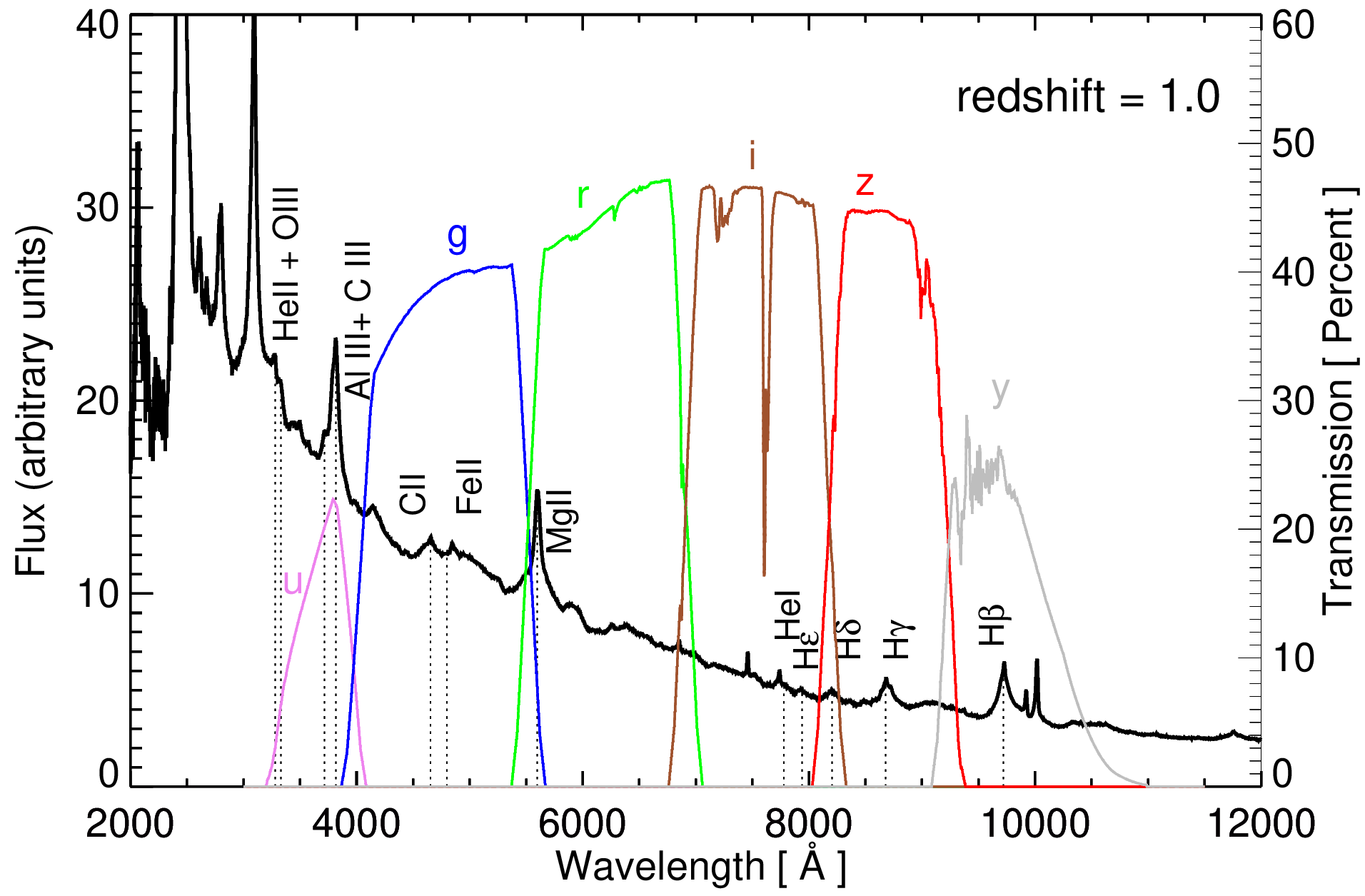}
   \includegraphics[width=0.67\columnwidth]{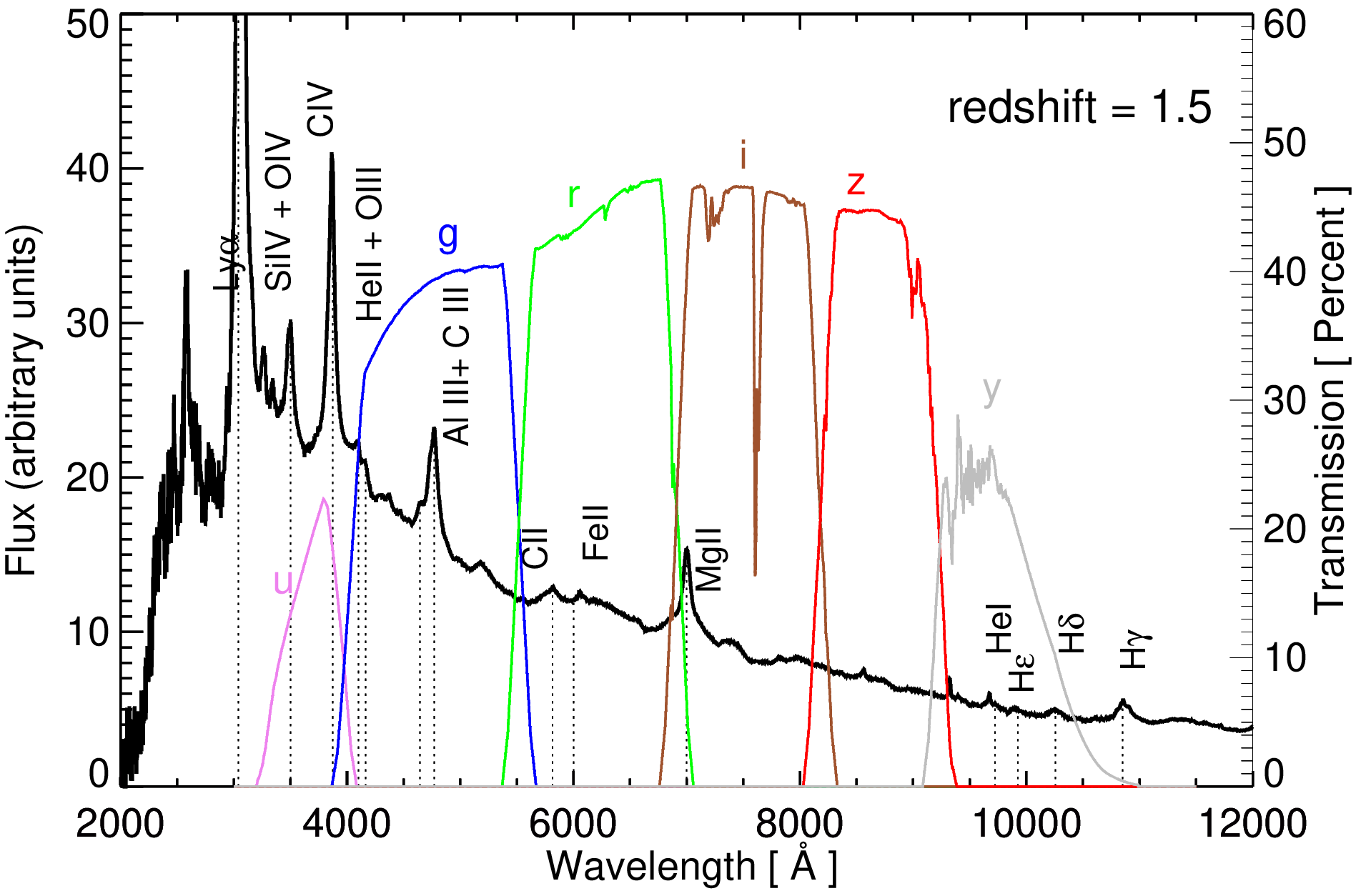}
   \includegraphics[width=0.67\columnwidth]{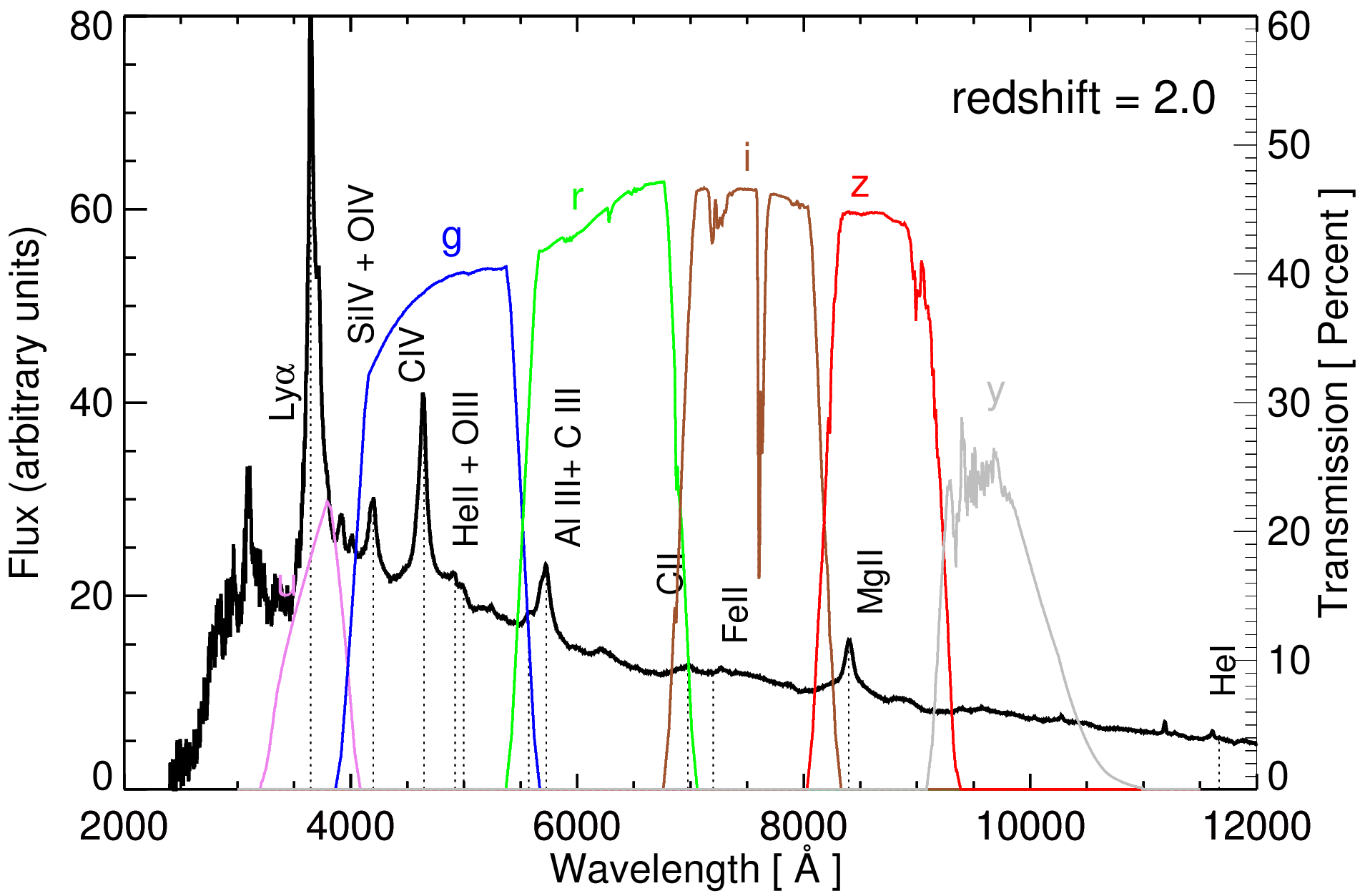}
   \includegraphics[width=0.67\columnwidth]{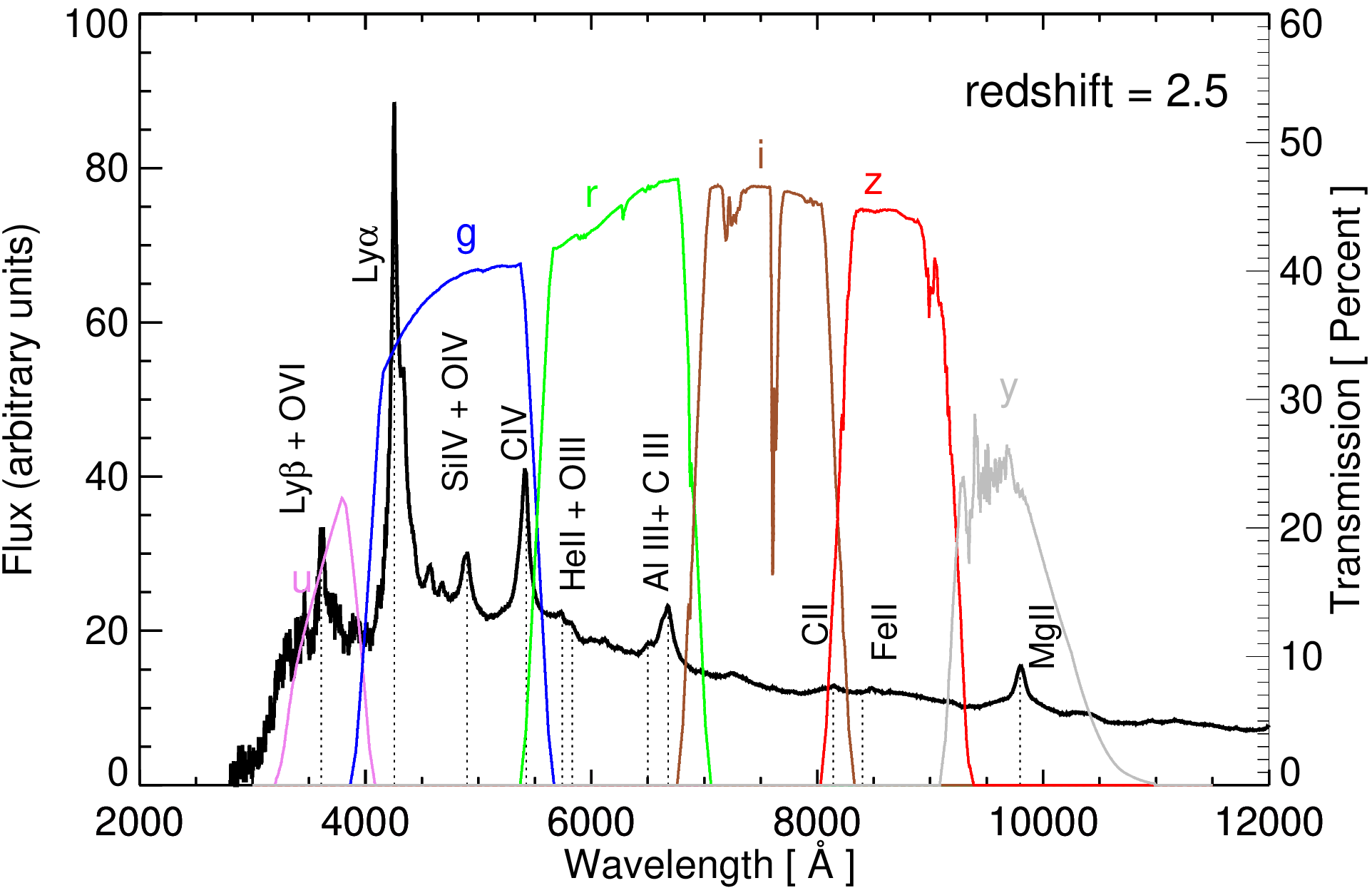}
   \includegraphics[width=0.67\columnwidth]{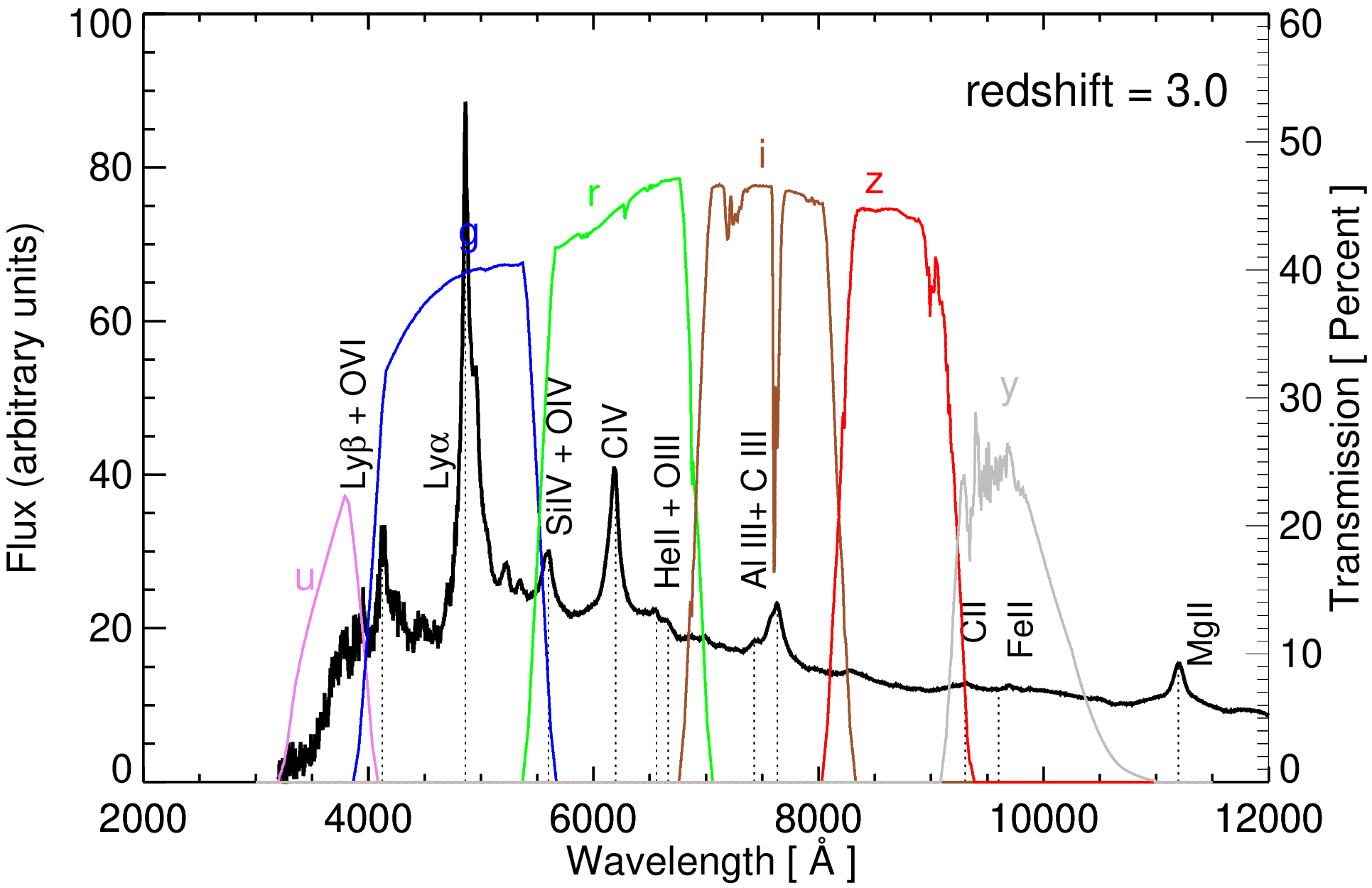}
   \includegraphics[width=0.67\columnwidth]{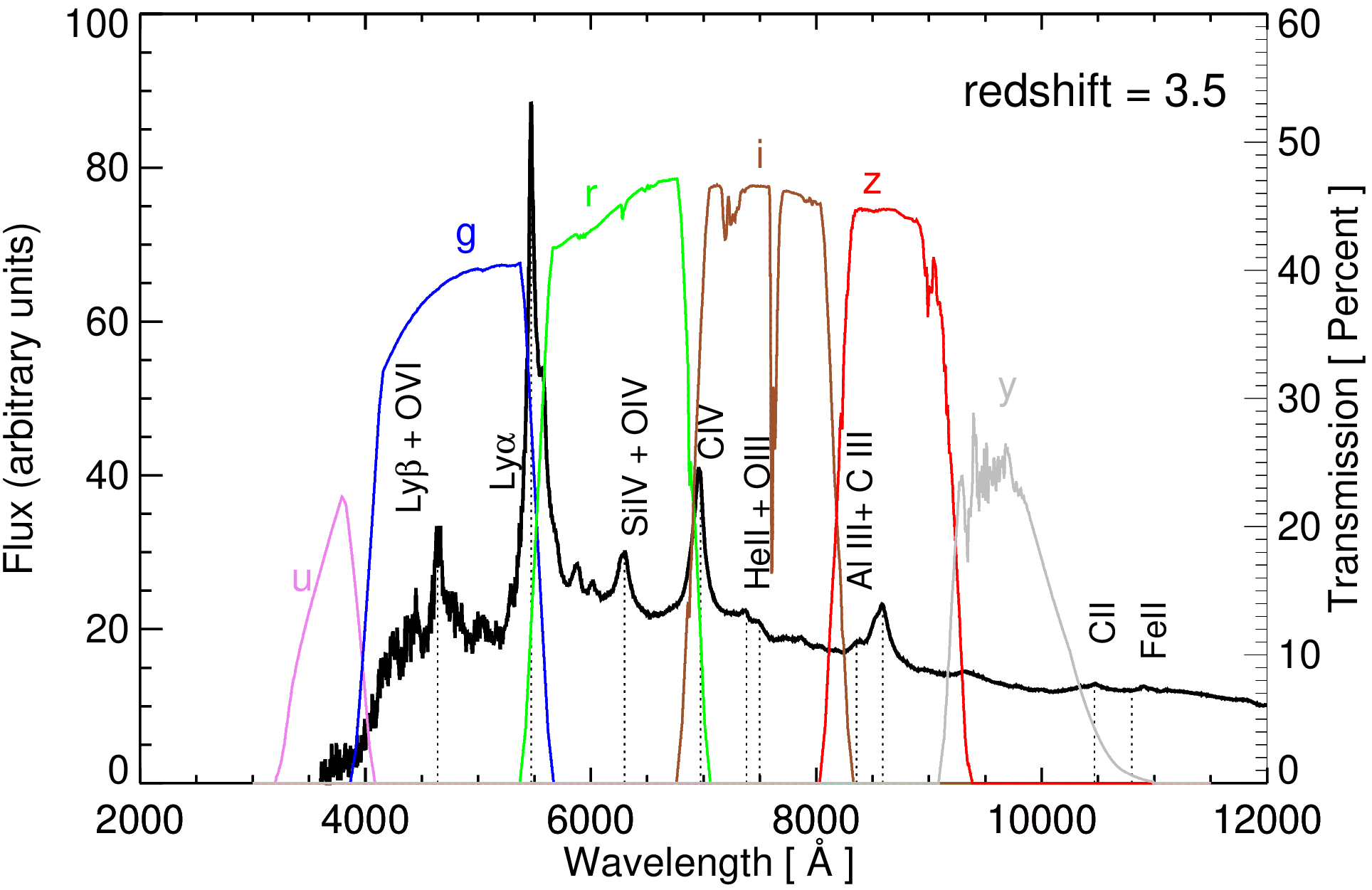}
   \includegraphics[width=0.67\columnwidth]{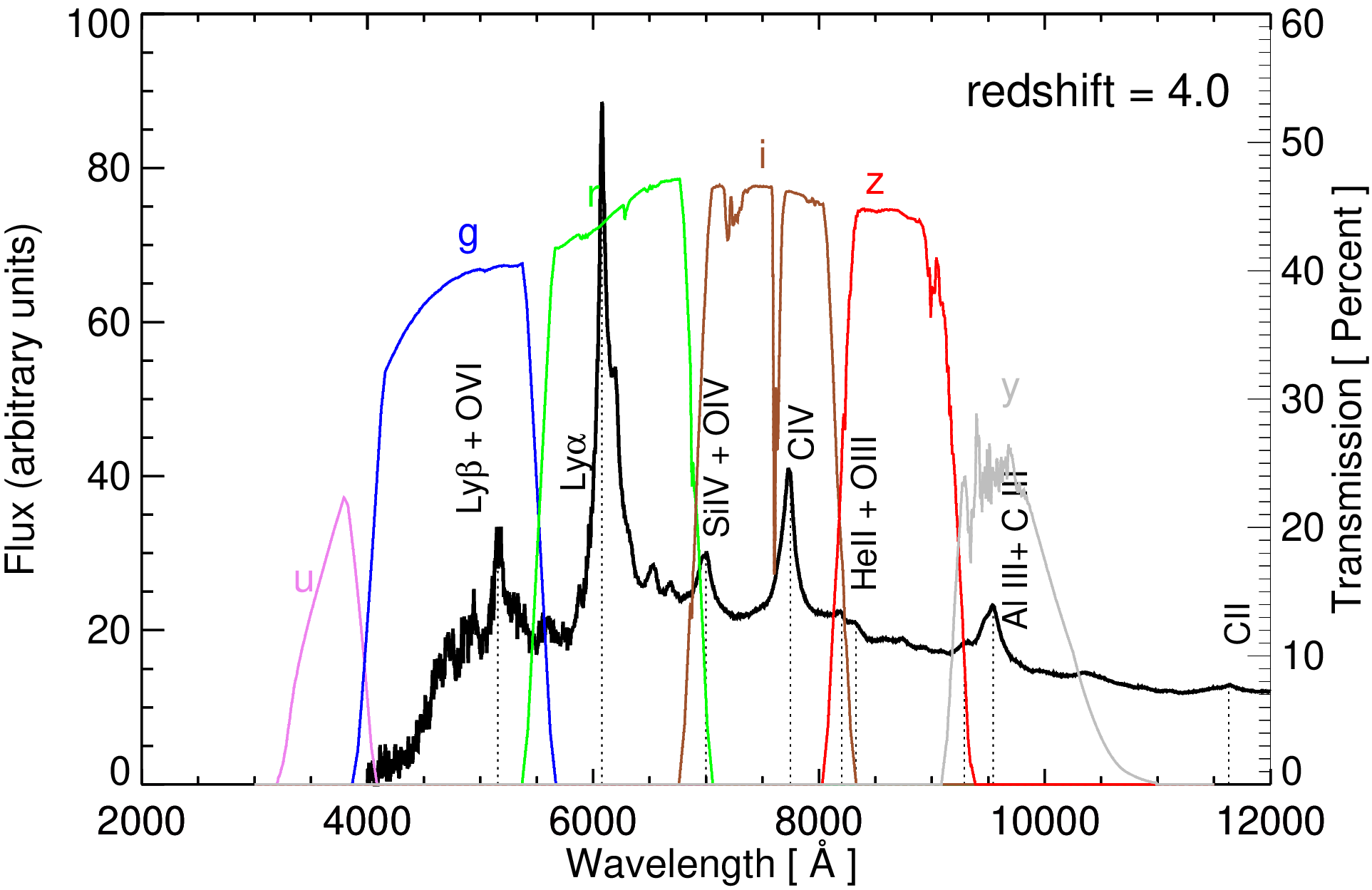}
   \includegraphics[width=0.67\columnwidth]{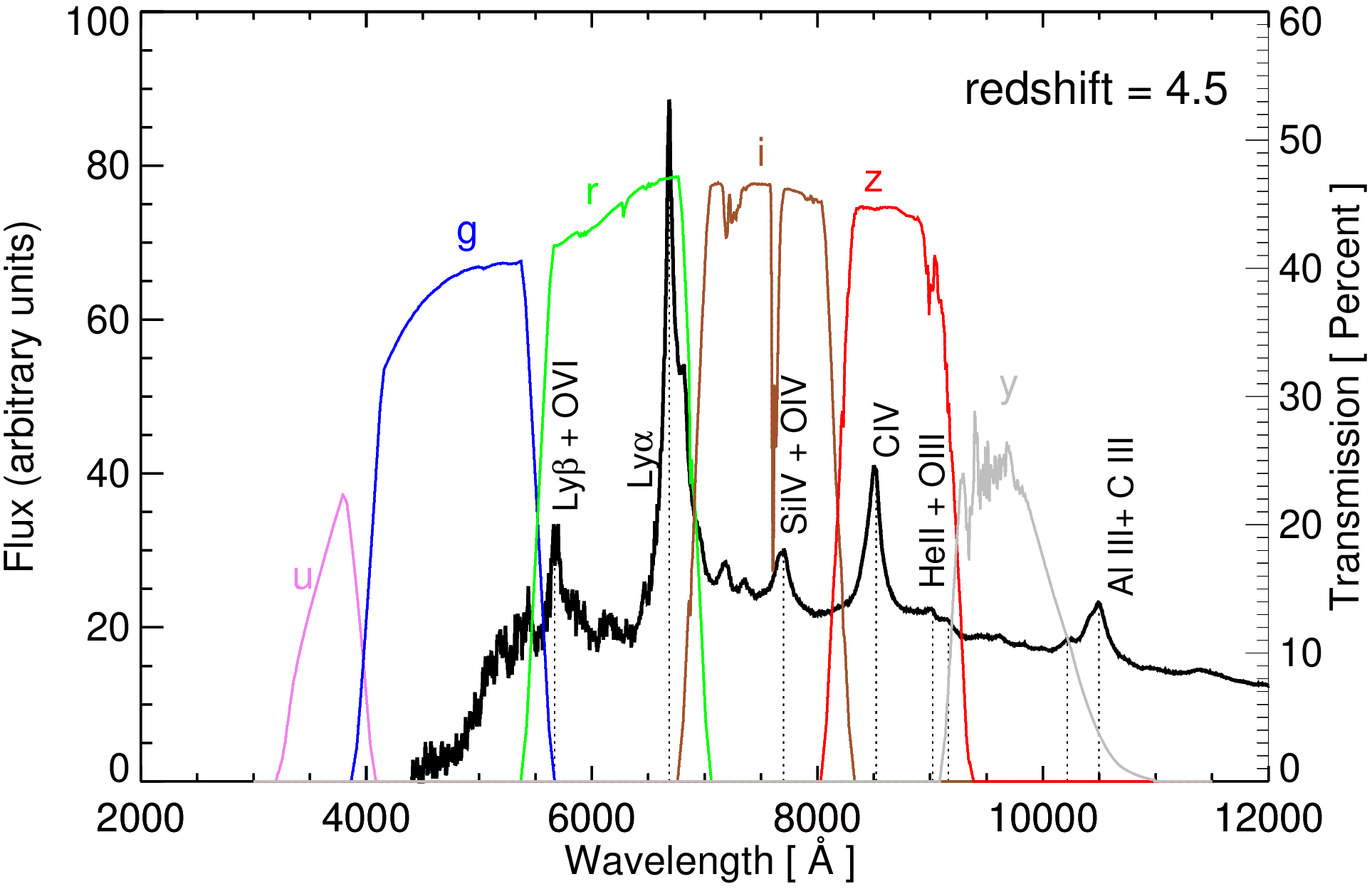}
   \includegraphics[width=0.67\columnwidth]{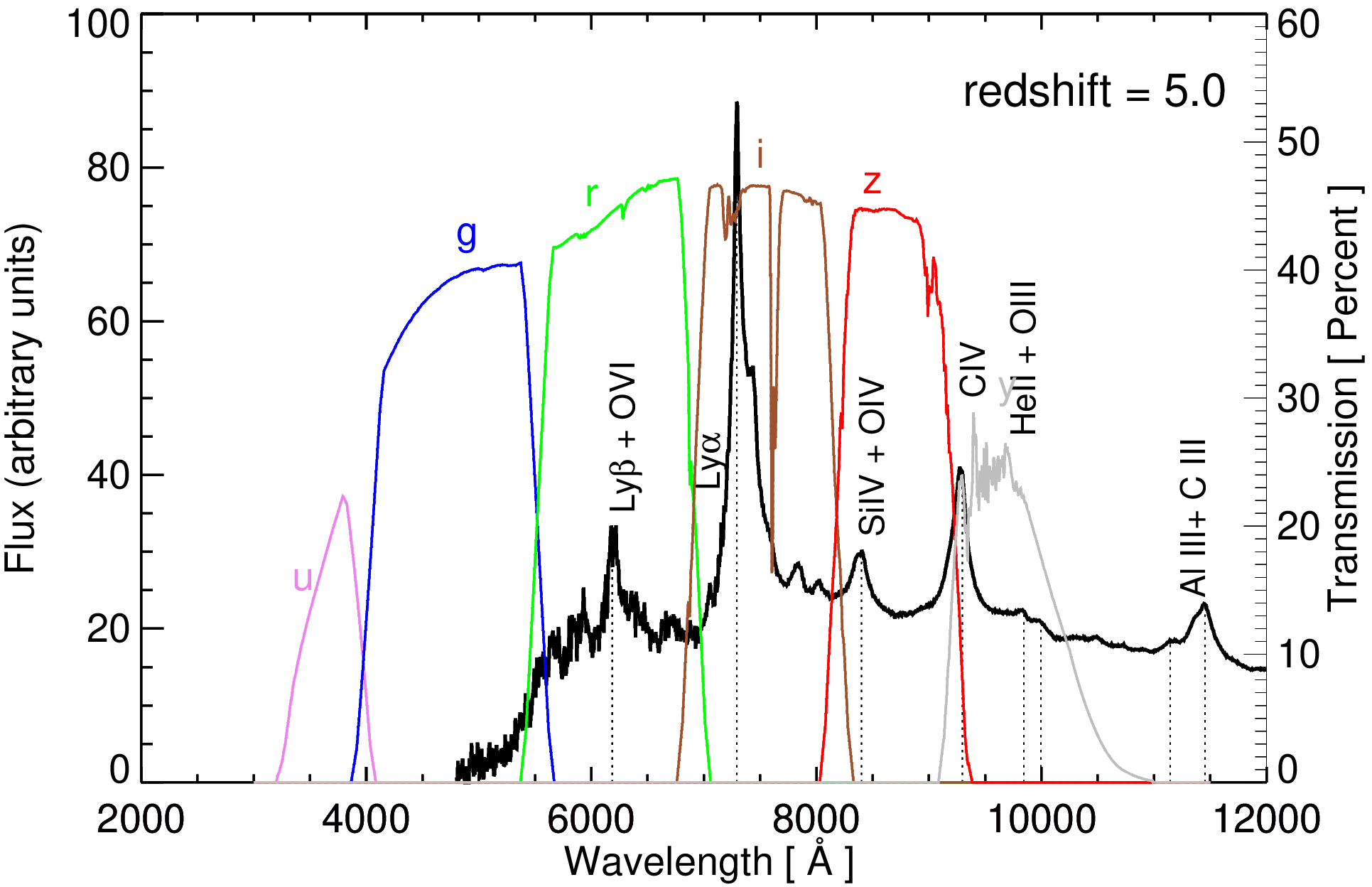}
   \includegraphics[width=0.67\columnwidth]{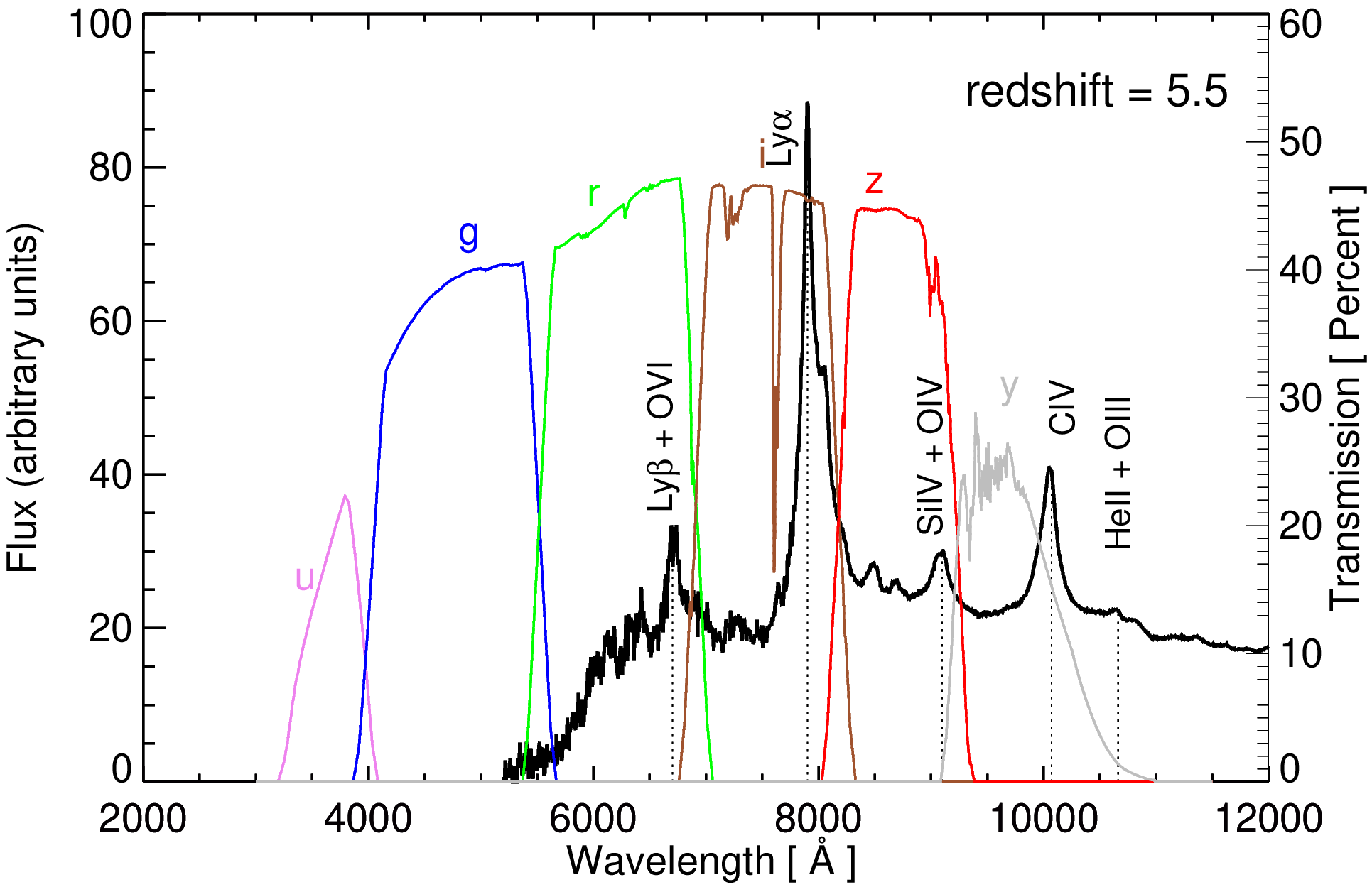}
   \caption{Composite quasar spectrum of \citet{2001AJ....122..549V} and \citet{2006ApJ...640..579G} overlaid with the transmission curves of the LSST photometric bands (ugrizy). For illustration, the spectrum is shown at different redshifts (0.0 $\leq$ z $\leq$ 5.5) including several emission features whose degree of contribution to the photometric bands depends on the selected redshift range.}
\label{speclines}
\end{figure*}

\begin{figure*}
  \centering
  \includegraphics[width=18cm,clip=true]{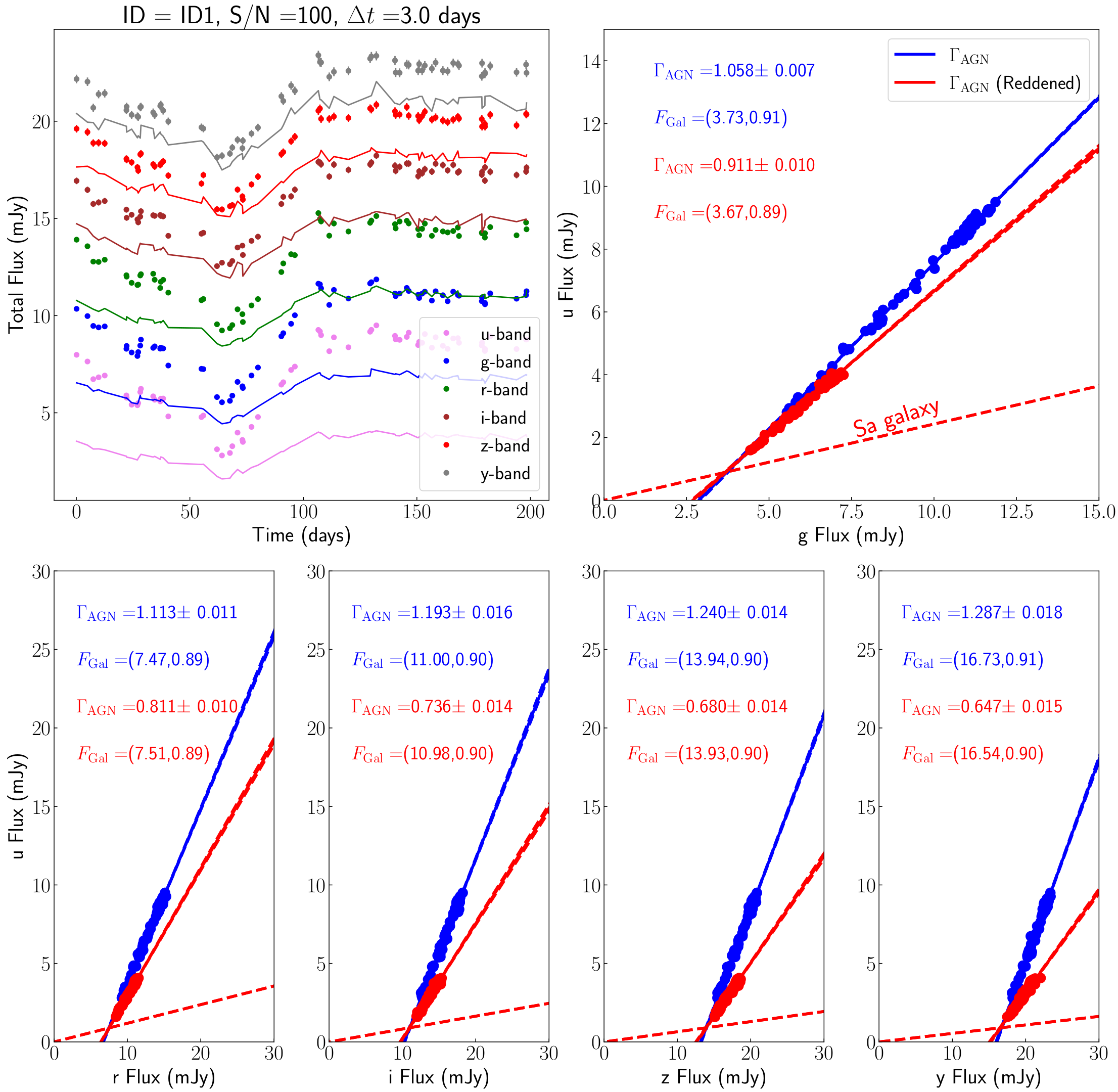}
  \caption{FVG for the LSST-simulated light curves from our mock catalog. The coloured solid lines represent the light curves including the host galaxy's contribution and internal reddening, whereas the dotted lines represent the light curves after corrections.}
\label{fvgext}
\end{figure*}


\bsp	
\label{lastpage}
\end{document}